\documentclass[prx,twocolumn,amsmath,amssymb,aps,longbibliography,superscriptaddress,citeautoscript,bibnotes,nofootinbib]{revtex4-1}

\usepackage[dvipsnames]{xcolor}
\usepackage[colorlinks=true,
			linkcolor=NavyBlue,
			citecolor=NavyBlue]{hyperref}         
\usepackage{tikz}

\newcommand{\ket}[1]{| #1 \rangle} % for Dirac bras
\newcommand{\bra}[1]{\langle #1 |} % for Dirac kets
\begin{document}

	\title{Superuniversality from disorder at two-dimensional topological phase transitions}
	
	\author{Byungmin Kang}
	\thanks{Electronic Address: bkang119@kias.re.kr}
	\affiliation{School of Physics, Korea Institute for Advanced Study, Seoul 02455, Korea}
	
	\author{S. A. Parameswaran}
	\affiliation{Rudolf Peierls Centre for Theoretical Physics,  Clarendon Laboratory, Oxford OX1 3PU, United Kingdom}
	
	\author{Andrew C. Potter}
	\affiliation{Center for Complex Quantum Systems, University of Texas at Austin, Austin, Texas 78712, USA}
	
	\author{Romain Vasseur}
	\affiliation{Department of Physics, University of Massachusetts, Amherst, Massachusetts 01003, USA}
	
	\author{Snir Gazit}
	\thanks{Electronic Address: snir.gazit@mail.huji.ac.il}
	\affiliation{Racah Institute of Physics and The Fritz Haber Research Center for Molecular Dynamics, Hebrew University of Jerusalem, Jerusalem 91904, Israel}
	
	\date{\today}
	
	\begin{abstract} 
		We investigate the effects of quenched randomness on  topological quantum phase transitions in strongly interacting two-dimensional systems. We focus first on transitions driven by  the condensation of a subset of fractionalized quasiparticles (`anyons') identified with `electric charge' excitations of a phase with intrinsic topological order. All other  anyons have nontrivial mutual statistics with the condensed subset  and hence become confined at the anyon condensation transition. Using a combination of microscopically exact duality transformations and asymptotically exact real-space renormalization group techniques applied to these two-dimensional disordered gauge theories, we argue that the resulting critical scaling behavior is `superuniversal' across a wide range of such condensation transitions, and is controlled by the same infinite-randomness fixed point as that of the 2D random transverse-field Ising model. We validate this claim using large-scale quantum Monte Carlo simulations that allow us to  extract  zero-temperature critical exponents and correlation functions in (2+1)D disordered interacting systems. We discuss generalizations of these results to a large class of ground-state and excited-state topological transitions in systems with intrinsic topological order as well as those where topological order is either protected or enriched by global symmetries. When the underlying topological order and the symmetry group are Abelian, our results provide prototypes for topological phase transitions between distinct many-body localized phases.
	\end{abstract}
	
	\maketitle

	\section{Introduction}
	Understanding the interplay of interparticle interactions and quenched disorder is particularly relevant to the study of  transitions into or between topologically ordered phases of matter.  These intrinsically interacting phases  cannot be characterized in terms of a local order parameter, and are instead described at long wavelengths by a topological quantum field theory that encodes the properties of the gapped fractionalized quasiparticles or {\it anyons}~\cite{kitaev2006anyons, RevModPhys.80.1083}, that are their most distinctive feature. Since they lack a local order parameter, topological orders are quite stable to quenched disorder, as the only natural variable to which disorder can couple is the energy density~\cite{PhysRevB.89.144201}. The existence of a bulk energy gap makes topological  phases perturbatively stable to this type of randomness (analogous to `bond disorder' in a magnet). The situation is more delicate and interesting at quantum critical points between trivial and topologically ordered phases: while the {\it perturbative} relevance of disorder is determined by the Harris criterion~\cite{harris1974effect, chayes1989correlation, chayes1986finite}, sufficiently strong randomness can completely change the universality class of the transition. In this paper, we argue that, remarkably, some topological phase transitions become analytically tractable in the limit of very strong disorder, even though the corresponding clean transitions are in general described by complicated, strongly-coupled gauge theories~\cite{burnell2018anyon, PhysRevB.96.155127, PhysRevB.100.245125}.
	
	Our treatment of the  strong-disorder limit is enabled by a combination of renormalization-group techniques, duality mappings, and exact reformulations that can be  tackled conveniently by state-of-the-art numerical algorithms. Specifically, we consider discrete gauge theories related to exactly solvable `quantum double models'~\cite{kitaev2006anyons} that describe phases with both Abelian and non-Abelian topological orders. (The latter nomenclature refers to the nature of the braid group that characterizes the anyon statistics in 2+1 space-time dimensions.) 
	
	A  generic topologically ordered phase is an emergent gauge theory with both dynamical `electric' gauge charge and `magnetic' gauge flux excitations, both of which are point objects in 2D. Confinement transitions can be driven either by condensation of gauge charges while gauge fluxes remain massive, or vice-versa~\cite{burnell2018anyon}. In the case of Abelian topological order, these two transitions are  equivalent and related by electric-magnetic-  duality (or `$S$-duality'), but in  non-Abelian topological orders charge and flux sectors behave differently~\cite{buerschaper2013electric}. In these cases, for reasons explained below, our approach remains tractable only when the confinement transition is driven by gauge charge condensation. 
	
	For this reason, we work in the sector with zero gauge flux, and add terms that give a finite bandwidth for the hopping of (a subset of) the anyons. This leads to a model that we describe in Sec.~\ref{sec:models}.
	When the  hopping bandwidth exceeds the anyon gap, the anyons condense, leading to confinement. When the couplings are highly random, we show that the critical physics is correctly captured via {a} strong-disorder real-space renormalization group (RSRG) procedure~\cite{PhysRevB.22.1305, PhysRevLett.69.534, PhysRevB.50.3799, PhysRevB.51.6411, Motrunich_2000, PhysRevB.82.054437, PhysRevX.4.011052, PhysRevLett.114.217201, PhysRevB.93.104205, PhysRevB.95.024205}. The RSRG, described in Sec.~\ref{sec:RSRG}, successively `decimates' individual local terms of the Hamiltonian in descending order of their magnitude while determining how this feeds back on the remaining couplings, to arrive at a fixed-point that characterizes the scaling behavior.  
	
	We find that our   Hamiltonian, despite being the most natural one for such discrete gauge theories --- and frequently employed in the literature --- is a special ``high-symmetry point'' in the space of gauge theories. As shown in Sec.~\ref{sec:duality} this point is related via a duality transformation to a $Q$-state random quantum Potts model, where $Q = |G|$ is the size of the finite group $G$ that characterizes the gauge theory.  Similar strong-disorder RSRG arguments suggest that the critical properties of the random quantum Potts model are $Q$-independent and identical to those controlled by the infinite-randomness critical point of the random transverse-field Ising model~\cite{Motrunich_2000, PhysRevB.82.054437}. In Sec.~\ref{sec:qmc} we  provide supporting evidence in favor of this long-standing conjecture by means  of high-precision stochastic series expansion quantum Monte Carlo (SSE-QMC) simulations~\cite{PhysRevB.43.5950, sandvik1992generalization} of strongly disordered lattice models at $Q=2,3$. We analyze the stability of the fixed-point behavior to perturbations of the model away from the special $|G|$-state Potts point in Sec.~\ref{sec:pert}.
	
	Our results indicate that the underlying algebraic properties of $G$ --- including whether it is Abelian or not --- are irrelevant to the critical properties of the deconfinement transitions we study here. Intuitively, this is because the transition is driven by the geometrical structure of the `clusters' generated by the strong-disorder RG scheme rather than the microscopic details of the onsite degrees of freedom. It is this underlying geometrical structure that is shared by all the topological phase transitions we study in this paper.   
	
	Viewing superuniversality as a consequence of  this geometrical structure naturally raises the question of whether it also emerges at  other topological phase transitions with quenched randomness. Global symmetries enlarge the phase structure of quantum systems, by both {\it protecting} and {\it enriching} topological structures. Symmetry-protected topological phases  (SPTs)~\cite{PhysRevB.80.155131,PhysRevB.84.235128,PhysRevB.83.075102,PhysRevB.83.075103,Chen2011b,Pollmann2012,YuanMing2012,PhysRevB.86.115109,Chen1604,PhysRevB.87.155114,doi:10.1146/annurev-conmatphys-031214-014740} are phases of matter that lack intrinsic topological order (i.e., they do not exhibit fractionalization) but are nevertheless distinct from trivial paramagnets in the presence of the eponymous symmetries. This is by virtue of their nontrivial local entanglement structure, which obstructs their deformation into trivial product states by local unitary transformations as long as a protecting global symmetry is unbroken. Symmetry-enriched topological phases (SETs)~\cite{PhysRevB.87.155115, PhysRevB.94.235136, PhysRevB.96.115107} are topologically ordered phases whose fractionalized quasiparticles carry  quantum numbers that capture their transformation properties under the global symmetries. SPTs/SETs typically also have gapless modes on symmetry-preserving boundaries with distinct SPTs/SETs or the trivial phase. When the global symmetries are broken, there is no longer any sharp distinction between these phases and their symmetry-less parents (for SPTs, this is the trivial paramagnetic vacuum state, and for SETs, the underlying topologically ordered phase). Therefore, phase transitions out of SPTs/SETs via spontaneous breaking of global symmetries involve a change in the short range entanglement  of their ground state, that encodes the symmetry-protected/enriched topological structure. This makes them formally distinct from `trivial' symmetry-breaking transitions, though the relevant distinguishing features are likely subtle and visible only in the boundary-critical behavior~\cite{PhysRevLett.118.087201,PhysRevX.7.041048,2019arXiv190506969V}. We find that our techniques can be applied to this class of symmetry-breaking transitions by leveraging a description of SETs and SPTs in terms of so-called `Dijkgraaf-Witten' theories~\cite{dijkgraaf1990topological, PhysRevB.87.125114}. These are `twisted'  discrete gauge theories linked to SPTs and SETs by a gauging procedure~\cite{PhysRevB.87.155115} applied to fluxes of the global symmetry. 

On a purely formal level, the universality class of the transition in  twisted gauge theories can be  obtained without  extra effort, via a  unitary transformation that maps between a Dijkgraaf-Witten gauge theory and its conventional, untwisted counterparts~\cite{2020arXiv1506.00592}\footnote{Strictly speaking, this is only true in the bulk, or on manifolds without boundaries; see the discussion in Sections~\ref{sec:DW-models} and ~\ref{sec:summ}.}. Since we have already analyzed the untwisted discrete gauge theory, and a unitary transformation does not affect the scaling of thermodynamic quantities, the result follows.  However, as we have emphasized, several interesting aspects of of infinite-randomness fixed points are tied to the underlying geometric structure produced by the RSRG decimations. \textit{A priori}, the twisting unitaries have nontrivial interplay with the flow of lattice geometry induced by the RSRG. Furthermore, local observables can be transformed nontrivally by the unitary, and hence correlation functions (rather than thermodynamics) can be difficult to access using the mapping. Therefore,  while one can infer the critical scaling behavior based on such arguments, more complex questions are often more transparently addressed by a direct implementation of the RSRG in terms of the original degrees of freedom. 

In fact, a similar situation arises already in the untwisted case: while it is possible to infer critical bulk scaling behavior by using  the Potts duality to the magnetic model, implementing the RSRG directly in the gauge language is necessary to address various questions ---  such as the stability to perturbations and the scaling of local observables ---  in a physically transparent manner. In that case, we show that individual RSRG decimations and duality transformations form a ``commuting diagram’’ so that one can freely map between gauge and symmetry-breaking models at any stage of the RG scheme (see Sec.~\ref{sec:duality} for details). This  gives us a way to implement RSRG directly on the gauge theory side, while also allowing us to lift existing results on the transverse field Ising model and simulation techniques tailored for the Potts/magnetic side. Similarly, it is desirable to directly implement the RSRG on the twisted Dijkgraaf-Witten model, rather than using the trick off unitary `untwisting', in order to establish RSRG as a technique for studying these models in the presence of strong disorder. 

	Accordingly, in Sec.~\ref{sec:DW-models}, we construct a generalization of the strong-disorder RG treatment that correctly tracks the nontrivial phase factors that encode the twisted Dijkgraaf-Witten gauge structure. Despite a more complicated decimation procedure, we find that the RG flows are identical to the untwisted case. We thereby provide a \textit{direct} argument that the superuniversal infinite-randomness scaling also applies to global symmetry-breaking phase transitions in SPTs/SETs, complementing the simpler but less instructive unitary untwisting trick. [Note that we do {\it not} study a different class of  quantum critical points between  SPTs/SETs that are distinguished only in the presence of the global symmetry. This remains an active question of research even in the clean case \cite{2020arXiv200806509D}, but we conjecture that disorder would play a similarly rich role at such transitions.] 
	
	While extremely general, the discrete gauge theories studied here do not encompass all possible topological orders in two dimensions. These are more generally captured in the language of modular tensor categories which can be cast into a Hamiltonian framework in so-called ``Levin-Wen'' models~\cite{PhysRevB.71.045110, PhysRevB.94.235136, PhysRevB.96.115107} in the case of nonchiral topological orders. In Sec.~\ref{sec:LW}  we discuss the possibility of generalizing our approach to apply to these models. Although we are unable to fully implement the RG scheme, we flag  the emergence of some universal structure as an avenue for further work. We also note that chiral phases such as integer and fractional quantum Hall states will not be considered here. It has been recently shown that disorder and interactions in chiral phases show some interesting superuniversal criticality~\cite{2020arXiv2006.11862K}, albeit different from ours.
	
	A final generalization that we explore (in Sec.~\ref{sec:RSRG-X}) is to excited-state phase transitions between distinct many-body localized  (MBL) topological phases of matter. {In an isolated quantum system, strong disorder can prevent thermalization, leading to MBL states that can support quantum orders in highly excited states. While the existence of MBL has been firmly established in one dimension~\cite{imbrie2016many}, its fate in two dimensions remains controversial~\cite{PhysRevB.95.155129, PhysRevB.99.205149}: the rare thermal inclusions believed to drive the 1D MBL transition have a more drastic role in $d>1$ where they have been argued to  destabilize MBL even with arbitrarily strong disorder~\cite{PhysRevB.95.155129} (though quasiperiodic MBL systems may evade this fate~\cite{PhysRevB.87.134202}). However, this mechanism operates on a timescale that is double-exponentially long in the disorder strength. Therefore, although MBL may sharply exist only in the limit of infinite disorder strength or zero correlation length in $d>1$, it can provide an excellent approximation of the dynamics of $2$D systems up to extremely long timescales. 

	We argue that this MBL-controlled regime may be understood by studying extensions of the disordered  lattice gauge models to excited state dynamics. We focus on the Abelian case and examine excited-state properties using the generalization of the RSRG procedure that targets excited states, termed RSRG-X~\cite{PhysRevX.4.011052, PhysRevLett.114.217201, PhysRevB.93.104205, PhysRevB.95.024205}. The RSRG-X rules in the Abelian cases are essentially identical to RSRG rules, which is in accordance with the belief that MBL is in general compatible with Abelian topological order~\cite{PhysRevB.94.224206}. The corresponding infinite-randomness fixed point characterizes the excited-state transition (or sharp dynamical crossover, if MBL is indeed destroyed in $d>1$) between trivial and topological MBL phases. The stability of MBL in two dimensions against thermalization remains a subject of debate; this is perhaps even more true of the question of whether excited-state transitions between MBL phases can be nonthermal~\cite{2020arXiv200808585S,2020arXiv200809113M}. However, RSRG-X controls at least the physics of  intermediate length scales and timescales at sufficiently strong disorder, and hence remains a useful tool even when transitions are rounded to crossovers. For non-Abelian models, the excited states are expected on general grounds to exhibit topologically protected degeneracies that cannot be localized in the tensor product fashion necessary for the conventional definition of MBL~\cite{PhysRevB.94.224206}  (Such phases could nevertheless exhibit more complicated critical nonthermal behavior~\cite{PhysRevB.85.161301, PhysRevB.85.224201, PhysRevLett.114.217201, PhysRevB.95.024205}; while intriguing, we do not explore this possibility as it lies beyond the scope of the present work).
		
		We close in Sec.~\ref{sec:summ} with a summary of results and a discussion of future directions motivated by this work. Four appendices  provide brief primers on group cohomology and Dijkgraaf-Witten models, as well as technical details of lattice isomorphisms, Levin-Wen models, and numerical methods.

%\tableofcontents		

		\section{Gauge Model}
		\label{sec:models}
		In this paper, we will focus primarily on {discrete gauge theories}~\cite{kitaev2006anyons} defined on a planar graph $\mathcal{G}=(\mathcal{V},\mathcal{L})$ consisting of vertices $v\in  \mathcal{V}$ connected by directed links $l\in \mathcal{L}$. We denote $l= \langle v, v' \rangle$ if the link $l$ is directed from the vertex $v \in \mathcal{V}$ to the vertex $v' \in \mathcal{V}$. The degrees of freedom (sometimes termed the \textit{computational basis}) are elements $g$ of a discrete group $G$ placed on the links. We adopt a convention where traversing $l$ along (against)} its orientation gives a factor of $g_l$ ($g_l^{-1}$) respectively.
		We consider the following Hamiltonian:
		\begin{equation}
		H = - \sum_{v\in \mathcal{V}} J_v A_v - \sum_{l \in \mathcal{L}} h_l C_l.
		\label{eq:Ham_topological}
		\end{equation}

\begin{figure}[tt]
			\centering
			\includegraphics[width=0.9\columnwidth]{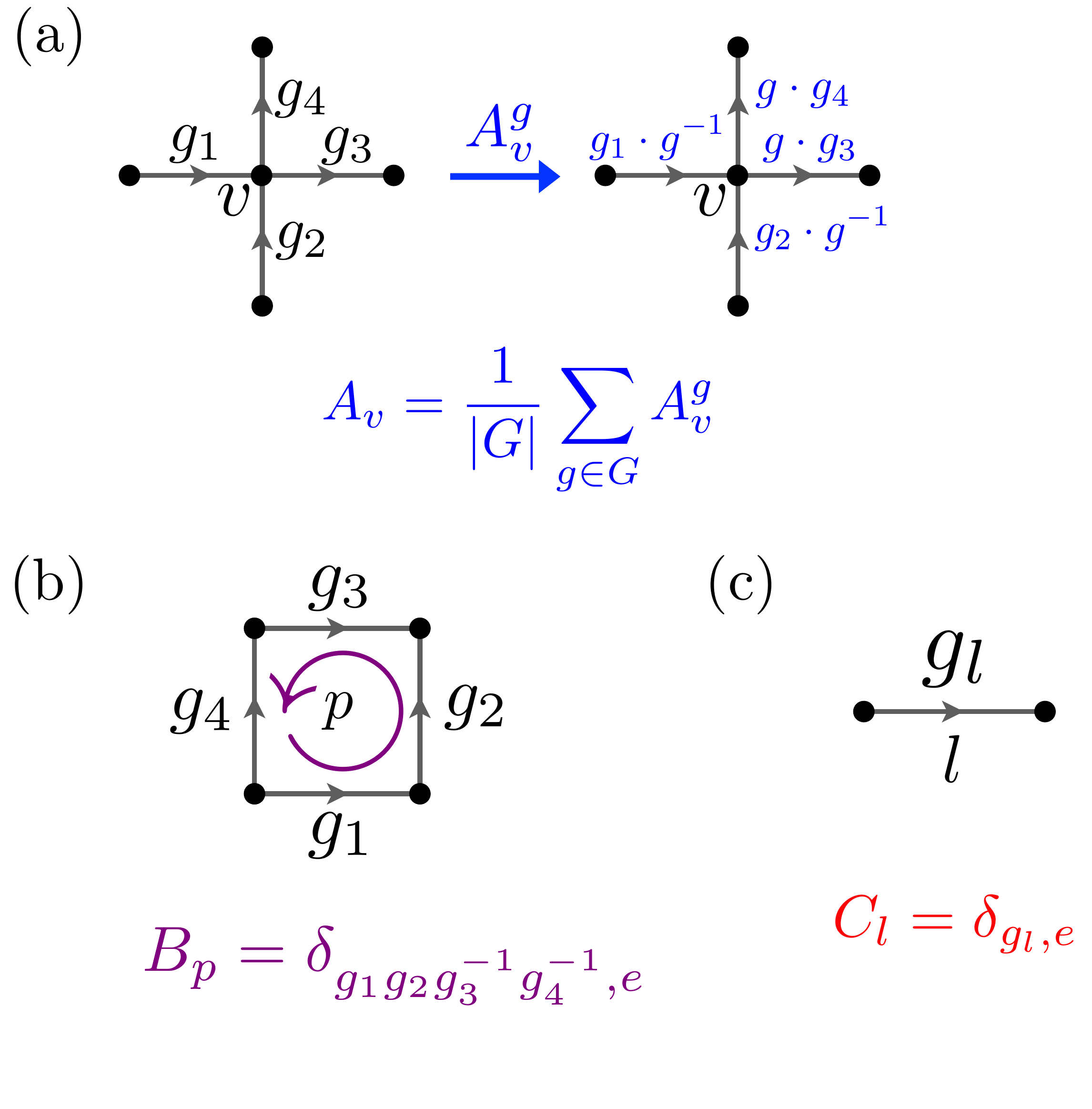}
			\caption{Terms in the Hamiltonian \eqref{eq:Ham_topological} and the gauge constraints acting on group elements $g\in G$ on links. (a) The vertex term $A_v = \sum_g \frac{1}{|G|} A_v^g$ symmetrically permutes  group elements around vertex $v$. (b) The plaquette operator  $B_p$ projects onto `zero flux' configurations in which the oriented product of group elements around plaquette $p$ is the identity $e$. (c) The link term $C_l$ favors $g_l=e$ on link $l$.  We impose gauge constraints that restrict to zero-flux configurations ({$\Phi_p=e$ on all plaquettes}) and take $H = - \sum_v J_v A_v - \sum_l h_l C_l$.}
			\label{fig:QD-basics}
		\end{figure}

		Here, the non-negative coupling constants $J_v, h_l\geq 0$ are independent and identically distributed (i.i.d.) random variables, with finite variance. The vertex term is given by
		\begin{equation}A_v = \sum_{g\in G} \frac{1}{|G|} A_v^g,
		\end{equation}
		where the operators $A_v^g$ act on each link $l$ with state $g_l$ that leaves (enters) the vertex $v$ by multiplying $g_l$ on the left by $g$ (right by $g^{-1}$), see Fig.~\ref{fig:QD-basics} (a). Effectively, $A_v$  implements a symmetric permutation of group elements on the links adjacent to vertex $v$.  Each link term
		\begin{equation}
		C_l \vert g \rangle_l=\delta_{g,e} \vert g \rangle_l
		\end{equation}
		projects the state on link $l$ to the identity element $e$. More generally, we can define $C_l^g$ as a link projector to the group element $g$, in particular, $C_l^e=C_l$.
		
		Physically, we can view $A_v$ as a projector onto the state with zero gauge charge on vertex $v$, where the gauge charges correspond to the irreducible representations (irreps) of the gauge group, and the zero charge sector is identified with the trivial irrep. Similarly, the term $C_l$ on a link $l$ between $v$ and $v'$ creates a superposition of all possible charges on $v$ with corresponding opposite charge on $v'$, with an amplitude proportional to the quantum dimension of the charge~\cite{kitaev2006anyons, PhysRevB.78.115421}. In the presence of an existing charge on a vertex $v$, $C_l$ acts as a tunneling operator that moves this charge to a neighboring site. In the absence of gauge charge, it creates a pair of opposite charges at $v$ and $v'$. Note that since both $A_v$ and $C_l$ are projectors, they satisfy $A_v^2 = A_v$, $C_l^2 =C_l$. We will make  extensive use of this property below.

		To properly define a gauge theory, we restrict the total Hilbert space to the set of physical states $\ket{\Psi}$ satisfying the {zero-gauge flux constraint 
		\begin{equation}\label{eq:nofluxconstraint}
		B_p\ket{\Psi} = \ket{\Psi}
		\end{equation}
		for all plaquettes $p$. Here, { $B_p = \delta_{\Phi_p, e}$} where the operator 
		\begin{equation}
		\Phi_{p} = \prod_{\l \in p} g_l^{\sigma_l}
		\end{equation}
		measures the $G$-flux through plaquette $p$}, which is computed by moving around  $p$ counter-clockwise and picking up a multiplicative factor ${g_l}^{\sigma_l}$ from each link with $\sigma_l = +1$ $(-1)$ if the link is traversed parallel (antiparallel) to its orientation [see Fig.~\ref{fig:QD-basics} (b) for an example]. The constraint Eq.~\eqref{eq:nofluxconstraint}  thus projects onto states where the $G$-flux through the plaquette equals the identity element. In other words, it is a Gauss law restricting the set of physical configurations to those with zero flux through every plaquette: enforcing it forbids anyonic flux excitations. All possible charge excitations, on the other hand, are allowed.
		
		We note that our model, Eq.~\eqref{eq:Ham_topological}, can be viewed as a limiting case of the exactly solvable
		quantum double models introduced by Kitaev~\cite{kitaev2006anyons} to describe systems with topologically ordered ground states. This can be seen by setting $h_l=0$ and weakening the Gauss law to allow finite-energy flux excitations by adding a term of the form $\sum_p J_p B_p$ with $B_p = \delta_{\Phi_p,e}$. This is permitted since the plaquette term commutes with all vertex and link terms.

		For large, positive $J_v\gg h_l$, the system is in a deconfined phase where gauge charge excitations are gapped and deconfined (gauge fluxes are explicitly excluded by the constraint Eq.~\eqref{eq:nofluxconstraint}). In the opposite limit, $J_v\ll h_l$, a confined phase is expected, whose ground state is a  condensate that quantum-fluctuates between different charge configurations. The physical manifestation of confinement is that any pair of test flux excitations (i.e., plaquettes carrying nonvanishing gauge flux) experience linear confinement due to their nontrivial mutual statistics with the charge condensate.\footnote{Note that this diagnostic is strictly speaking only valid in the limit where the fluxes are non-dynamical; if there are dynamical fluxes, more subtle probes may be necessary~\cite{Gregor_2011}.}
		
		We now comment briefly on our choice of model Eq.~\eqref{eq:Ham_topological} and its generality.
		
		First, our imposition of a zero-flux constraint is slightly unconventional. Typically, the `pure gauge' limit of the quantum double model is obtained by a zero-{\it  charge} constraint, which is implemented by imposing the local constraint $A_v\ket{\Psi} = \ket{\Psi}$ at every vertex $v$ on physical states~\cite{RevModPhys.51.659}. For Abelian $G$, the two choices are formally equivalent via an analog of electromagnetic duality, but for non-Abelian theories this is not the case~\cite{buerschaper2013electric}. In the absence of such a duality relation, our model can describe only confinement transitions that are driven by a gauge charge condensation. [In gauge theory language, we are describing a transition to a Higgs phase with flux confinement, rather than to a phase with charge confinement.] As will become clear later, considering a zero-flux constraint is crucial in order to render the RSRG scheme tractable. Since this is what allows us to make analytical progress,  we restrict ourselves to this choice from the outset.
		
		Second, we note that in Equation~\eqref{eq:Ham_topological} one can consider more complicated gauge-invariant vertex terms of the form $A^\chi_v\equiv \sum_{g\in G} \frac{\chi(g)}{|G|} A_v^g$, where the sum is over different group characters $\chi$.  Eq.~\eqref{eq:Ham_topological} is the simplest choice, involving only the trivial character $\chi(g)=1$. More complicated choices would correspond to assigning different energy penalties to different gauge charges on a vertex. (A Hamiltonian corresponding to an arbitrary assignment of energy penalties to the gauge charges can be expressed as a linear combination of the $A^\chi_v$s.) The stability of the infinite-disorder fixed point with respect to vertex terms with nontrivial characters is discussed in Sec.~\ref{sec:pert}. We find that, at least for the Abelian case, such terms do not appear to change universal properties of the confinement transition.

		\section{Real-Space Renormalization and Scaling at  Strong Disorder} 
		\label{sec:RSRG}
		\subsection{Real-Space Renormalization Group}
		For strongly random $J_l, h_v$, we may  access ground-state properties of $H$ by leveraging the strong disorder real-space renormalization group (RSRG)~\cite{PhysRevB.22.1305, PhysRevLett.69.534, PhysRevB.50.3799, PhysRevB.51.6411, Motrunich_2000}. The implementation of the scheme for our model proceeds iteratively as follows. At any stage of the RG, we identify the local term with the strongest remaining coupling in $H$. We \textit{decimate} this term by projecting the associated degrees of freedom into the configuration that minimizes its energy in isolation, and determine a new  effective $H$ by perturbatively  computing how virtual excitations of the frozen degrees of freedom mediate interactions between the remaining ones. In addition, as we detail below, we must also renormalize the lattice structure, $\mathcal{G}\mapsto\mathcal{G}'$, by removing a link or a vertex or both locally, while preserving the planar nature of our problem. This perturbative scheme is self-consistently justified if (as we expect on physical grounds) this RG flows to an infinite-disorder fixed point~\cite{Motrunich_2000, PhysRevB.82.054437}; our results and scalings arguments  then become asymptotically exact as this fixed point is approached.
		
		As will become clear below, the decimation procedure generates `long-link' terms associated with pairs of vertices, $v$ and $v'$ that are {\it not} nearest-neighbors.  In order to define $C_l$ for such long links, $l = \langle v, v' \rangle$, we first construct a directed path $\pi_l$ from $v$ to $v'$ that consists entirely of links $l_i\in \mathcal{L}$ and define $g_{\pi_l}= \prod_{l_i\in\pi_l} g_i^{\sigma_i}$, where $g_i$ is the link state at $l_i$, $\sigma_i=+1 (-1)$ if the orientation of $l_i$ is parallel (antiparallel) to the direction of $\pi_l$, and take $C_l$ to project onto configurations where $g_{\pi_l}=e$. Though the choice of  $\pi_l$ is ambiguous, $C_l$ is well-defined because of the zero-flux condition\footnote{For simplicity, we assume that the holonomy along a nontrivial cycle is always trivial.} on each plaquette in $\mathcal{G}$ (as elaborated in Appendix~\ref{app:lattice-iso}). 
		
		To keep track of long-link terms, we must generalize the previously introduced set of  short links $\mathcal{L}$ belonging to the planar graph $\mathcal{G}$.  For a given vertex list, $\mathcal{V}$, we define the set of ordered vertex pairs $\mathcal{L}_\mathcal{V}=\{l = \langle v, v' \rangle\ | v, v' \in \mathcal{V}\}$, which also includes long links that are not present in the planar graph $\mathcal{G}$.  With the above definitions, we generalize the Hamiltonian in Eq.~\eqref{eq:Ham_topological} to,
		\begin{equation}
		H = - \sum_{v\in \mathcal{V}} J_v A_v - \sum_{l \in\mathcal{L}_\mathcal{V}} h_l C_l.
		\label{eq:Ham_topological_mod}
		\end{equation}
		
		Instead of treating a long link as a path defined on the planar graph $\mathcal{G}$, one can imagine a direct link connecting a pair of higher-neighbor vertices. This point of view deviates from the planar structure of the lattice.  We emphasize that to properly define our gauge theory all degrees of freedom must reside on the planar graph $\mathcal{G}$. Specifically, the set $\mathcal{L}_\mathcal{V}$ is used only to keep track of Hamiltonian terms and does not introduce spurious degrees of freedom associated with nonplanar links. At the initial stage of our RG procedure, the coupling constants $h_l$ are nonvanishing only on short links ($l\in\mathcal{L}$), but as the RG proceeds, higher-neighbor interactions are generated, such that $h_l$ is potentially finite also on long-links.
		
				A key step of the decimation procedure is the construction of an effective low-energy Hamiltonian for the residual degrees of freedom, that includes new interactions mediated by virtual fluctuations of the high-energy degrees of freedom that have been `integrated out' in the decimation step. The relevant perturbative calculation is most conveniently accomplished via a Schrieffer-Wolff transformation. Let us define $P$  to be a projector ($C_\ell$ or $A_v$) onto the ground state of the decimated terms in $H$, and $V$ the terms (assumed small)  that couples the decimated degrees of freedom to the remainder of the system. A standard application of the Schrieffer-Wolff procedure (see, e.g., Ref.~\cite{PhysRevX.4.011052}) then shows that terms appearing at first order in perturbation theory will be proportional to $PVP$, and second-order terms will be proportional to $PV(1-P)VP$. Higher-order contributions will be asymptotically irrelevant, since variance of the coupling distributions will grow without bound under the RSRG flow at an infinite-randomness fixed point, so that in late stages of the RSRG (which determine the universal properties), the decimated coupling is asymptotically infinitely stronger than local competitors. For our purposes, it will be sufficient to consider only those second-order contributions of the form $PVVP$, since for the models in this paper, unless explicitly stated otherwise the operator $PVPVP$ acts trivially on the ground-state manifold.

		We now turn to construct our RSRG rules. We first consider decimating a nearest-neighbor link term $h_l C_l$,  see Fig.~\ref{fig:H_and_decimations} (a). Projecting the associated link degree of freedom,  $g_l$, to the trivial group element $e$ renormalizes coupling constants belonging to adjacent vertex terms. The first order contribution in perturbation theory, 
		\begin{equation}
		C_l ( A_v ) C_l = \frac{1}{|G|} \sum_{g \in G} C_l ( A_v^g ) C_l = \frac{1}{|G|} C_l,
		\label{eq:Cl-Av-Cl}
		\end{equation}
		gives only  a trivial constant term. 
		
		To see how nontrivial terms are generated from second-order perturbation theory, we consider a product of two vertex terms $A_v$ and $A_{v'}$ acting on the vertex pair $v$ and $v'$, connected by $l$, which yields
		\begin{align}
			&C_l ( A_v A_{v'} ) C_l = \frac{1}{|G|^2} \sum_{g,g' \in G} C_l ( A_v^g A_{v'}^{g'} ) C_l   \nonumber \\
			&= \frac{1}{|G|^2} \sum_{g \in G} C_l ( A_v^g A_{v'}^{g} ) C_l = \frac{1}{|G|} \bigg( \frac{1}{|G|} \sum_{g \in G} A_v^g A_{v'}^g \bigg) C_l,
			\label{eq:sec_ord_link} 
		\end{align}
		where we demand $g = g'$ in order to preserve $\vert e \rangle$ at the strong-link $l$ and use the fact that $A_v^g A_{v'}^g$ commutes with $C_l$. As the final stage of the link decimation step we renormalize the planar graph $\mathcal{G}\to\mathcal{G}'$ by removing the link $l$ and merging the adjacent vertices $v, v'$  into a new vertex $\tilde{v}$ with a renormalized coupling constant,
		\begin{equation}
		J_{\tilde{v}} A_{\tilde{v}} = \frac{2}{|G|}\frac{J_v J_{v'}}{h_l} A_{\tilde{v}}.
		\label{eq:J_renorm}
		\end{equation}
		Link decimation does not introduce any higher-neighbor couplings.
		
		Next, we consider decimating a vertex term $J_v A_v$, see Fig.~\ref{fig:H_and_decimations} (b). The first-order contribution in perturbation theory originates from link terms $C_l$, where $l$ starts on the vertex $v$, and is proportional to
		\begin{align}
			A_v (C_l) A_v &= \frac{1}{|G|^2}\sum_{g,g'\in G} A_v^{g'} C_l^e A_v^{g}
			 =  \frac{1}{|G|^2}\sum_{g,g'\in G} A_v^{g'\cdot g} C_l^{g^{-1}} 
			\nonumber\\&= 
			 \frac{1}{|G|^2}\sum_{h,h'\in G} A_v^{h} C_l^{h'} 
			=	\frac{1}{|G|}A_v
			\label{eq:Av-Cl-Av}
		\end{align}
		where in the second line we have relabeled $h=g'\cdot g$, and $h'=g^{-1}$, and used $\sum_{g \in G} C_l^g=\openone_l $.

An identical result holds also for link terms $C_l$ with $l$ ending on $v$. We see that, as with link term decimation, the first order contribution generates only trivial terms.
		
		To compute terms generated in second-order perturbation theory, we first identify all distinct link pairs, $l_i,l_j$ adjacent to $v$. Each such pair defines a long link $l_{ij}$ associated with the planar graph path $\pi_{l_{ij}}$ built up from the union of $l_i$ and $l_j$, see Fig.~\ref{fig:H_and_decimations} (b). Since we can always modify the path $\pi_{l_{ij}}$ so that it never passes $v$, it follows that $[A_v, C_{l_{ij}}] = 0$. Using the relation $C_{l_i} C_{l_j} = C_{l_{ij}} C_{l_j}$, we obtain,
		\begin{equation}
		A_v ( C_{l_i} C_{l_j} ) A_v = C_{l_{ij}} A_v ( C_{l_j} ) A_v = \frac{1}{|G|} C_{l_{ij}} A_v,
		\label{eq:sec_ord_vert}
		\end{equation}
		so that second-order perturbation theory gives rise to a long-link term
		\begin{equation}
		h_{l_{ij}} C_{l_{ij}} =\frac{2}{|G|}\frac{h_{l_i} h_{l_j} }{J_v} C_{l_{ij}} .
		\label{eq:h_renorm}
		\end{equation}
		We conclude the vertex decimation step by renormalizing the planar graph, $\mathcal{G}\to\mathcal{G}'$. This is achieved by removing the vertex $v$ from $\mathcal{V}$ and all its adjacent links from $\mathcal{L}$. {In addition, we  add to $\mathcal{L}$ all newly generated links that preserve the planar structure of our graph. While this procedure is not unique, the associated ambiguity will not affect the RG flow, i.e., the flow of coupling terms is invariant to the specific choice of added links. The above two steps correspond to the lattice isomorphisms $\hat{T}_3$ and $\hat{T}_2$, respectively, as we detail in Appendix~\ref{app:lattice-iso}. Newly generated links may appear to introduce spurious degrees of freedom. To see why this is not the case, note that one can always use the zero-flux constraint to fix the associated link variables, so that they are not dynamical degrees of freedom. The link decimation step, unlike a vertex decimation, generically produces higher-neighbor link terms. As previously commented, this necessitates introducing the generalized model in Eq.~\eqref{eq:Ham_topological_mod}, which allows for such long-link terms. 
			
			\begin{figure}[tt]
				\centering
				\includegraphics[width=0.8\columnwidth]{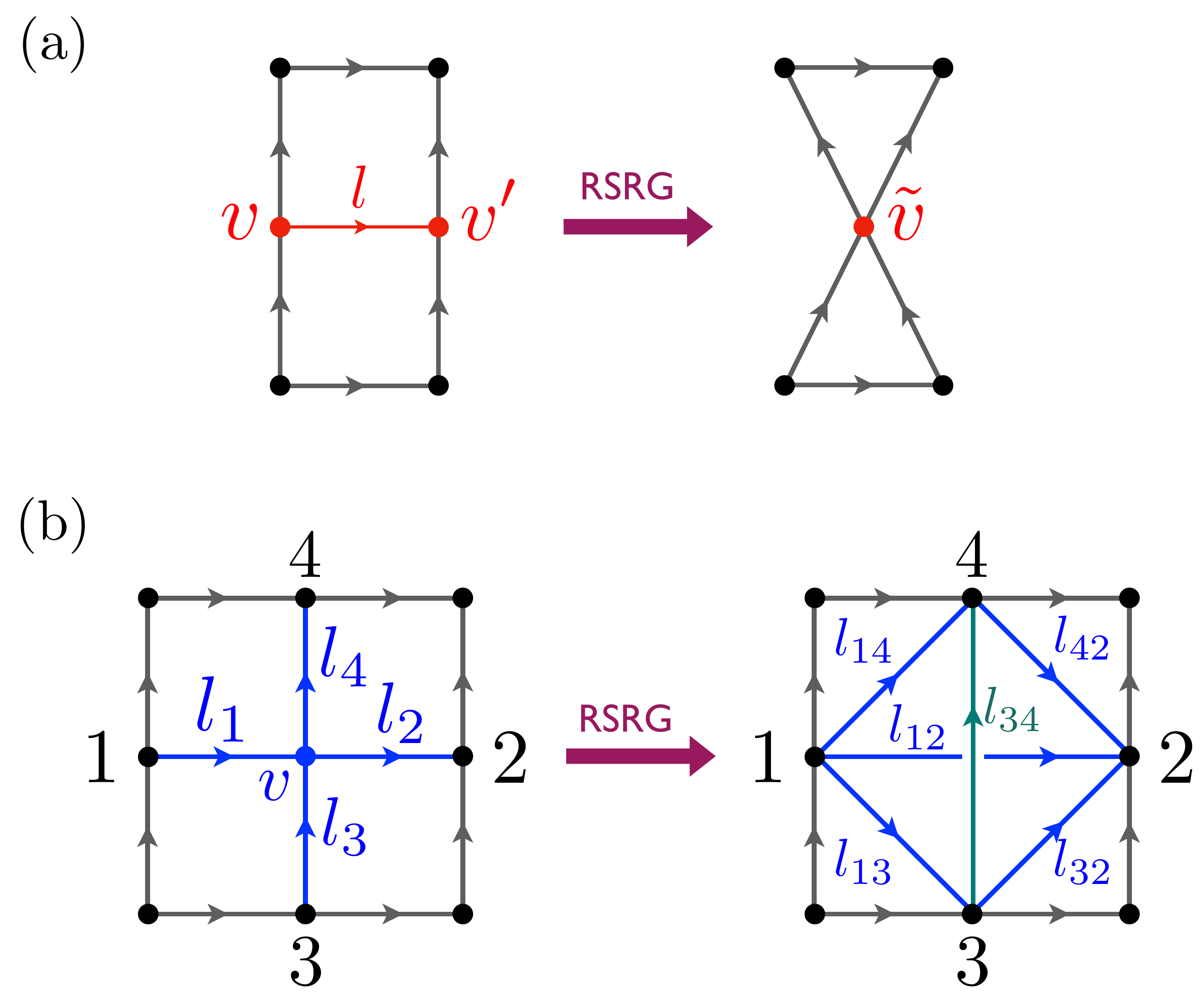}
				\caption{Our RSRG scheme involves two types of decimations for strongly random $J_v, h_l$: (a)~decimating link $l$ adjacent to vertices $v, v'$ yields a new vertex $\tilde{v}$ with $J_{\tilde{v}} = \frac{2}{|G|}\frac{J_v J_{v'}}{h_l}$; (b) decimating vertex $v$ adjacent to links $l_1,\ldots, l_4$ yields  $h_{l_{ij}}=\frac{2}{|G|}\frac{h_{l_i} h_{l_j} }{J_v}$ on both blue and cyan links. While blue links ($l_{12}, l_{13}, \cdots, l_{42}$) are short links in the renormalized planar graph $\mathcal{G}'$, the cyan link ($l_{34}$) is a long link and appears only through the term in the renormalized Hamiltonian. While the specific assignment of short versus long links is not unique, the specific choice does not affect the resulting RG flow. }
				\label{fig:H_and_decimations}
			\end{figure}
			
			As the RG proceeds, we eventually need to take into account long-link terms when decimating either vertex or link terms. A long-link $l$ is potentially affected by decimating a vertex term,  $A_v$,  when (1) it passes $v$ as an intermediate vertex or (2) ends or starts on $v$. For (1), one can always use the zero-flux constraints to deform the path $\pi_l$ so that it never passes $v$ and the links adjacent to $v$. For (2), one can explicitly check that the vertex decimation procedure remains identical as before. 
			
			In order to decimate a long-link term, we initially perform a series of lattice isomorphisms under which it is mapped to a nearest-neighbor link term, as we explain in Appendix~\ref{app:lattice-iso}. (A lattice isomorphism is a map that modifies the connectivity of the graph and thus the associated Hilbert space, and preserves the terms in the Hamiltonian while altering the precise link structure. It thus relates two different representations of the same gauge model.) Following that, we can simply use the nearest-neighbor link decimation rule outlined above. We emphasize that the Hilbert space mapping induced by the lattice isomorphism leaves the coupling constants intact.
			
			Each decimation renormalizes both the planar graph $\mathcal{G}$ and terms in $H$ according to the rules outlined above. Iterating this procedure generates a flow in the space of couplings. For sufficiently strong disorder, we anticipate that the flow will be to stronger disorder, i.e., the distributions of couplings get progressively broader as the RG proceeds, which makes the decimation procedure asymptotically exact at the critical point. We now turn to the universal scaling behavior implied by this expectation.

			\subsection{Super-universal scaling at strong disorder}
			Owing to the multiplicative nature of the RSRG updates, it is convenient to work with  logarithmic couplings
			\begin{align}
			 	h_v = \Omega e^{-\beta_v},~~ J_l = \Omega e^{-\zeta_l}
				\label{eq:rg_variables}
			\end{align}
defined at RG energy scale $\Omega = \Omega_0e^{-\Gamma}$, where $\Omega_0$ is a microscopic energy scale~\cite{PhysRevB.22.1305, PhysRevLett.69.534, PhysRevB.50.3799, PhysRevB.51.6411, Motrunich_2000}. Building on  previously studied examples of infinite-randomness criticality, we will argue that the following scaling properties hold at the infinite-randomness confinement-deconfinement critical point: (1)~ the coupling distributions $R(\beta;\Gamma), P(\zeta;\Gamma)$ of the $h$ and $J$ couplings flow to broad power-law scaling forms; (2)~critical fluctuations are governed by a dynamical scaling exponent $z=\infty$, corresponding to a logarithmic length-time scaling $\ell \sim \left(\log t\right)^{1/\psi}$, where $\psi$ is the {\it infinite-randomness exponent}; and (3)~typical and average correlation functions scale distinctly, with the former exhibiting stretched-exponential behavior while the latter shows power-law scaling, indicating that they are dominated by rare disorder realizations with anomalously strong long-range correlations.

Already at this stage, we can make some predictions regarding the critical properties of the infinite-randomness confinement transition without a numerical implementation of the above outlined RG scheme: First, all the $G$-dependence of the RSRG procedure is encoded in the $1/|G|$ prefactor appearing in the RG rules, which is not expected to affect leading scaling behavior. This  is similar to the case of the disordered $Q$-state 1D quantum Potts model~\cite{Senthil_1996}, where the RSRG analysis leads to a similar prediction that universal properties at the infinite-randomness critical point are $Q$-independent. Also, second, as we show more explicitly in the next section, the well-known duality mapping~\cite{RevModPhys.52.453} between confinement and symmetry breaking in 2+1 dimensions holds not only at the Hamiltonian level, but also as an exact mapping between the respective RSRG rules~\cite{Motrunich_2000}. Strong-disorder RSRG arguments suggest that the critical properties of the random quantum Potts model are $Q$-independent also in two dimensions, and identical to those controlled by the infinite-randomness critical point of the random transverse-field Ising model~\cite{Motrunich_2000, PhysRevB.82.054437}.
			
		Together, these indicate that at strong disorder and for model Eq.~\eqref{eq:Ham_topological} all  gauge groups $G$ are equivalent at criticality up to sub-leading corrections to scaling; in other words, we establish that the scaling is superuniversal. This is one of the central points of this paper, which relies on exact mappings and RSRG results, and must ultimately be confirmed by microscopic numerical simulations. We devote the remainder of this paper to providing analytical and numerical evidence in favor of superuniversality, and to exploring the range of its applicability.
			
			The arguments above also suggest that we may gain insights into superuniversal scaling and the RG structure by considering the {\it dual} description of Eq.~\eqref{eq:Ham_topological} as a model of global symmetry-breaking, to which we now turn.
			
			\section{Duality to quantum Potts Model} 
			\label{sec:duality}
			\begin{figure}[tt]
				\centering
				\includegraphics[width=0.9\columnwidth]{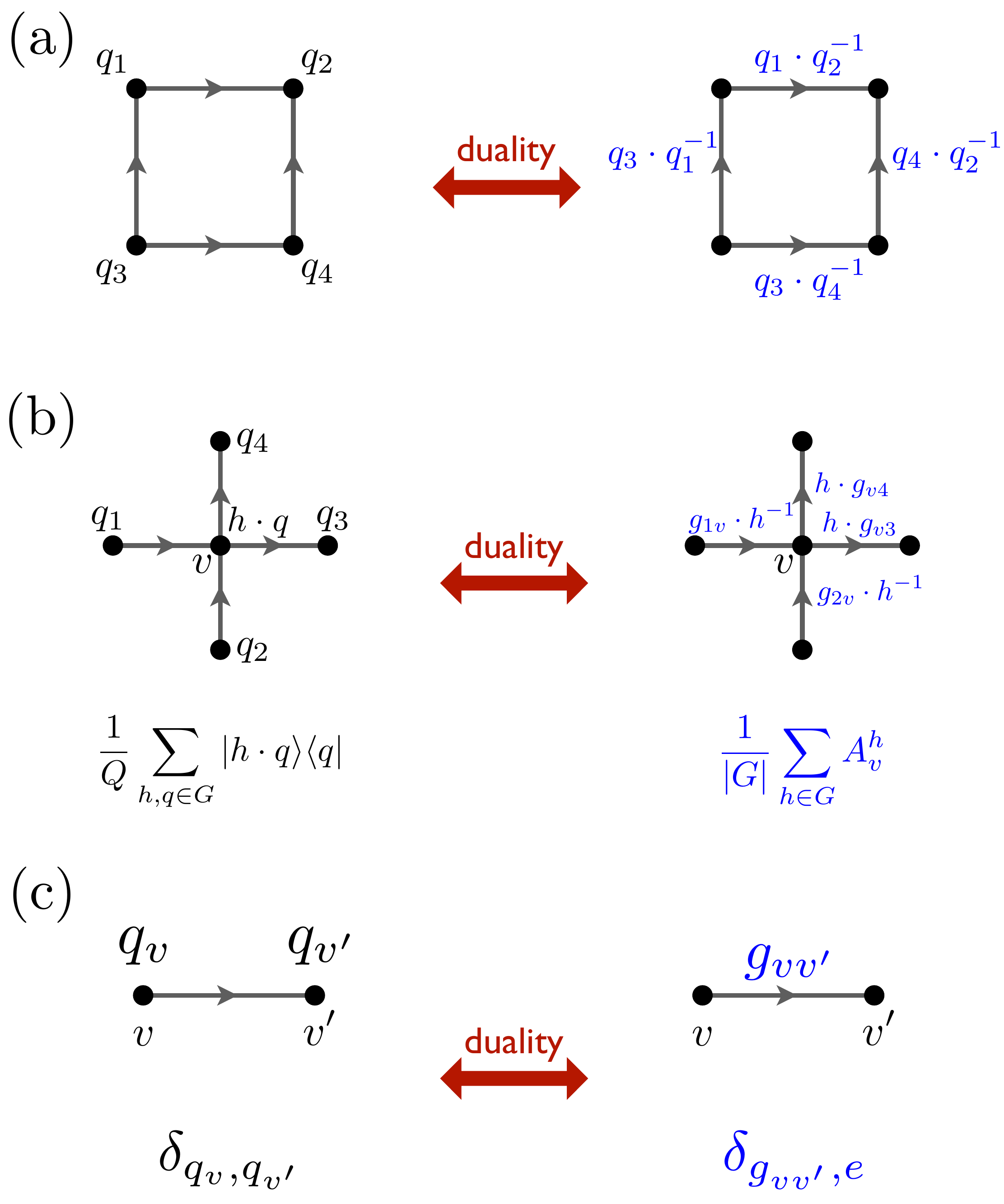}
				\caption{Duality mapping between a spin configuration in the $Q$-state Potts model (with $Q=|G|$) (left)  and  a link configuration (gauge field configuration) in the gauge theory (right) under the duality. (a)  The mapping automatically preserves the zero-flux gauge constraint. (b) The transverse-field term on vertex $v$ in the $Q$-state Potts model maps to a vertex term $A_v = \frac{1}{|G|} \sum_{h \in G} A_v^h$ under the duality. (c) The ferromagnetic interaction in the $Q$-state Potts model corresponds to the link term under the duality. }
				\label{fig:gauge-Potts-duality}
			\end{figure}
			\begin{figure*}[tt]
				\centering
				\includegraphics[width=1.8\columnwidth]{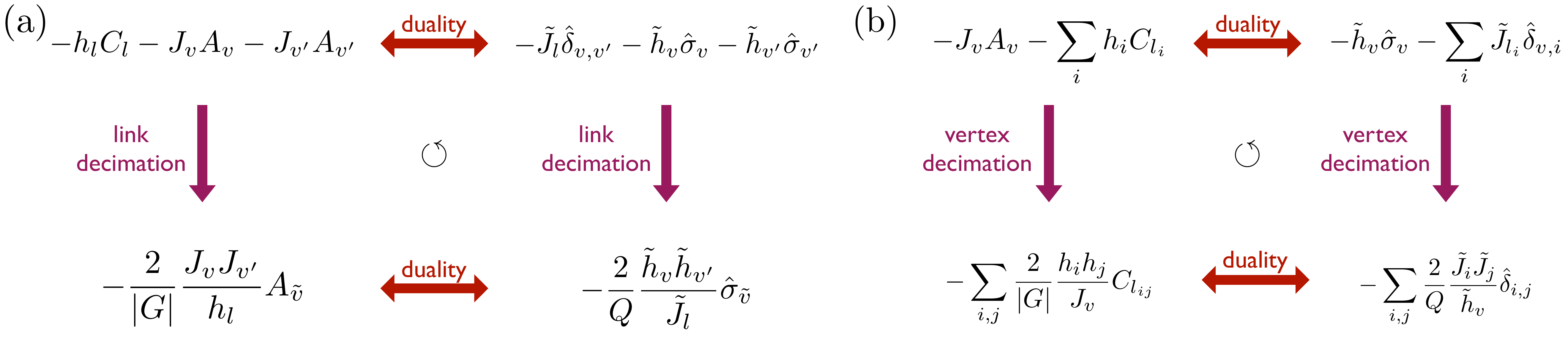}
				\caption{The RSRG of the $Q$-state Potts model and the $G$-gauge theory and the duality mappings (where $Q=|G|$). Here, we denote the Ising-type interaction and the transverse field term in the Potts model as $\hat{\delta}_{v,v'}$ and $\hat{\sigma}_v$. (a) Link decimation and (b) vertex decimation both commute with the duality mapping where the corresponding changes in the graph $\mathcal{G} \mapsto \mathcal{G}'$ can be found in Fig.~\ref{fig:H_and_decimations}. A long link term $C_l$ such as $C_{l_{34}}$ in Fig~\ref{fig:H_and_decimations} (b) corresponds to a dual higher-neighbor ferromagnetic interaction term between spin states at vertex $3$ and $4$ with respect to a planar graph $\mathcal{G}'$.
				}
				\label{fig:RSRG-duality}
			\end{figure*}
			In this section, we construct a duality mapping between the discrete lattice gauge theory model appearing in Eq.~\eqref{eq:Ham_topological} and the quantum $Q$-state Potts model. Crucially, the duality mapping relates not only  the degrees of freedom and Hamiltonian terms but also the RSRG rules of the two theories. The dual degrees of freedom are $Q$-state spins placed at each vertex $v$ of $\mathcal{G}$, where $q=1,2, \cdots, Q$ are identified with the elements of the discrete group $G$, so that $Q = |G|$  is the order of the gauge group $G$. The link degrees of freedom, $g_l$, are then represented by $g_l = q_{v} \cdot q_{v'}^{-1}$, where the directed link $l$ is directed from the vertex $v$ to the vertex $v'$. Crucially, with the above definition, the link variables, $g_l$, automatically satisfy the zero-flux constraint on each plaquette, while the spin values $q_v$ remain unrestricted.\footnote{Technically, the exact duality only holds when we restrict to global $G$-singlet states in the dual $Q$-state Potts model and to the states with the gauge constraint imposed on every plaquette as well as having zero holonomies along non-trivial cycles in the gauge theory.} The action of a vertex $A_v^g$ is equivalent to left-multiplying $q_v$ by $g$ in the dual $Q$-state Potts model such that we can identify
			\begin{equation}
			A_v^g \mapsto \sum_{q_v=1}^{Q} \ket{g\cdot q_v}\bra{q_v}.
			\end{equation}
			Averaging the above over all group elements yields
			\begin{equation}
			A_v \mapsto \sum_{q, q'=1}^Q \frac1{Q} \ket{q}\bra{q'}.
			\end{equation}
			Turning now to the the link terms  $C_l$, a projection of a link variable to the identify element is equivalent to a ferromagnetic coupling in the Potts model between the vertices connected by $l$. Similarly to the gauge-theory case, we anticipate that during the RSRG flow of the dual Potts model, terms describing higher-neighbor ferromagnetic interactions will be generated. For that reason, akin to Eq.~\eqref{eq:Ham_topological_mod}, we allow for ferromagnetic interactions defined on long links $l\in\mathcal{L}_\mathcal{V}$.  When $C_l$ is a long-link term its corresponding dual ferromagnetic interaction can also be regraded as a direct link connecting distant vertices as part of a nonplanar graph structure, as was implemented in the RSRG scheme presented in Ref. \onlinecite{Motrunich_2000}.  With the duality mapping of both vertex and link terms established, we can write the dual Potts model Hamiltonian as,
			\begin{equation}
			\tilde{H} = - \sum_{v \in \mathcal{V}} \tilde{h}_v \hat{\sigma}_{v} - \sum_{l=\langle v,v'\rangle \in \mathcal{L}_\mathcal{V}} \tilde{J}_l \hat{\delta}_{v,v'}.
			\label{eq:Ham_Potts}
			\end{equation}
			Here, $\hat{\sigma}_v \equiv \sum_{q, q'=1}^Q \frac1{Q} \ket{q} \bra{q'}_v$ represents a transverse-field term at $v$ and $\hat{\delta}_{v,v'} \vert q \rangle_v \vert q' \rangle_{v'} \equiv \delta_{q, q'} \vert q \rangle_v \vert q \rangle_{q'}$ induces ferromagnetic interactions between $v$ and $v'$. $\tilde{H}$ is precisely a ferromagnetic $Q$-state quantum Potts model, where the random couplings, $\tilde{h}_v$ and $\tilde{J}_l$ are, respectively, identified with $J_v$ and  $h_l$ defined in the original gauge field model. In Fig.~\ref{fig:gauge-Potts-duality}, we summarize the duality relations. Note that Eq.~\eqref{eq:Ham_Potts} is invariant under the group of permutations $S_Q$ between the $Q$ states of the Potts model.  In physical terms, the duality mapping identifies domain walls of the Potts model with electric field lines of the gauge theory, generalizing the well-known duality mapping between the Ising Lattice gauge theory and the transverse field Ising model~\cite{RevModPhys.52.453}, in two special dimensions.

			We now explain how the different RSRG decimation steps map under the duality transformation. We first note that Hamiltonian terms defined on the Potts model side remain invariant under lattice isomorphisms applied on the gauge theory side.  In addition, the decimation rules for newly generated Hamiltonian terms and their corresponding coupling constants are identical, as presented in Fig.~\ref{fig:RSRG-duality}. Specifically, vertex terms generated during link term decimations, in the gauge theory side, are directly mapped to transverse field terms that are generated during ferromagnetic term decimations, in the Potts model description. In the same way, vertex term decimation maps to the corresponding transverse field term decimation. In sum, we can move freely between the gauge theory Eq.~\eqref{eq:Ham_topological_mod} and its dual $Q$-state Potts model \eqref{eq:Ham_Potts} at any stage of the RG.  
			
			For the special model in Eq.~\eqref{eq:Ham_topological} this exact duality mapping of each RG step in the gauge model to a corresponding one on the Potts side removes the need to explicitly implement RSRG rules on the gauge side. This is because the Potts-side RG rules have already been implemented numerically for the $Q=2$ case~\cite{Motrunich_2000, PhysRevB.82.054437} (corresponding to the random transverse-field Ising model), where they flow to an infinite-randomness fixed point at criticality. Analyzing the flow equations indicates that different $Q$ values yield identical exponents. Combined with the duality mapping, this observation suggests that the critical  fixed point to which the Hamiltonian Eq.~\eqref{eq:Ham_topological} flows under RG is `superuniversal' and specifically, $G$-independent --- in sharp contrast to the clean case. In the next section, we use fully microscopic numerical simulations to verify that the critical exponents for the magnetic model are  $Q$-independent, which is a more stringent test than simply implementing the RG.
			
		 Adding terms of the from $A_v^\chi$ to the Hamiltonian \eqref{eq:Ham_topological}, involving nontrivial characters $\chi\neq 1$, would lower the large permutation symmetry $S_{|G|}$ of the dual model. For Abelian $G \cong \mathbb{Z}_k$ for some $k \in \mathbb{Z}^+$, the symmetry of the dual model is reduced to $\mathbb{Z}_{|G|}=\mathbb{Z}_k=G$, corresponding to the symmetry of a quantum clock model. Reasoning in analogy with the 1D case~\cite{Senthil_1996}, we do not expect this symmetry reduction $S_k \to \mathbb{Z}_k$ to change critical properties of the strong-disorder fixed point {since the difference between the two symmetry groups is merely a numerical prefactor in the RG rules. On the other hand, in the non-Abelian case, determining the effect of such perturbations requires a numerical implementation of a modified set of RSRG rules (see Sec.~\ref{sec:pert} for more details).
			
			Finally, the  mapping also allows us to identify dual operators that are predicted to share common universal properties. An important example is the spin susceptibility, which is expected to follow a power-law scaling form. Under the duality mapping, the spin susceptibility is identified with a nonlocal string operator known as the Polyakov loop \cite{RevModPhys.52.453}, which allows us to diagnose the presence or absence of confinement. The Polyakov loop is thus expected to follow the same scaling form as the spin susceptibility near criticality. 
			
			\section{Numerical Simulations} 
			\label{sec:qmc}
			We now turn to numerical simulations of the disordered $Q$-state quantum Potts model \eqref{eq:Ham_Potts}. Previous work \onlinecite{Pich_1998} has already considered the $Q=2$ case and  found numerical evidence for an infinite-randomness critical fixed point. Below, we present a refined numerical analysis of $Q=2$ and $Q=3$. In the clean limit, the $Q=3$ quantum Potts model in two dimensions undergoes a first-order symmetry breaking transition and hence the predicted flow of both models to the {\it same} infinite-randomness fixed point is nontrivial. Most importantly, we test the superuniversality conjecture by comparing the critical exponents of the $Q=2$ and $Q=3$ models.
			
			Although direct quantum Monte Carlo (QMC) simulations on the model in Eq.~\eqref{eq:Ham_topological} are possible~\cite{Assaad2008, gazit2017emergent}, the accessible system sizes are limited by the lack of efficient nonlocal updates. This becomes an especially acute problem in disordered systems because of the need to simulate a large number of disorder realizations, particularly in cases (such as ours) where rare events are important~\cite{PhysRevB.22.1305, PhysRevLett.69.534, PhysRevB.50.3799, PhysRevB.51.6411, Motrunich_2000, PhysRevB.82.054437}. Therefore, we instead perform QMC on the dual Potts model \eqref{eq:Ham_Potts}, for which efficient nonlocal cluster updates exist~\cite{PhysRevLett.58.86, PhysRevLett.62.361, ding2017monte}. Owing to the fact that the duality is microscopically exact the Potts model results can be readily translated back to the gauge theory language.
			
In order to motivate our specific choice of QMC technique, we comment on discrepancies between our  work and a previous QMC study of the random transverse-field Ising model~\cite{Pich_1998}.  According to our numerical findings below, the largest inverse temperature, $\beta^{\text{Pich}}_\textrm{max} \le 2\cdot10^3$, (in units of $1/J_\textrm{max}$, see below for our convention) and the number of disorder realizations, $N_{\text{dis}}\sim512$, considered in Ref.~\onlinecite{Pich_1998} do not accurately capture ground-state properties near the critical point. Furthermore, Ref.~\onlinecite{Pich_1998} attempted an extrapolation to the zero-temperature limit by considering a series of finite temperature transitions $h_c(T)$. The highly anisotropic space-time scaling ($z=\infty$, corresponding to $\ell \sim \left(\log t\right)^{1/\psi}$) of the infinite-randomness fixed point  leads to a prohibitive (exponential) numerical sensitivity, which ultimately makes such zero-temperature extrapolation uncontrolled. 
						
		In light of this, we employ the stochastic series expansion (SSE) QMC technique~\cite{PhysRevB.43.5950, sandvik1992generalization, ding2017monte}, which is formally numerically exact (up to statistical errors).  SSE is particularly suited to the present problem as it is free of discretization (Trotter) errors,  and  because it adaptively samples local operators based on their weight, ensuring fast convergence. Crucially,  SSE-QMC  evades the complications associated with  $z=\infty$ dynamical scaling since it does not treat space and time on an equal footing: going to lower temperature (larger $\beta$) corresponds to keeping more terms in the series expansion,  but because of the nature of the algorithm we sample the most important terms relevant to the physics. To overcome the critical slowing down phenomenon near criticality, we implement the Swendsen-Wang cluster update~\cite{PhysRevLett.58.86, ding2017monte}. As we have mentioned already, given the  $z=\infty$ critical scaling,  it is crucial to carefully take the $T \to 0$ limit for each system size $L$ in order to perform a reliable finite-size scaling analysis. 
	To speed up the QMC thermalization times we used the ``$\beta$-doubling''~\cite{Sandvik_2002, PhysRevLett.114.155301, PhysRevLett.114.255701} SSE scheme. The largest $\beta_\textrm{max}$ (hence the smallest temperature) is chosen to ensure that disorder-averaged physical observables have converged to their $T=0$ values. In what follows, we present QMC data evaluated at the largest $\beta = \beta_\textrm{max}$ considered, as a proxy for the ground state physics. We provide additional technical details concerning our numerical scheme in Appendix~\ref{append:QMC}. 			
			
			The coupling constants in Eq.~\eqref{eq:Ham_Potts} are drawn from a positively supported uniform distribution  $h_l \sim U[0,h_\textrm{max}]$, and similarly $J_v \sim U[0,J_\textrm{max}]$. For simplicity, throughout, we fix $J_\textrm{max}=1$, and measure all energy scales in units of $J_\textrm{max}$. Both for the $Q=2$ and $Q=3$ Potts models, we studied systems sizes $L=12,14,16$ with $\beta_\textrm{max}=2^{11}$ and $L=18$ with $\beta_\textrm{max}=2^{12}$.  We averaged over up to $50,000$ independent disorder realizations for each $L$ and $Q$ \cite{SIBIDANOV2017299,James2020}, finding this to be essential in order to arrive at reliable estimates for critical exponents.
			\begin{figure}[!b]
				\includegraphics[width=1.0\columnwidth]{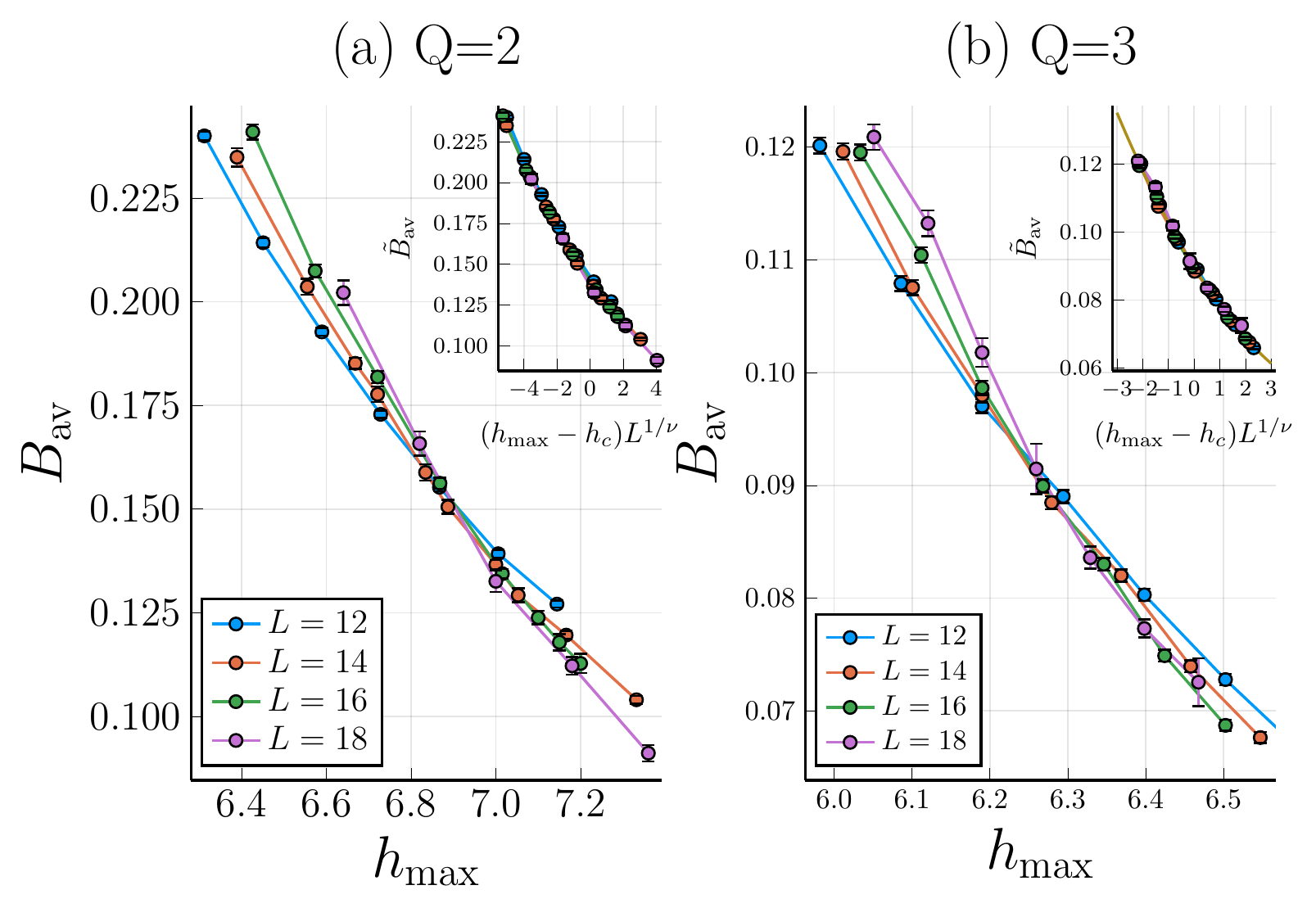}
				\caption{ Disorder-averaged zero-temperature binder ratio $\textrm{B}_\textrm{av}$ as a function of $h_\textrm{max}$ for (a) $Q=2$ and (b) $Q=3$. Different curves correspond the different system sizes. In the insets, we plot the universal scaling functions $\tilde{B}_{\textrm{av}}(x=(h_\textrm{max}-h_c)L^{1/\nu})$ obtained from a curve collapse analysis; see main text. Solid lines depicts a numerical fit to second order polynomial in the scaling variable $x$.} 
				\label{fig:Bindercollapse}
			\end{figure}
			
			\begin{figure}[!t]
				\includegraphics[width=0.96\columnwidth]{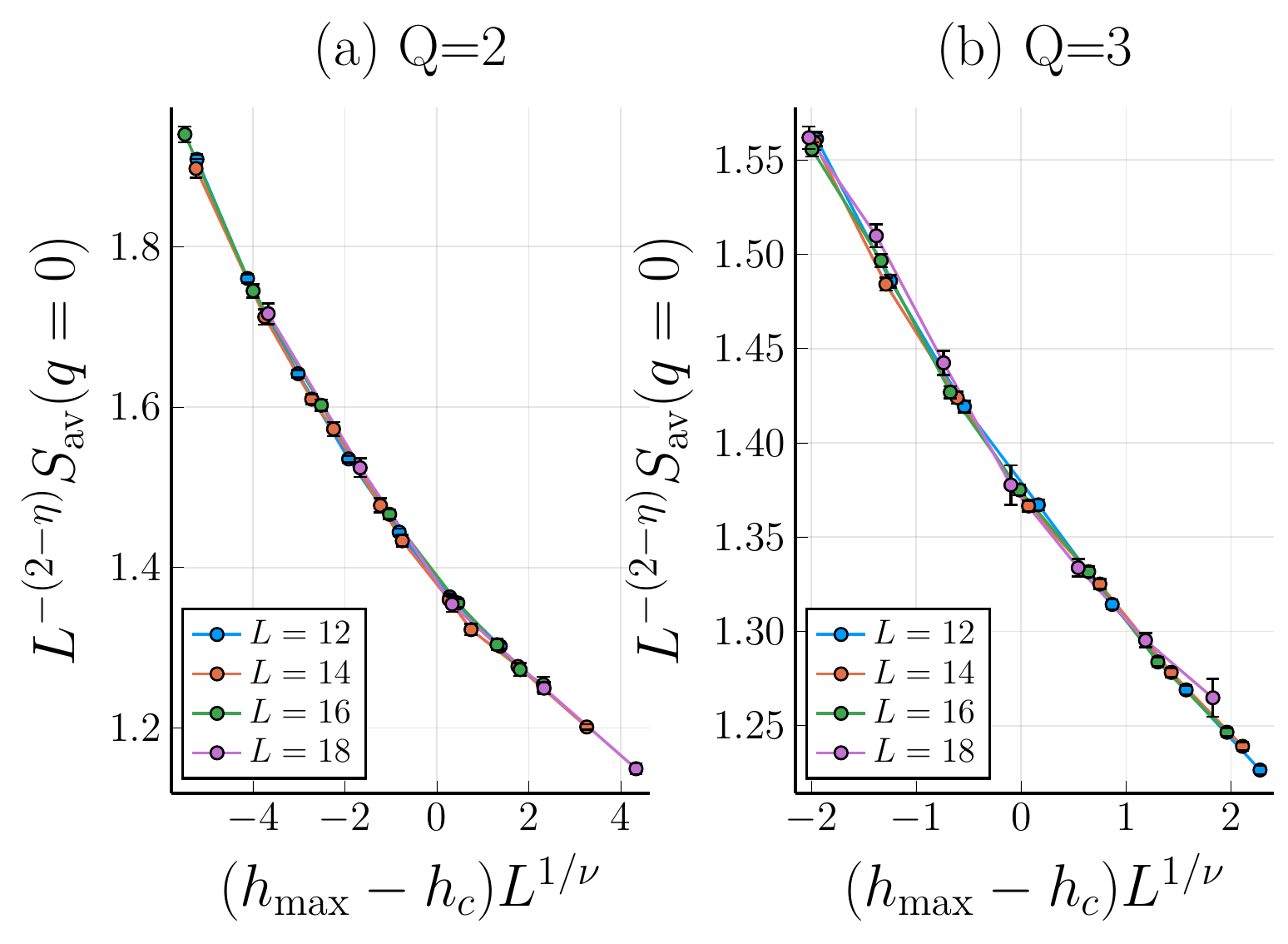}
				\caption{The universal scaling function $\tilde{S}_\textrm{av} \big(x = (h_\textrm{max}-h_c)L^{1/\nu} \big)$ (see main text) obtained from a curve collapse analysis for (a) $Q=2$ (b) $Q=3$. We use this analysis to extract the critical exponent $\eta$ controlling the scaling of the magnetization squared.}
				\label{fig:eta}
			\end{figure}
			
			Physically, for small $h_\textrm{max}$, a ferromagnetic state is expected (note that $J_v \ge 0$), where the global $Q-$flavor permutation symmetry is broken. With increase in $h_\textrm{max}$, quantum fluctuations restore the symmetry at a critical disorder strength $h_c$. We begin our analysis of the critical properties of the disordered $Q$-state quantum Potts model by studying the disorder-averaged Binder ratio at zero temperature:
			\begin{equation}
			{B}_{\text{av}} = \frac{1}{1 + \delta_{Q,2}}\left[2 + \delta_{Q, 2}- \frac{\langle M^4\rangle}{\langle M^2\rangle^2}\right]_{\text{dis}} . 
			\end{equation}
			Here, $\left\langle \mathcal{O} \right\rangle$ ($\left[ \mathcal{O}\right]_{\mathrm{dis}}$) denotes a quantum (disorder) average of the operator $\mathcal{O}$, $M^2$ is the square of the total magnetization  which is defined as $|\sum_{\boldsymbol{r}} e^{2 \pi i q_{\boldsymbol{r}}/Q}|^2$, where the sum is over all sites $\boldsymbol{r}$, and $M^4 \equiv (M^2)^2$. Conveniently, $B_\textrm{av}$ has a vanishing scaling dimension and hence is expected to follow a simple scaling form ${B}_{\textrm{av}}(L,\delta h)=\tilde{{B}}_{\textrm{av}}(\delta h L^{1/\nu})$, near criticality, where $\delta h=h_\textrm{max} - h_c$ measures the detuning from criticality, $\nu$ is the correlation length exponent, and $\tilde{{B}}_\textrm{av} (x)$ is a universal scaling function. Note also that the normalization in $B_\textrm{av}$ is chosen such that $B_\textrm{av}$ approaches  1 (0) deep in the ferromagnetic (paramagnetic) phase.
			
			Following the standard finite-size scaling approach, we fit our QMC data to the above scaling form. The result of this analysis is shown in Fig.~\ref{fig:Bindercollapse}. We find an excellent collapse of the data into a single curve identified with the universal scaling function $\tilde{B}_\textrm{av}(x)$. We estimate the critical couplings to be $h_c=6.97(3)$ for $Q=2$ and $h_c=6.27(2)$ for $Q=3$, and the correlation length exponent to be $\nu = 1.2(1)$ for $Q=2$ and $\nu = 1.3(1)$ for $Q=3$. Crucially, we observe that $\nu$ is independent of $Q$ (within  error bars). This key result signifies the insensitivity of universal data to the value of $Q$, providing further evidence for the superuniversal nature of the strong disorder fixed point. Our estimate for $\nu$ is also in agreement with the value extracted from by implementing the RSRG scheme for the 2D random transverse-field Ising model~\cite{PhysRevB.82.054437}, $\nu_\text{RSRG}= 1.24(2)$.

			We now proceed to extract the anomalous scaling dimension of the magnetization. To that end, we compute the disordered averaged equal-time spin-spin correlation function \begin{equation}\chi_{\rm av}(r-r')=\left[\left\langle{e^{i\frac{ 2\pi }{Q} (q_r-q_{r'})}}\right\rangle\right]_{\mathrm{dis}},
			\end{equation} and the associated structure factor,
			\begin{equation}
			S_{\rm av} (q) = \sum_{r} \chi_{\rm av} (r) e^{i q \cdot r}  .
			\end{equation}
			Using the expected power-law behavior in $\chi_{\rm av}(r) \sim r^{-\eta}$ at $h_c$, we can extract $\eta$ through a curve collapse analysis of the scaling ansatz $S_{\rm av} (q=0) = L^{(2-\eta)}\tilde{S}_{\textrm{av}} (\delta h L^{1/\nu})$, for some universal scaling function $\tilde{S}_{\textrm{av}} (x)$. In our fitting procedure, we use  $h_c$ and $\nu$ values obtained above and account for their statistical error through a standard bootstrap analysis. We find $\eta\approx 1.75(2)$ for $Q=2$ and $\eta\approx 1.80(1)$ for $Q=3$. These values are in reasonable agreement with the RSRG calculations of Ref.~\onlinecite{PhysRevB.82.054437}, $\eta_\text{RSRG} = 1.96(3)$. The closeness of $\eta$ for $Q=2,3$ gives a measure of added support for superuniversality. In Fig.~\ref{fig:eta}, we use our numerical estimates for critical exponents to plot the universal scaling function $\tilde{S}_{\textrm{av}} (x)$ by a rescaling of our finite size data. We indeed obtain the expected collapse of data points belonging to different system sizes into a single universal curve. The estimated values of $\eta$ display a mild drift as a function of the system size used in the numerical fit; a more robust calculation of $\eta$ will require simulations on larger lattices, beyond our current numerical capabilities. 
			\begin{figure}[!t]
				\includegraphics[width=1.\columnwidth]{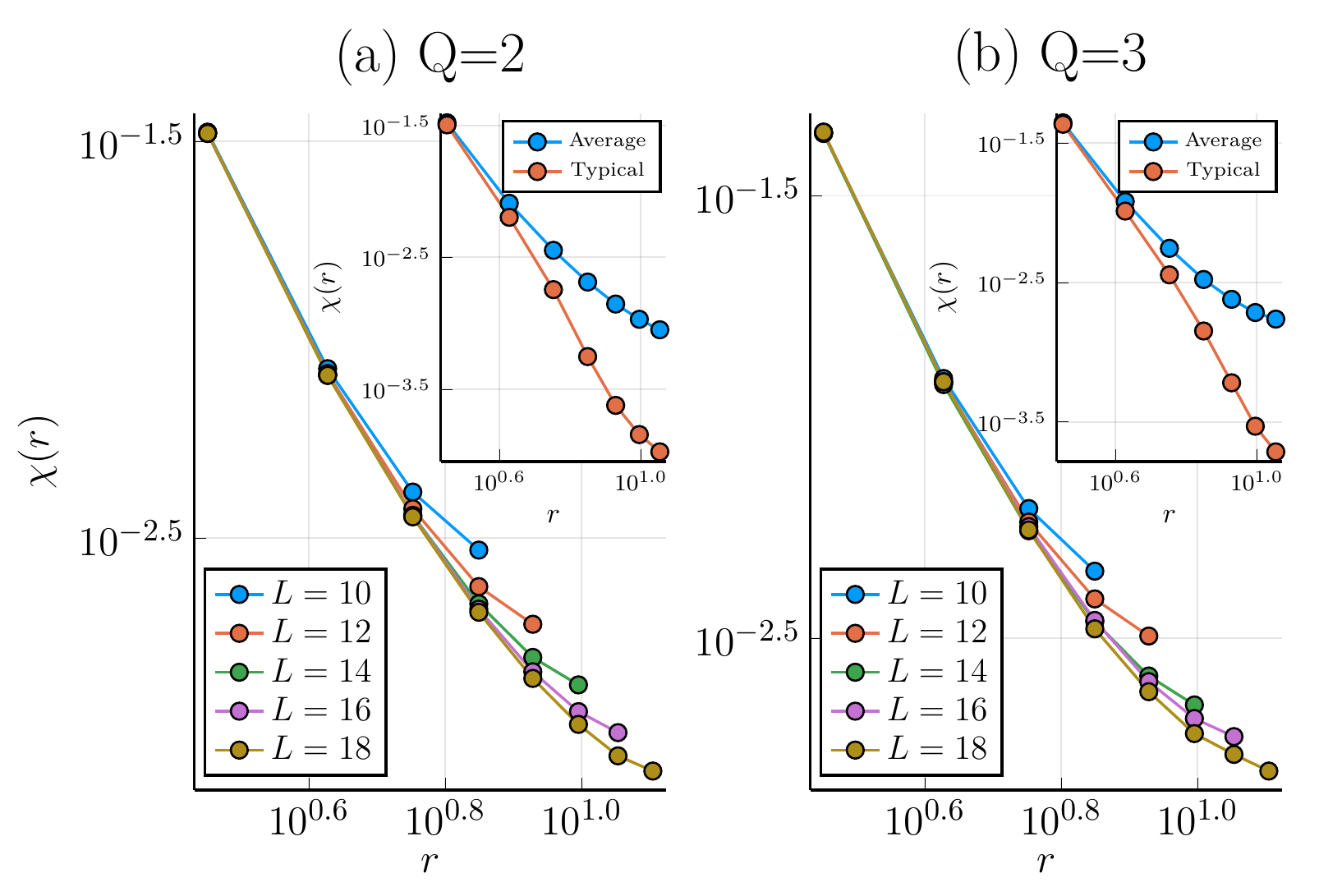}
				\caption{(a) Disorder averaged spin-spin correlation function at the critical point for various system sizes for (a) $Q=2$ and (b) $Q=3$. (inset) Comparison between averaged and typical spin-spin correlation functions for $L=18$. Typical spin-spin correlations exhibit a faster decay as expected at an infinite-randomness fixed point.} 
				\label{fig:correlationsdata}
			\end{figure}
			
			Finally, in Fig.~\ref{fig:correlationsdata}, we compare the typical and average spin-spin correlation functions at $h_c$. At an infinite-randomness fixed point, the typical spin-spin correlation function shows a streched exponential decay~\cite{PhysRevB.22.1305, PhysRevLett.69.534, PhysRevB.50.3799, PhysRevB.51.6411, Motrunich_2000}: $-\log [\chi_{\rm typ} (r)] \sim r^\psi$, which is very different from the average correlations showing power-law behavior. We were unable to reliably infer the value of the infinite randomness critical exponent, $\psi$, due to the exponential sensitivity of the associated scaling form. However, we do find that the typical correlations decay faster than any power law, in accord with expectations for an infinite-randomness fixed point.
			
\section{Perturbative stability}
\label{sec:pert}
			So far, we have mainly focused on the model Eq.~\eqref{eq:Ham_topological} which, albeit natural, is a highly symmetric point in the space of gauge theories as we have noted. The special choice of vertex terms in Eq.~\eqref{eq:Ham_topological} leads to a  degeneracy between all  possible nontrivial gauge charge assignments on vertices.  In more physical terms,  all the elementary `electric' gauge charges have the same `mass gap' above the vacuum state, and the same nearest-neighbor hopping amplitude on the lattice. It is important to consider how perturbations away from the high-symmetry point  might affect the superuniversal infinite-randomness scaling.
			
			To that end, we analyze perturbations to Eq.~\eqref{eq:Ham_topological} that remove the symmetry between different gauge charges, while maintaining the underlying gauge symmetry. The most natural way to do this is to introduce generalizations of the vertex term that involve combinations of the $A_v^{g}$s  with nontrivial group characters. Explicitly, we consider the following generalization of the Hamiltonian in Eq.~\eqref{eq:Ham_topological_mod}:
			\begin{equation}
			H = - \sum_{\substack{v\in \mathcal{V}, \\\Gamma \in \textrm{irrep}(G)}} J_v^\Gamma \mathcal{A}_v^\Gamma - \sum_{\substack{l \in \mathcal{L}_\mathcal{V},\\ [g] \in \textrm{class}(G)}} h_l^{[g]} \mathcal{C}_l^{[g]}.
			\label{eq:Ham_pert_topological_mod}
			\end{equation}
			Here, $\mathcal{A}_v^\Gamma \equiv \frac{d_\Gamma}{|G|} \sum_{g \in  G} \chi_\Gamma^*(g) A_v^g$, with $d_\Gamma$ and $\chi_\Gamma$ the dimension and the character of an irreducible representation (irrep) $\Gamma$ of $G$, and $\mathcal{C}_l^{[g]} = \sum_{h \in [g]} C_l^h = \sum_{h \in [g]} \vert h \rangle \langle h \vert_l$ is a projector to the conjugacy class $[g] \in \textrm{class} (G)$ (where in case that $l$ is a long link, the projection is defined along the oriented path, similarly to $C_l$). As before, the random and non-negative coupling constants $J_v^\Gamma$ and $h_l^{[g]}$ are, e.g., distributed uniformly, $J_v^\Gamma\sim U[0,J^\Gamma_{\text{max}}]$ and $h_l^{[g]} \sim  U[0,h^{[g]}_{\text{max}}]$\footnote{We can always make the coupling constants $J_v^\Gamma$ and $h_l^{[g]}$ positive by adding identity operators: $\openone_v = \sum_{\Gamma \in \textrm{irrep}(G)} \mathcal{A}_v^\Gamma$ and $\openone_l = \sum_{[g] \in \textrm{class}(G)} \mathcal{C}_l^{[g]}$.}. The generalized model reduces to the symmetric one (Eq.~\eqref{eq:Ham_topological_mod}) by setting $J^\Gamma_{\text{max}}=0$ and $h^{[g]}_{\text{max}}=0$ for all nontrivial irreps $\Gamma \ne 1$ and conjugacy classes $[g]\ne [e]$. In particular, we have $A_v=\mathcal{A}_v^{\Gamma=1}$ and $C_l=\mathcal{C}^{[e]}_l$.  Note also that in order to allow for long-link terms, we consider the generalized set of links $\mathcal{L}_\mathcal{V}$, similarly to Eq.~\eqref{eq:Ham_topological_mod}.
			
			Within our perturbative analysis, we will assume that terms breaking the permutation symmetry are small, namely, the condition $J^{\Gamma\ne 1}_{\text{max}1},h^{[g]\ne e}_{\text{max}}\ll J^{\Gamma=1}_{\text{max} },h^{[e]}_{\text{max}}$ holds. Thus, at least in the initial steps of RG, we need to understand only how the terms in Eq.~\eqref{eq:Ham_pert_topological_mod} renormalize upon decimating the permutation-symmetric terms appearing in Eq.~\eqref{eq:Ham_topological_mod}. 
			
We first consider decimating link terms $ \mathcal{C}^{[e]}_l (= C_l)$. As before, the first-order contribution in perturbation theory,
			\begin{equation}
			\mathcal{C}^{[e]}_l \big( \mathcal{A}_v^\Gamma \big) \mathcal{C}_l^{[e]} = \frac{d_\Gamma}{|G|} \sum_{g \in G} \chi^*_\Gamma (g) \mathcal{C}_l^{[e]} (A_v^g) \mathcal{C}_l^{[e]}= \frac{d_\Gamma^2}{|G|} \mathcal{C}_l^{[e]},
			\end{equation}
			gives only trivial constant terms. To treat products of vertex terms, we will employ the following group theory identity~\cite{hamermesh2012group}:  
			\begin{equation}
			\chi_\Gamma (g) \chi_{\Gamma'} (g) = \sum_{\Lambda \in \textrm{irrep}(G)} (\Gamma, \Gamma' | \Lambda) \chi_\Lambda(g) ,
			\end{equation} 
			where $\Gamma$, $\Gamma'$, and $\Lambda$ are irreducible representations of $G$, $g$ is a group element in $G$, and $(\Gamma, \Gamma'| \Lambda) \in \mathbb{Z}^+$ is the Clebsch-Gordan coefficient among $\Gamma$, $\Gamma'$, and $\Lambda$.  Using the above identify and following similar arguments used in deriving Eq.~\eqref{eq:sec_ord_link}, we show that terms generated in second-order perturbation theory will take the following form,
			\begin{align}
				\mathcal{C}^{[e]}_l \big( \mathcal{A}_v^\Gamma \mathcal{A}_{v'}^{\Gamma'} \big) \mathcal{C}^{[e]}_l &= \frac{d_\Gamma d_{\Gamma'}}{|G|^2} \sum_{g, g' \in G} \chi^*_\Gamma (g) \chi^*_{\Gamma'} (g') \mathcal{C}^{[e]}_l \big( A_v^g A_{v'}^{g'} \big) \mathcal{C}^{[e]}_l \nonumber \\
				&= \frac{d_\Gamma d_{\Gamma'}}{|G|^2} \sum_{g \in G} \chi^*_\Gamma (g) \chi^*_{\Gamma'} (g) \mathcal{C}^{[e]}_l \big( A_{\tilde{v}}^{g} \big) \mathcal{C}^{[e]}_l \nonumber \\
				&= \Big(\sum_{\Lambda \in \textrm{irrep}(G)} \frac{d_\Gamma d_{\Gamma'} (\Gamma, \Gamma'| \Lambda)}{d_\Lambda |G|} \mathcal{A}_{\tilde{v}}^\Lambda \Big) \mathcal{C}^{[e]}_l.
				\label{eq:vertex_gamma_merge}
			\end{align}
			Since in addition $[\mathcal{A}_{\tilde{v}}^\Lambda, \mathcal{C}^{[e]}_l]=0$, the renormalized vertex terms generated within the RSRG link decimation step preserve the algebraic structure of Eq.~\eqref{eq:Ham_pert_topological_mod}, as required.
			
			Next, we consider the decimation of a vertex term $\mathcal{A}_v^{1} (=A_v)$ in Eq.~\eqref{eq:Ham_pert_topological_mod}. Following similar arguments used in deriving Eq.~\eqref{eq:Av-Cl-Av}, we can show that the first order contribution in perturbation theory yields a trivial term,
			\begin{equation}
			\mathcal{A}_v^1 \big( \mathcal{C}_l^{[g]} \big) \mathcal{A}_v^1 = \frac{|[g]|}{|G|} \mathcal{A}_v^1,
			\label{eq:RSRG-X-Av-Clg-Av}
			\end{equation}
			where $|[g]|$ is the order of the conjugacy class $[g]$. Next, we consider two links $l_1$ and $l_2$ which are both adjacent to the vertex $v$ at which $A_v$ acts. Without loss of generality, we may assume that $l_1$ is directed from $v_1$ to $v$ and $l_2$ is directed from $v$ to $v_2$.  For a long link $\tilde{l} = \langle v_1, v_2 \rangle$, the identity 
			\begin{equation}
			C_{l_1}^{h_1} \cdot C_{l_2}^{h_2} = C_{\tilde{l}}^{h_1 \cdot h_2} \cdot C_{l_2}^{h_2}
			\end{equation}
			holds and we may choose a path $\pi_{\tilde{l}}$ for $C_{\tilde{l}}^{h_1 \cdot h_2}$ such that it never passes through $v$ by the virtue of zero-flux constraints. This implies that $[\mathcal{A}_v^1, C_{\tilde{l}}^{h_1 \cdot h_2}] = 0$. In second-order perturbation theory, we get
			\begin{align}
				&\mathcal{A}_v^1 \big( \mathcal{C}_{l_1}^{[g_1]} \cdot \mathcal{C}_{l_2}^{[g_2]} \big) \mathcal{A}_v^1 = \sum_{h_1 \in [g_1], h_2 \in [g_2]} \mathcal{A}_v^1 \big( C_{l_1}^{h_1} \cdot C_{l_2}^{h_2} \big) \mathcal{A}_v^1 \nonumber \\
				&= \sum_{h_1 \in [g_1], h_2 \in [g_2]} \mathcal{A}_v^1 \big( C_{\tilde{l}}^{h_1 \cdot h_2} \cdot C_{l_2}^{h_2} \big) \mathcal{A}_v^1 \nonumber \\
				&= \Big( \sum_{h_1 \in [g_1], h_2 \in [g_2]} C_{\tilde{l}}^{h_1 \cdot h_2} \Big) \frac{1}{|G|} \mathcal{A}_v^1 \nonumber \\
				&= \Big( \sum_{[h] \in \textrm{class} (G)} \big([g_1], [g_2] \big| [h] \big) \mathcal{C}_{\tilde{l}}^{[h]} \Big) \frac{1}{|G|} \mathcal{A}_v^1 ,
			\end{align}
			where from the second line to the third line, we use the fact that $C_{\tilde{l}}^{h_1 \cdot h_2}$ commutes with $\mathcal{A}_v^1$ and the algebraic manipulations used in deriving Eq.~\eqref{eq:RSRG-X-Av-Clg-Av}. In the last line, $\big([g_1], [g_2] \big| [h] \big) \in \mathbb{Z}^+$ is the fusion coefficient among conjugacy classes~\cite{hamermesh2012group}. We see that the generated link terms maintain the algebraic structure of Eq.~\eqref{eq:Ham_pert_topological_mod} and hence do not spoil the self-similar property of the RSRG scheme.
			
			Summarizing the above analysis, we indeed find that the generalized set of RSRG rules admits a self-similar structure. Determining the IR relevance of these perturbations will require a numerical implementation of the RSRG rules. Nevertheless, for the special case of an Abelian gauge group $G$, we can conclude that the universality class remains intact even when the perturbation is large (either initially or during the RSRG flow). This can be understood most easily through the duality mapping of Sec.~\ref{sec:duality}; When $G \cong \mathbb{Z}_k$ for some $k \in \mathbb{Z}^+$, breaking the permutation symmetry reduces the $S_{|G|}$ symmetry group of the dual Potts model down to $\mathbb{Z}_{|G|}$, as realized in quantum clock model. Following the arguments of Ref.~\onlinecite{Senthil_1996}, the RSRG structure and, relatedly, the infinite-randomness critical behavior of both models are identical. We leave the interesting question regarding the stability of the critical point for non-Abelian $G$ to future work.
			
			\section{Extension to Dijkgraaf-Witten Models of SET and SPT Phases}
			\label{sec:DW-models}
			\begin{figure}[!t]
				\includegraphics[width=1.0\columnwidth]{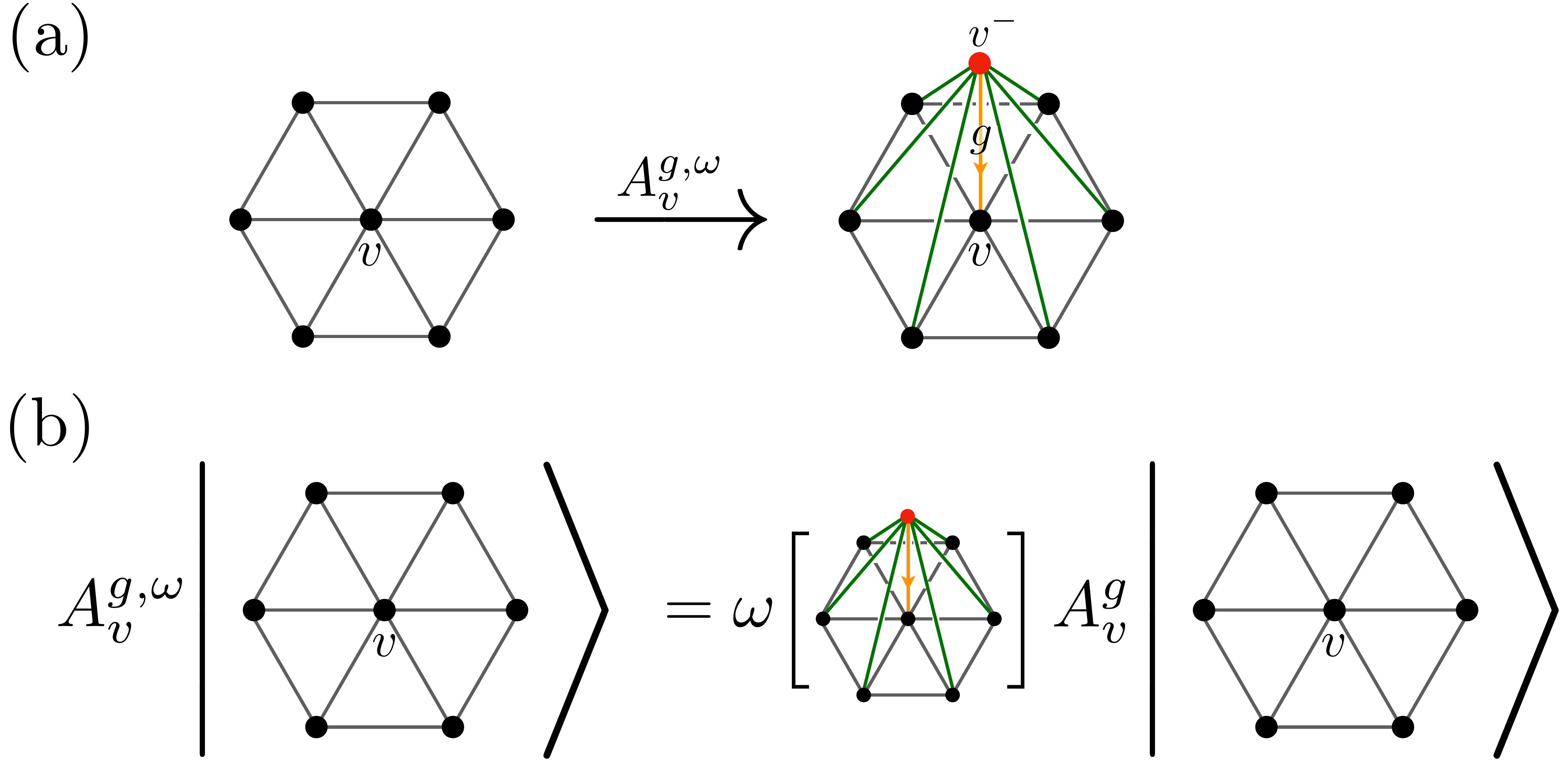}
				\caption{(a) To compute the $U(1)$ phase factor of $A_v^{g, \omega}$ when acting at the vertex $v$, we construct a three dimensional ``tent'' manifold by adding a new vertex $v^-$ and links connecting $v^-$ and other vertices. The index of $v^-$ is set by a number that is infinitesimally smaller than the index of $v$, which induces the orientation of links in the tent. (Orientations of links other than the yellow link connecting $v^-$ and $v$ are not drawn explicitly.) (b) The $U(1)$ phase factor $e^{i \phi_g (\omega)}$ of $A_v^{g, \omega}$ operator is given by the $U(1)$ phase factor $\omega[M]$ associated with the ``tent'' manifold $M$. The computation of this phase factor is explained in Appendix~\ref{app:top-via-group-coho}; a crucial point is that for a lattice of triangular plaquettes, any tent manifold $M$ can be decomposed in terms of tetrahedrons $T$ so that $M = \cup_{T \in M} T$, and  $\omega[M] = \prod_{T\in M} \omega[T]$. The tetrahedrons are the fundamental objects that are employed to properly track phase factors under decimations. The remaining operator in $A_v^{g, \omega}$ is $A_v^g$, which is identical to the one in Fig.~\ref{fig:QD-basics} (a).} 
				\label{fig:DW}
			\end{figure}
			
			The first extension that we consider concerns strong-disorder confinement transitions of Dijkgraaf-Witten (DW) models~\cite{dijkgraaf1990topological, PhysRevB.87.125114, PhysRevB.87.155115}, which provide physical realizations of twisted lattice gauge theories. This family of models encompasses a large (though not exhaustive) class of systems exhibiting intrinsic nonchiral topological order in two spatial dimensions, and goes beyond the ones realizable by Eq.~\eqref{eq:Ham_topological}. In a broader context, a systematic construction of SPT and SET phases~\cite{PhysRevB.87.155114, PhysRevB.87.155115} can be accomplished though a gauging--ungauging procedure~\cite{PhysRevB.94.235136, PhysRevB.96.115107} (either partial or complete) of DW models. {In this regard, as we explain below, our results naturally carry over to dual order-disorder phase transitions in SPT and SET phases, at which the protecting/enriching symmetry is spontaneously broken.}
			
			For concreteness, we focus on lattice models of twisted gauge theories~\cite{PhysRevB.87.125114, PhysRevB.87.155115} that are closely related to Kitaev's quantum double model (introduced in Eq.~\eqref{eq:Ham_topological}). We fix a discrete group $G$ and an element $[\omega] \in H^3\big(G, U(1) \big)$ from the third group cohomology, which classifies the distinct phases of twisted discrete $G$-gauge theories in (2+1) dimensions~\cite{dijkgraaf1990topological}. For convenience, we impose the following normalization conditions on $\omega$~\cite{PhysRevB.87.125114, PhysRevB.87.155115}: $\omega(e, g, h) = \omega(g, e, h) = \omega(g, h, e)=1$, where $e$ is the identity element in $G$. Due to the rather rigid structure needed to define a DW model, a proper definition of vertex terms is natural only on lattices comprising solely triangular plaquettes. While in the untwisted case [Eq.~\eqref{eq:Ham_topological}] link orientations can be chosen arbitrarily, here they are encoded through a predefined vertex ordering: a given link is always directed from a small to large vertex index. As before, the total Hilbert space is a tensor product of local $|G|$-dimensional Hilbert spaces. We enforce the gauge constraint by projecting to states obeying the zero-flux condition: $B_p \vert \Psi \rangle = \vert \Psi \rangle$ on every plaquette $p$. 
			
			The DW Hamiltonian then reads~\cite{PhysRevB.87.125114, PhysRevB.87.155115}
			\begin{equation}
			H_\textrm{DW} = - \sum_{v \in \mathcal{V}} J_v A_v^\omega - \sum_{l \in \mathcal{L}_\mathcal{V}} h_l C_l ,
			\label{eq:DW-Ham}
			\end{equation}
			where $A_v^\omega \equiv \frac{1}{|G|} \sum_{g\in G} A_v^{g, \omega} \equiv \frac{1}{|G|} \sum_{g\in G} e^{i \phi_g(\omega)} A_v^{g}$, with a pure phase factor $e^{i \phi_g (\omega)}$ depending on both the group cohomology element $\omega$ and the quantum state on which $A_v^g$ acts. $C_l$ projects the state at $l$ to the identity element of $G$, as in the untwisted case. 
			
			The $U(1)$ phase factor $e^{i \phi_g(\omega)}$ is defined through the so-called ``tent'' construction~\cite{PhysRevB.87.125114, PhysRevB.87.155115} by constructing an associated tent manifold $M$ on the vertex $v$. In order to define the tent manifold $M$, we include an additional vertex $v^-$  (Fig.~\ref{fig:DW}) `above' the vertex. $M$ is then defined to be the $3$-manifold whose vertex set consists of $v, v^-$, and all vertices adjacent to $v$ in $\mathcal{G}$, and with a link orientation given by combining the predetermined ordering in $\mathcal{G}$  with the assignment to $v^-$ of an infinitesimally smaller index relative to $v$.  As is evident from this construction, $M$ is a \textit{triangulated} $3$-manifold, i.e., it is obtained by gluing tetrahedrons. Tetrahedrons are thus the fundamental topological object in DW models, as they are the objects naturally associated with the $U(1)$ phase factor needed to `twist' the theory. While we relegate the precise definition of the $U(1)$ phase factor to Appendix~\ref{app:top-via-group-coho}, here we just note the fact that $e^{i \phi_{g=e}(\omega)}=1$, due to the normalization condition imposed on $\omega$, which is relevant for our discussions below. The 3-cocycle condition of $[\omega]$ ensures the commutativity between vertex operators: $[A_v^{g, \omega}, A_{v'}^{g', \omega}] = 0$ for all distinct vertex pairs $v, v'$ and arbitrary group elements $g, g' \in G$. Also, $A_v^{g, \omega}$ forms a group representation since $A_v^{g, \omega} A_v^{g', \omega} = A_v^{g \cdot g', \omega}$. These properties imply that the vertex operators in Eq.~\eqref{eq:DW-Ham} are commuting projectors as in the untwisted gauge theories. 
			
			By controlling the relative strength of $h$ and $J$, one can drive a phase transition between topologically ordered and trivial phases.  Here, we are interested in studying these confinement transitions in the presence of quenched disorder. As with the untwisted case, we construct a set of RSRG rules for decimating vertex and link terms and perturbatively computing newly generated Hamiltonian terms. The specific structure of twisted gauge theories, however, renders the RSRG procedure more intricate than the one considered above for the simpler case of untwisted theories. Remarkably, despite these complications, under the RSRG the DW models flow to the same superuniversal fixed point as the untwisted gauge theories presented in Sec.~\ref{sec:RSRG}. 
			
		As we noted in the Introduction, this result could have been anticipated from the fact that the twisted and untwisted theories are linked (on  closed manifolds) by a unitary transformation which leaves the critical properties invariant. However, such an indirect argument provides little insight into the geometric structures that emerge at strong disorder. A second subtlety is that although the {\it bulk} properties of the twisted and untwisted theories are identical, they differ  on boundaries. Twisted gauge theories host nontrivial gapless boundary modes that are absent in the untwisted case. Therefore, the unitary map between the two breaks down on manifolds with a boundary, where they are physically distinct. In order to address these geometrical questions and to lay the groundwork for future investigations of the interplay of strong disorder with the gapless boundary modes, it is clearly desirable to formulate an RSRG procedure that can be implemented  {\it directly} on the DW models. While we defer many details to appendices, in this section we devise such an RSRG scheme that treats the topological structure in a consistent manner.

			As we detail in Appendix~\ref{app:lattice-iso-DW}, a central issue in implementing the RG scheme in the DW model  is that the set of allowed lattice isomorphisms that are compatible with these models is more restrictive than in untwisted case. This leads to a modified RG flow of the graph $\mathcal{G}$. The key difference relative to the untwisted case lies in the requirement that at each RG stage, $\mathcal{G}$ must preserve the vertex ordering information in order to properly define link orientations needed to compute the phase factor $e^{i \phi_{g}(\omega)}$.
			
			To highlight the above issue, we first discuss the simplest case of a short-link term decimation starting from the Hamiltonian in Eq.~\eqref{eq:DW-Ham}. Using similar calculations as in the untwisted case and the fact that $A_v^{g, \omega}$ is the identity operator when $g=e$, we find that the first order contribution is trivial. Terms generated at second order take an identical form to Eq.~\eqref{eq:sec_ord_link}, following the substitution $A_v \mapsto A^\omega_v$:
			\begin{align}
				&C_l \big(A_v^\omega A_{v'}^\omega \big) C_l = \frac{1}{|G|} \Big(\frac{1}{|G|} \sum_{g \in G} A_v^{g, \omega} A_{v'}^{g, \omega} \Big) C_l .
				\label{eq:J_renorm}
			\end{align}
			Crucially, when acting on the projected Hilbert space ($C_l=1$), the operator $A_{v v'}^\omega\equiv\frac{1}{|G|} \sum_{g \in G} A_v^{g, \omega} A_{v'}^{g, \omega} $ satisfies all the desired algebraic properties of a regular vertex term outlined above. This fact is important in maintaining a self-similar structure of Hamiltonian terms in our RG procedure.
			
			For the untwisted case, at this point we would continue the decimation process by eliminating the link variable $l$ and merging the vertices $v$ and $v'$ into a single vertex, thereby renormalizing the planar graph $\mathcal{G}'$. However, in the twisted case, this would lead to an ambiguity in assigning a vertex index for the merged vertex. We emphasize that while the projected link variable on $l$ is no longer a dynamical degree of freedom, we must still keep memory of the associated geometrical information, encoded in $\mathcal{G}$, to properly define the DW theory. In order to address  this complication, we modify our RG scheme as follows. First, we  generalize our definition of a vertex term to encompass a set of multiple microscopic vertices connected by decimated links, as in Eq.~\eqref{eq:J_renorm}. Second, we modify the link decimation, so that, rather than fuse vertices into a single new vertex, we merge the associated vertex sets into new, larger set. Since the substructure of the microscopic vertices (including, crucially, their indexing) is preserved in this procedure, at all stages we can always compute the necessary phase factors. Finally, in the vertex decimation, we are allow to remove the associated vertices and its connected links from the planar graph unlike in the link decimation. We now elaborate on the technical details that these modifications entail.
			
			To that  end we introduce the collection of renormalized vertex sets $\mathcal{V}_R$. Formally, $\mathcal{V}_R$ is a collection of equivalence classes, each comprising vertices defined on the planar graph $\mathcal{G} = (\mathcal{V}, \mathcal{L})$. In other words, elements in $\mathcal{V}_R$ are mutually disjoint and the union of all elements in $\mathcal{V}_R$ equals $\mathcal{V}$. Each element $[w] \in \mathcal{V}_R$ consists of vertices belonging to $\mathcal{V}$: $[w]=\{v_{1}, v_{2}, \ldots, v_{k}\} \subset \mathcal{V}$. Correspondingly, we define a collection of distinct vertex set pairs $\mathcal{L}_{\mathcal{V}_R} = \{ l = \langle [w], [w'] \rangle \vert [w],[w']\in \mathcal{V}_R \textrm{ and } [w] \ne [w'] \}$. As before, the Hilbert space $\mathcal{H}_\mathcal{G}$ is defined in terms of $\mathcal{L}$ in $\mathcal{G}$ with the zero-flux constraint imposed on every plaquette. However, here, we impose additional constraints on the Hilbert space by projecting to $C_{\langle v, v' \rangle}=1$ for every distinct vertex pair $\langle v, v' \rangle$ belonging to the same equivalence class in $\mathcal{V}_R$, i.e., $v,v'\in[w]$. We denote the subspace defined by these projections as $\mathcal{P}_R$.
			
			Following the above discussion, we can define a generalized Hamiltonian as
			\begin{equation}
			H = - \sum_{[w] \in \mathcal{V}_R} J_{[w]} A_{[w]}^\omega -  \sum_{l \in \mathcal{L}_{\mathcal{V}_R}} h_l C_l ,
			\label{eq:Ham-DW-gen}
			\end{equation}
			where $A_{[w]}^\omega$ is a generalized vertex term defined as
			\begin{equation}
			A_{[w]=\{v_1, \ldots, v_k\}}^\omega \equiv \frac{1}{|G|} \sum_{g \in G} A_{v_1 \ldots v_k}^{g, \omega} \equiv \frac{1}{|G|} \sum_{g \in G} A_{v_1}^{g, \omega} \cdots A_{v_k}^{g, \omega} 
			\end{equation}
			and the link term for $l = \langle [v], [v'] \rangle \in \mathcal{L}_{\mathcal{V}_R}$ is defined as
			\begin{equation}
			C_{l = \langle [w], [w'] \rangle} \equiv C_{\langle v, v' \rangle} .
			\end{equation}
			for some representative vertices $v \in [w]$ and $v' \in [w']$. The operator $C_l$ is independent of the above choice, since all link variables connecting vertices in the same equivalence class are trivial. Using the commuting properties of the vertex operators, one can show that the generalized vertex terms in Eq.~\eqref{eq:Ham-DW-gen} are also mutually commuting projectors. The generalized vertex term $A_{[w]=\{v_1, \ldots, v_k\}}^\omega$ is well-defined on the subspace $\mathcal{P}_R$, since the product of vertex terms $A_{v_1}^{g, \omega} \cdots A_{v_k}^{g, \omega}$ preserves the subspace for every $g \in G$. On the other hand, an individual vertex term in $A_{v_1}^{g, \omega} \cdots A_{v_k}^{g, \omega}$ does not preserve the subspace $\mathcal{P}_R$}. 
		
		We now turn to construct our RSRG rules for the DW model. Initially, we start with a planar graph $\mathcal{G} = (\mathcal{V}, \mathcal{L})$ comprising triangular plaquettes, and set $\mathcal{V}_R = \mathcal{V}$. As the RG flows, we renormalize both $\mathcal{G}$ and $\mathcal{V}_R$, where the former defines the lattice structure and the latter keeps track of Hamiltonian terms. Both degrees of freedom are used to define the projected Hilbert space on which the renormalized Hamiltonian acts. 
		
		First, suppose that the strongest term in Eq.~\eqref{eq:Ham-DW-gen} is a link term $C_l$ for some link $l = \langle [w],[w'] \rangle \in \mathcal{L}_{\mathcal{V}_R}$. If necessary, we initially employ a series of lattice isomorphisms, introduced in Appendix~\ref{app:lattice-iso-DW}, that convert  $C_l$ into a short-link term. To simplify the calculation, we perform the computations on $\mathcal{H}_\mathcal{G}$ instead of the subspace $\mathcal{P}_R$; however, the end results hold when restricted to $\mathcal{P}_R$. To this end, we consider the link term $C_{l_0 \equiv \langle v, v' \rangle}$ in $\mathcal{H}_\mathcal{G}$, where $v \in [w]$ and $v' \in [w']$ are representative vertices. Other choices of representatives of $[w]$ and $[w']$ would give distinct link terms in $\mathcal{H}_\mathcal{G}$, while they all become $C_l$ in $\mathcal{P}_R$.
		
		We begin by considering the renormalization of a unique vertex operator $A_{[w]}^\omega$ satisfying $v\in[w]$. {If we also have $v'\in[w]$, the link $l = \langle [w],[w'] \rangle $ is trivial and hence does not appear in Eq.~\eqref{eq:Ham-DW-gen}.}  It is therefore convenient to introduce the  notation $[w]_v \equiv [w] \setminus \{v\}$ to denote the set of vertices that are equivalent to $v$ but distinct from it. Using the fact that $A_{q}^{g, \omega}$ commutes with $C_{l_0}$ for vertex $q \in [w]_v$ and $C_{l_0} \big( A_{v}^{g, \omega} \big) C_{l_0} = \delta_{g, e} C_{l_0}$, we get
		\begin{align}
			C_{l_0} \big( A_{[w]}^\omega \big) C_{l_0} &= \frac{1}{|G|} \sum_{g \in G} \Big( \prod_{q \in [w]_v} A_q^{g, \omega} \Big) C_{l_0} \big( A_{v}^{g, \omega} \big) C_{l_0} \nonumber \\
			&= \frac{1}{|G|} C_{l_0} .
			\label{eq:DW-Cl-A-vks-Cl}
		\end{align}
		Note that the computations are done in $\mathcal{H}_\mathcal{G}$, since as we commute $C_{l_0}$ through $A_q^{g, \omega}$, we no longer stay on the subspace $\mathcal{P}_R$. However, the end result holds in $\mathcal{P}_R$ and can be written as $C_l \big( A_{[w]}^\omega \big) C_l = \frac{1}{|G|} C_l$, which implies that only trivial terms are generated in first-order perturbation theory.
		
		Moving to second order, we consider an additional vertex operator $A^\omega_{[w']}$ with $v' \in [w']$. Using $C_{l_0} \big( A_{v}^{g, \omega} A_{v'}^{g', \omega} \big) C_{l_0} = \delta_{g, g'} A_{v}^{g, \omega} A_{v'}^{g', \omega} \, C_{l_0}$ and $[A_q^{g, \omega}, C_{l_0}] = [A_{q'}^{g, \omega}, C_{l_0}] = 0$ for all $q \in [w]_v$, $q' \in [w']_{v'}$, and $g \in G$, we get
		\begin{align}
			&C_{l_0} \big( A_{[w]}^\omega A_{[w']}^\omega \big) C_{l_0} \nonumber \\
			&= \frac{1}{|G|^2} \sum_{g,g' \in G} \Big( \prod_{q\in [w]_v} \big( A_{q}^{g,\omega} \big) \prod_{q' \in [w']_{v'} } \big( A_{q'}^{g',\omega} \big) \Big) \nonumber \\
			& \qquad \qquad \qquad \times C_{l_0} \big( A_{v}^{g, \omega} A_{v'}^{g', \omega} \big) C_{l_0} \nonumber \\
			&= \frac{1}{|G|^2} \sum_{g \in G} \Big( \prod_{q \in [w]_{v} } \big( A_{q}^{g,\omega} \big) \prod_{q' \in [w']_{v'} } \big( A_{q'}^{g,\omega} \big) \Big) A_{v}^{g, \omega} A_{v'}^{g, \omega} \, C_{l_0} \nonumber \\
			&= \frac{1}{|G|} A_{[w]\cup[w']}^{\omega} C_{l_0},
		\end{align}
		where the generated vertex term $A_{[w'']=[w]\cup[w']}^{\omega}$ commutes with $C_{l_0}$. As before, while the computations are done in $\mathcal{H}_\mathcal{G}$, the end result also holds in $\mathcal{P}_R$ and can be written as $C_{l} \big( A_{[w]}^\omega A_{[w']}^\omega \big) C_{l} = \frac{1}{|G|} A_{[w]\cup[w']}^{\omega} C_{l}$}. This motivates us to renormalize $\mathcal{V}_R$ by merging $[w]$ and $[w']$ into a single set $[w'']=[w]\cup[w']$. In addition, the decimation procedure introduces an additional projection, $C_{l_0}=1$, in $\mathcal{P}_R$.
	
	Second, we consider a vertex term decimation $A_{[w]}^\omega$ for some $[w] \in \mathcal{V}_R$.  As is the case for the link term decimations, computations are done on $\mathcal{H}_\mathcal{G}$ instead of the projected subspace $\mathcal{P}_R$. Since all vertices $\{v_1,\dots,v_k\}$ in $[w]$ were generated by decimating short-link terms, there exist a path $\pi_{v_1 \ldots v_k}$ connecting all and only the vertices in $\{v_1, \cdots, v_k\}$.\footnote{For $\{v \} \in \mathcal{V}_R$, $\pi_{v}$ is just a trivial constant path staying at $v$.} Prior to the decimation step, we employ a series of lattice isomorphisms to isolate $\{v_1, \cdots, v_k\}$ by a triangular plaquette; see Appendix~\ref{app:lattice-iso-DW} for a detailed explanation. 
	
	If a link term does not start or end on $[w]$, one can choose its defining path so that it never intersects $\pi_{v_1 \ldots v_k}$ by using the path deformations explained in Appendix~\ref{app:lattice-iso}. This allows us to conclude that such link term commutes with $A_{[w]}^\omega$. Next, we consider a link term $C_{\langle [w], [w'] \rangle}$, where $[w'] \ne [w]$.  To this end, we pick representative vertices  $v\in[w]$ and $v' \in [w']$ and consider the link term $C_{\langle v, v' \rangle}$ in $\mathcal{H}_\mathcal{G}$. Since $v' \notin [w]$, we can choose the defining path $\pi_{\langle v, v' \rangle}$ of $C_{\langle v,v' \rangle}$ such that $\pi_{\langle v, v' \rangle}$ and $\pi_{v_1 \ldots v_k}$ intersect only at a single vertex $v(\in[w])$. This implies that $[A_{q}^{g, \omega}, C_{\langle v, v' \rangle}] = 0$ for all $q\in[w]$ obeying $q \ne v$. Using a similar algebra employed in Eq.~\eqref{eq:Av-Cl-Av}, we obtain 
	\begin{align}
		&A_{[w]}^\omega \big( C_{\langle v, v' \rangle} \big) A_{[w]}^\omega \nonumber \\
		&= \frac{1}{|G|^2} \sum_{g, g' \in G} \Big( \prod_{q \in [w]_v} A_{q}^{g, \omega} \Big) \Big( \prod_{q' \in [w]_v} A_{q'}^{g', \omega} \Big) \Big( A_{v}^{g, \omega} \big( C_{\langle v, v'  \rangle} \big) A_{v}^{g', \omega} \Big) \nonumber \\
		&= \frac{1}{|G|^2} \sum_{g, g' \in G} \Big( \prod_{q \in [w]_v} A_{v}^{g \cdot g', \omega} \Big) \Big( A_{v}^{g \cdot g', \omega} C_{\langle v, v' \rangle}^{(g')^{-1}} \Big) \nonumber \\
		&= \frac{1}{|G|} \Big( \frac{1}{|G|} \sum_{h \in G}  \prod_{q \in [w]} A_{q}^{h, \omega}  \Big) \Big( \sum_{g' \in G} C_{\langle v, v' \rangle}^{(g')^{-1}} \Big) \nonumber \\
		&= \frac{1}{|G|} A_{[w]}^\omega,
		\label{eq:Aw-Cl-Aw_DW}
	\end{align}
	where from the second line to the third line, we used $C_{\langle v, v' \rangle} A_{v}^{g', \omega} = A_{v}^{g', \omega} C_{\langle v, v' \rangle}^{(g')^{-1}}$ and the fact that $A_{v}^{g, \omega} A_{v}^{ g' \omega} = A_{v}^{g \cdot  g', \omega}$, from the third line to the fourth line, we relabel $h = g \cdot g'$, and in the last equality, we used the fact that $\sum_{g' \in G} C_{\langle v, v' \rangle}^{(g')^{-1}}$ is the identity operator. This implies that the first order contributions are trivial. 
	
	To see how link terms are generated in second-order perturbation theory, we consider an additional link term $C_{\langle [w''], [w]\rangle}$, where $[w''] \ne [w]$ and $[w''] \ne [w']$. We consider link terms $C_{\langle v, v' \rangle}$ and $C_{\langle v'', v \rangle}$ in $\mathcal{H}_\mathcal{G}$, where $v$, $v'$, and $v''$ are representatives of $[w]$, $[w']$, and $[w'']$. We note that
	\begin{equation}
	C_{\langle v'', v \rangle} C_{\langle v, v' \rangle} = C_{\langle v'', v' \rangle} C_{\langle v'', v \rangle} , \nonumber
	\end{equation}
	where $C_{\langle v'', v' \rangle} $ commutes with $A_{[w]}^\omega$ since $v', v'' \notin [w]$. This implies that 
	\begin{equation}
	A_{[w]}^\omega \big( C_{\langle v'', v \rangle} C_{\langle v, v' \rangle} \big) A_{[w]}^\omega = \frac{1}{|G|} C_{\langle v'', v' \rangle} A_{[w]}^\omega ,
	\label{eq:Aw-Cl-Cl-Aw_DW}
	\end{equation}
	which shows that the link term $C_{\langle [w''], [w] \rangle}$ is generated in the subspace $A_{[w]}^\omega =1$. As before, Eqs.~\eqref{eq:Aw-Cl-Aw_DW} and \eqref{eq:Aw-Cl-Cl-Aw_DW} are derived in $\mathcal{H}_\mathcal{G}$, but the end result holds in $\mathcal{P}_R$.
	
	Remarkably, unlike the link decimation step, it is possible to completely eliminate the decimated vertex set $[w]$ by removing $[w]$ from $\mathcal{V}_R$ and all $v \in [w]$ and its adjacent links from the planar graph $\mathcal{G}$. The precise lattice isomorphism accomplishing the above step is explained in Appendix~\ref{app:lattice-iso-DW}.
	
	While performing the RSRG on the DW model, we keep track of both the collection of renormalized vertices $\mathcal{V}_R$ and also the planar graph $\mathcal{G}$. While the resulting RG flow of the planar graph structure deviates from the simpler untwisted case, the flow of coupling constants is {\it identical}. {Crucially, this leads to the prediction that critical properties (such as the energy-length scaling) of the DW models will  coincide with the untwisted gauge theory results --- a new manifestation of superuniversality distinct from that between different untwisted theories.}
	
We observe that our results also apply to order-disorder transitions in SPT and SET phases in the presence of strong quenched disorder. This  follows since an (either partial or complete) un-gauging duality transformation of DW models~\cite{PhysRevB.87.155115, PhysRevB.94.235136, PhysRevB.96.115107}  provides a systematic construction of a large class of SPT and SET theories in two spatial dimensions. Using similar arguments as in Sec.~\ref{sec:duality}, we can demonstrate that the RSRG rules are mapped one to the other under the duality. This implies that the resulting order-disorder phase transitions (on the SPT/SET side) belong to the same universality class as the corresponding confinement-deconfinement transition (on the twisted gauge side). Moreover, since the RSRG rules are identical and independent of the underlying gauge (or symmetry) group $G$ up to irrelevant constant factors, we conclude that all order-disorder phase transition in SET and SPT phases realizable by DW models belong to the same universality class as that of the random quantum Potts model and the family of discrete lattice gauge theories considered previously.

As a final remark, we note that  while the bulk critical properties of the twisted and untwisted theories are identical (as  indirectly inferred via the unitary untwisting trick and directly demonstrated using our RSRG arguments), and hence superuniversal, it would be interesting to also understand to what extent the {\it boundary critical physics} of the twisted theories exhibits superuniversal scaling. As we have mentioned above, a key distinction between the twisted and untwisted theories lies in the  fact that the former hosts nontrivial gapless boundary modes absent in the latter. Understanding how these modes {couple to critical bulk fluctuations} at strong-randomness critical points in two dimensions is an important problem for the future. Notably, the unitary mapping between twisted and untwisted theories fails in the presence of a boundary, leaving our RSRG approach as the only one suited to tackle such questions. Recent work involving one of the present authors has explored a similar question in the setting of 1D SPTs~\cite{RV_new_SPT}; results in higher dimensions are likely as rich or even richer.
	
	\section{Levin-Wen Models: Non-Abelian Anyons and Obstructions to RSRG}
	\label{sec:LW}
	As an attempt at a  second extension, we consider the generalization of our RSRG approach to Levin-Wen models~\cite{PhysRevB.71.045110}, which are believed to realize all nonchiral topological orders admitting commuting projector Hamiltonians~\cite{PhysRevB.94.235136, PhysRevB.96.115107}. As we demonstrate below, for non-Abelian topological orders, the RSRG of the Levin-Wen model deviates from the RSRG structure presented above, as nontrivial terms can be generated already at  first order in perturbation theory. Nevertheless, the RSRG structure in the Levin-Wen model exhibits interesting universal features that we present below, though a complete solution has eluded us to date.
	
	For concreteness, we focus on the doubled Fibonacci phase, which is the simplest example of non-Abelian topological order realized by the Levin-Wen model. Note however that our discussion below is quite generic and, unless stated otherwise, can be applied to other types of topological order. We focus, for the most part, on the honeycomb lattice structure but any trivalent planar graph would work equally well. For completeness, we provide a summary of the Levin-Wen model in Appendix~\ref{appendix:LW}.
	
\begin{figure}[!t]
		\includegraphics[width=1.0\columnwidth]{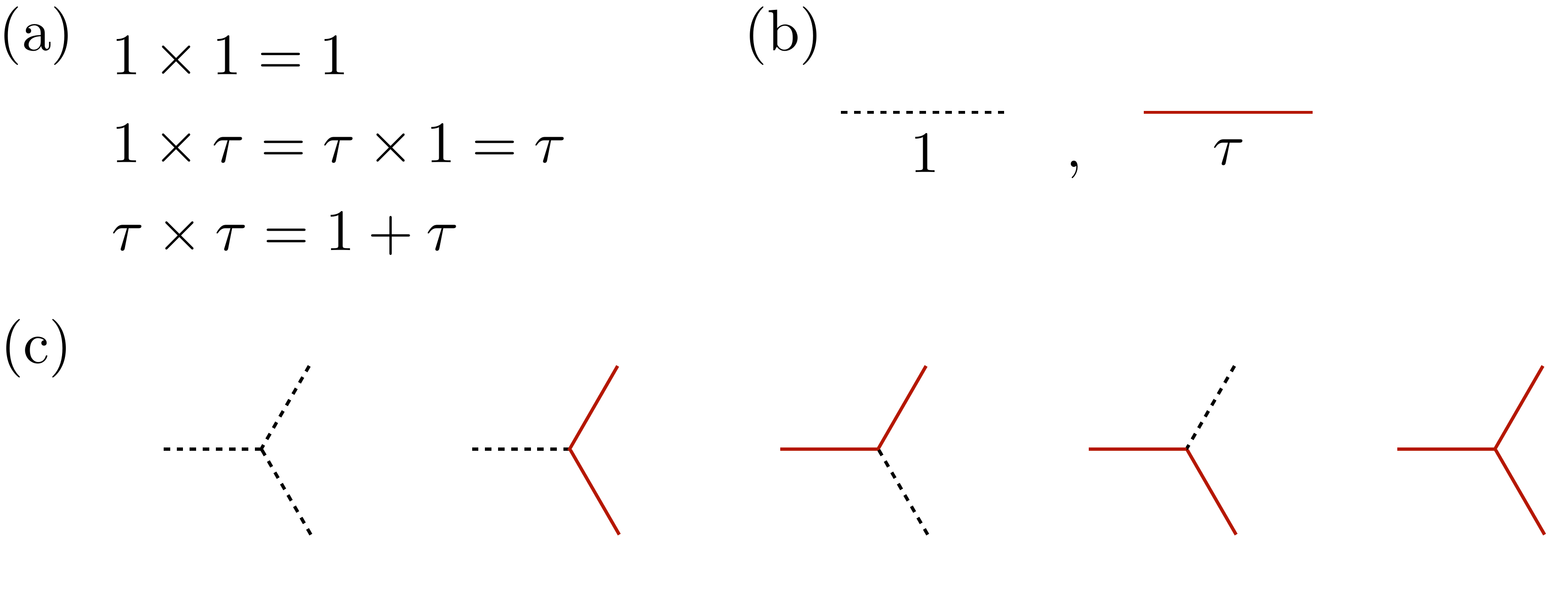}
		\caption{Summary of the Fibonacci anyonic system. (a) The fusion rule of the Fibonacci anyonic system, which has two anyons: the trivial/vacuum anyon ($1$) and a non-Abelian `Fibonacci' anyon  ($\tau$), which are (b)  respectively denoted by  a dashed line (equivalent to an empty line) and by a solid red line. (c) Possible fusion channels in the Fibonacci anyonic system, which are allowed set of configurations set by $A_v=1$ constraint in the Levin-Wen model.}
		\label{fig:Fibonacci}
	\end{figure}

\begin{figure}[!t]
		\includegraphics[width=0.8\columnwidth]{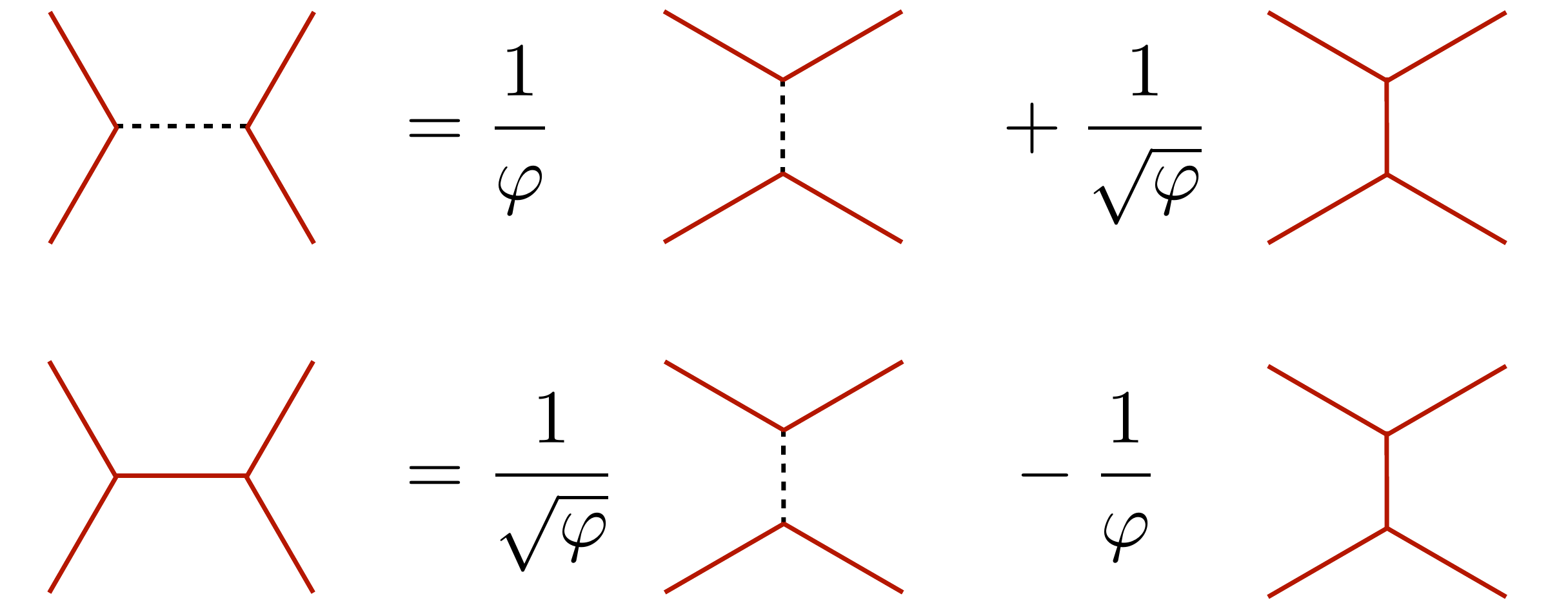}
		\caption{Change of basis in the fusion tree of the Fibonacci anyonic system from which the nontrivial $F$-matrix $[F_{\, \tau \tau}^{\tau \tau}]$ can be read off.}
		\label{fig:Fibonacci-F}
	\end{figure}

\begin{figure*}[!t]
		\includegraphics[width=2.0\columnwidth]{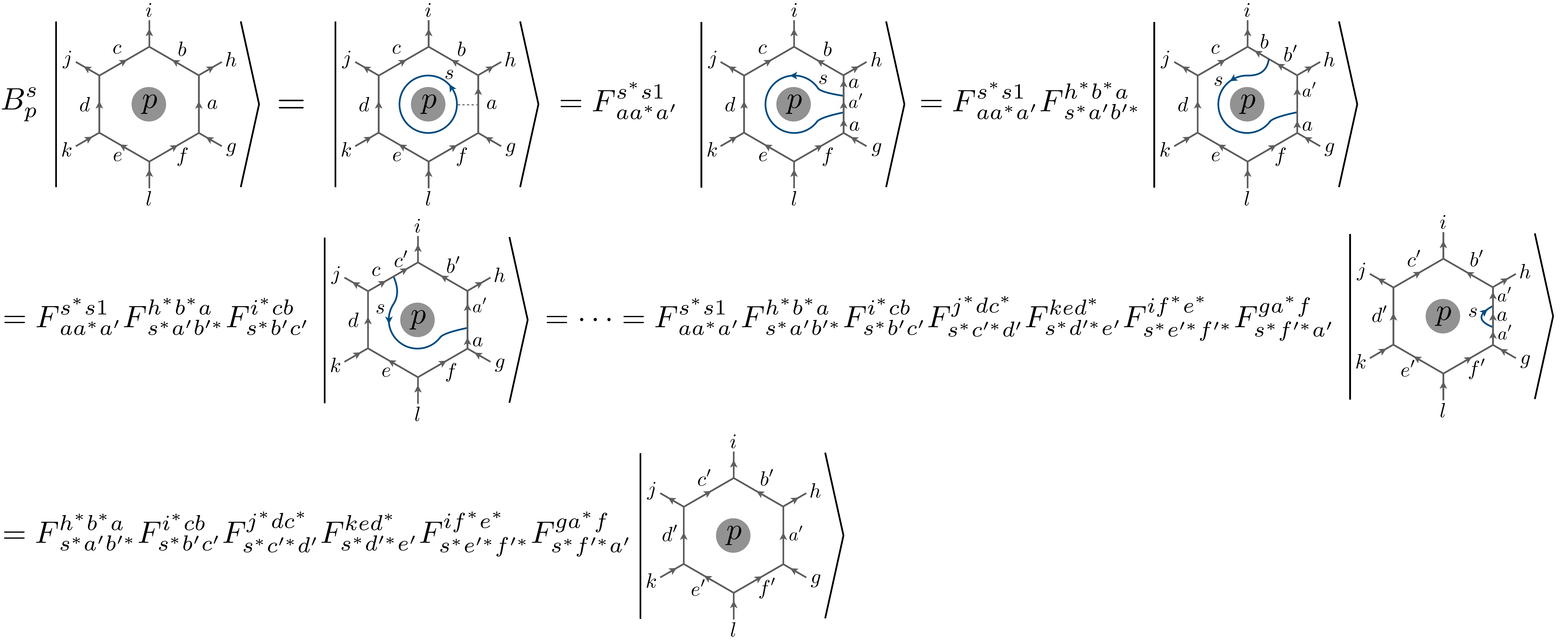}
		\caption{Definition of the $B_p^s$ term in the Levin-Wen model where we use the Einstein summation convention, i.e., summation is assumed for repeated indices. $B_p^s$ inserts an anyonic flux $s$ around the plaquette $p$; we use a series of $F$-matrices to compute the matrix element of $B_p^s$.  $B_p^s$ is a 12-spin interaction for a hexagonal plaquette $p$, and it acts diagonally on states on external legs denoted as states $i,j, \cdots, h$.}
		\label{fig:LW-B_p^s}
	\end{figure*}

	The local Hilbert space is defined on each link and is labeled by the trivial anyon $1$ and the Fibonacci anyon $\tau$, i.e., $\{\vert s \rangle : s \in\mathcal{C}= \{1, \tau\}\}$ [See Figs.~\ref{fig:Fibonacci} and \ref{fig:Fibonacci-F} for the summary of Fibonacci anyonic system and the nontrivial $F$-matrix]. We restrict the tensor product Hilbert space by imposing on each vertex a gauge constraint, which enforces the fusion rules on every vertex. The allowed configurations in the doubled Fibonacci case are presented in Fig.~\ref{fig:Fibonacci} (c). The Levin-Wen Hamiltonian~\cite{PhysRevB.71.045110} is given by a sum of commuting projectors, 
	\begin{equation}
	H_\textrm{LW} = - \sum_p J_p B_p = - \sum_p J_p \sum_{s \in \mathcal{C}} \frac{d_s}{D^2} B_p^s, 
	\label{eq:Ham-LW-original}
	\end{equation}
	where $p$ denotes the plaquettes of the honeycomb lattice, the coupling constant $J_p >0$, $d_s$ is the quantum dimension of the anyon $s$\footnote{We closely follow the convention in Ref.~\onlinecite{PhysRevB.71.045110} where the quantum dimension can be negative (e.g., $d_s=-1$ for the semion $s$). The additional minus factor comes from the Frobenius-Schur indicator $\varkappa_s$ of the anyon.} ($d_1 =1$ and $d_\tau = \varphi$, the golden ratio), and $\displaystyle D = \sqrt{\sum_{s \in \mathcal{C}} d_s^2}$ is the total quantum dimension. The plaquette term, $B_p^s$, inserts a flux loop generated by anyon $s$ to the plaqutte $p$, and is defined in Fig.~\ref{fig:LW-B_p^s}. In the honeycomb lattice, $B_p^s$ is a 12-link interaction term, which acts diagonally on states residing on the {\it external legs} (the links which are not part of the plaquette $p$). The Levin-Wen Hamiltonian \eqref{eq:Ham-LW-original} is exactly solvable as it is a sum of commuting projectors, i.e., $(B_p)^2 = B_p$ and $[B_p, B_{p'}] = 0$ for all $p, p'$. 
	
	To break the integrability of Eq.~\eqref{eq:Ham-LW-original} and potentially drive a confinement transition, we add link terms $C_l = \vert 1 \rangle \langle 1 \vert_l$ on every link $l$, which projects onto the trivial state $\vert 1 \rangle $. This leads to the Hamiltonian
	\begin{equation}
	\tilde{H}_\textrm{LW} = - \sum_p J_p B_p - \sum_l h_l C_l .
	\label{Ham-LW}
	\end{equation}
	As before, we introduce disorder by assuming that the coupling constants, $J_p$ and $h_l$, are random and non-negative. 
	\begin{figure*}[!t]
		\includegraphics[width=2.0\columnwidth]{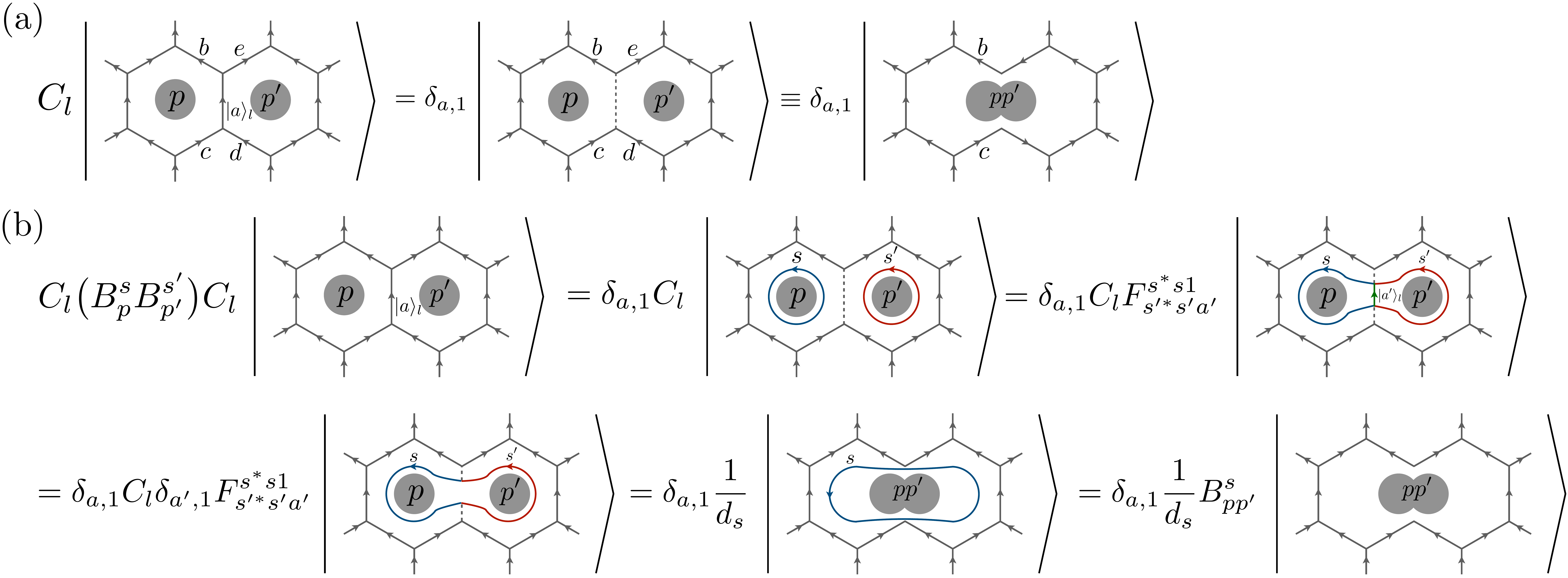}
		\caption{(a) For a link decimation $C_l=1$, we polarize the spin state at $l$ to the trivial state $\vert 1 \rangle$ and remove link $l$ from the lattice. Such link decimation naturally identifies the states $b = e^*$ and $c = d^*$, where ${}^*$ appears due to the direction of links. Accordingly, four adjacent links to $l$ become two links after the identifications as described in the RHS. (b) Computation of $C_l \big( B_p^s B_{p'}^{s'} \big) C_l$ on the $C_l^a (\equiv \vert a \rangle \langle a \vert_l)=1$ subspace. The summation over $a'$ is implicit in the RHS of the second equality. As a result of computation, we get the plaqutte term $B_{pp'}^s$ associated with the larger plaquette $p p'$.}
		\label{Fig:LW-link-decimation}
	\end{figure*}
	
	We now construct the RSRG rules of the Levin-Wen model. We begin with the rules pertinent to link decimation. We will first consider only nearest-neighbor links, and discuss the more involved case of long link decimations at the end of this section.  Suppose that the strongest term in the Hamiltonian is the link term $h_l C_l$ for some link $l$;  we then project onto the trivial anyon label on the link $l$.  Consequently,  to enforce the gauge constraints on the resulting quantum state, we must identify states in the remaining links up to possible changes in the corresponding link orientations, as described in Fig.~\ref{Fig:LW-link-decimation} (a). Similarly to previous cases, to first order in perturbation theory, we obtain only trivial terms, since
	\begin{equation}
	C_l \big( B_p \big) C_l = \sum_{s \in \mathcal{C}} \frac{d_s}{D^2} C_l \big( B_p^s \big) C_l = \frac{1}{D^2} C_l,
	\label{Eq:LW_Cl-Bp-Cl}
	\end{equation}
	where all the other terms except the term with $s=1$ in the summation vanish, as $s=1$ is the only term preserving the state $\ket{1}_l$ at $l$ under the insertion of the anyon flux $s$. To compute  nontrivial terms appearing in second-order perturbation theory, we consider the combination $C_l (B_p B_{p'} ) C_l$, where $p$ and $p'$ are two distinct plaquettes adjacent to the decimated link $l$. Following the manipulations depicted in Fig.~\ref{Fig:LW-link-decimation} (b), we get
	\begin{align}
		C_l (B_p B_{p'} ) C_l &= \frac{1}{D^4} \sum_{s,s' \in \mathcal{C}} d_s d_{s'} C_l \big( B_p^s B_{p'}^{s'} \big) C_l \nonumber \\
		&= \frac{1}{D^4} \sum_{s \in \mathcal{C}} d_s B_{p p'}^s = \frac{1}{D^2} B_{p p'} .
	\end{align}
	The resulting plaquette term, $B_{pp'}$, is defined on the merged plaquette $pp'$, which belongs to the renormalized lattice following the removal of the link $l$. Importantly, the renormalized Hamiltonian maintains the same form as the original model defined in Eq.~\eqref{Ham-LW}, supporting the self-similar structure of the RSRG link decimation step.
	
	\begin{figure*}[!t]
		\includegraphics[width=2.0\columnwidth]{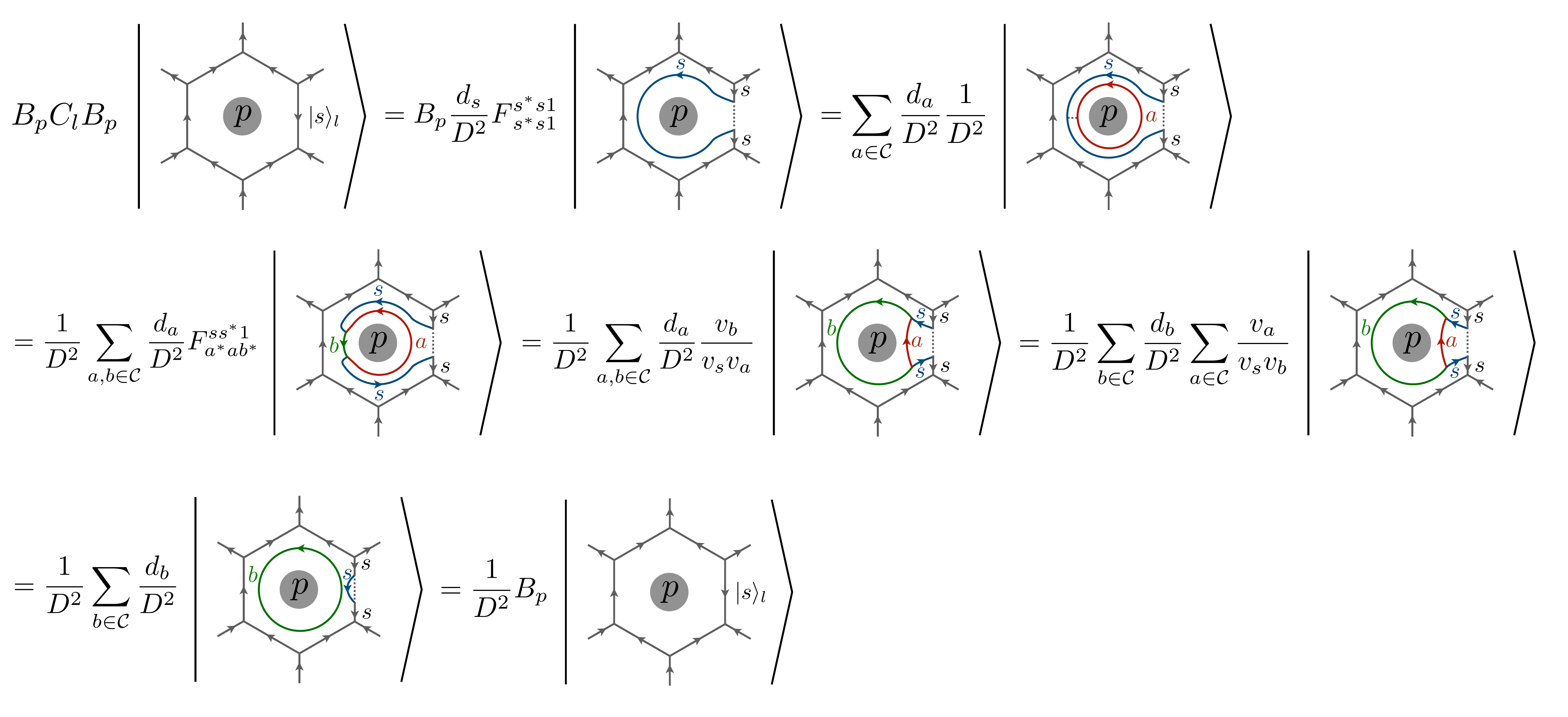}
		\caption{The proof of Eq.~\eqref{Eq:LW_Bp-Cl-Bp} on $C_l^s (\equiv \vert s \rangle \langle s \vert_l) =1$ subspace. Since $s$ is arbitrary, Eq.~\eqref{Eq:LW_Bp-Cl-Bp} holds in general. In the proof, we use the fact that $F_{a^* a s}^{b^* b 1} = \frac{v_s}{v_a v_b}$ where $v_s = \sqrt{d_s}$ and employ a handle slide from 3-manifold theory~\cite{kauffman1994temperley} in the second line.}
		\label{fig:LW-Bp-Cl-Bp}
	\end{figure*}
	
	Next, assuming that the strongest remaining term in the Hamiltonian is a plaquette term $J_p B_p$ for some plaquette $p$, we enforce $B_p = 1$ or equivalently place a trivial flux excitation at $p$. Doing so defines the ground state subspace as the set of states satisfying $B_p=1$ in addition to other gauge constraints. This projection renormalizes link terms acting on the boundary links of $p$. Explicitly, for a link $l$ belonging to the boundary of $p$, we get
	\begin{equation}
	B_p ( C_l ) B_p = \frac{1}{D^2} B_p ,
	\label{Eq:LW_Bp-Cl-Bp}
	\end{equation}
	as proven in Fig.~\ref{fig:LW-Bp-Cl-Bp}. Other link terms, including the link terms acting on the external legs to $p$, remain invariant since they commute with $B_p$. This implies that first-order perturbation theory gives only trivial terms. Moving to second-order perturbation theory, let us consider two distinct links $l_1$ and $l_2$ which belong to the boundary of the decimated plaquette $p$. To compute their contribution, we consider the following term,
	\begin{equation}
	B_p ( C_{l_1} C_{l_2} ) B_p  .
	\label{Eq:LW-Bp-Cl1-Cl2-Bp}
	\end{equation}
	If $l_1$ and $l_2$ share the common vertex $v$, there must exist a distinct link $l_3$ that shares the same vertex $v$, which is an external leg with respect to the plaquette $p$. As a result, the gauge constraint at $v$ dictates the relation $C_{l_1} C_{l_2} = C_{l_3} C_{l_2}$. Hence, for such links, we get
	\begin{align}
		B_p ( C_{l_1} C_{l_2} ) B_p &= B_p ( C_{l_3} C_{l_2} ) B_p = C_{l_3} B_p ( C_{l_2} ) B_p \nonumber \\ 
		&= \frac{1}{D^2} C_{l_3} B_p ,
	\end{align}
	where we used the fact that $B_p$ acts diagonally on the state at $l_3$ and  Eq.~\eqref{Eq:LW_Bp-Cl-Bp} to compute $B_p (C_{l_2} ) B_p$. Therefore, restricting to the $B_p=1$ subspace, the renormalized link term is identified with a regular link term $C_{l_3}$. 
	
	On the other hand, for links $l_1$ and $l_2$ that do not share any common vertex we must introduce a ``long-link'' term: 
	\begin{align}
		C_{l_1, l_2} \equiv D^2 B_p \big( C_{l_1} C_{l_2} \big) B_p ,
	\end{align}
	where $D^2$ is a normalization factor enforcing the projection condition $(C_{l_1, l_2})^2 = C_{l_1, l_2}$, see Eq.~\eqref{Eq:LW_Cl-Bp-Cl}. Note that the plaquette $p$ is implicit in the notation $C_{l_1, l_2}$. Moreover, due to presence of $B_p$ in the definition, $C_{l_1, l_2}$ acts not only on links $l_1$ and $l_2$, but also on links on the boundary of $p$ and on external legs of $p$. Below, we will demonstrate that the long-link terms in the Levin-Wen model {\it do not} commute with each other, i.e., $[C_{l_1 l_2}, C_{l_3 l_4}] \ne 0$, in the general case. 
	
	\begin{figure}[!t]
		\includegraphics[width=0.4\columnwidth]{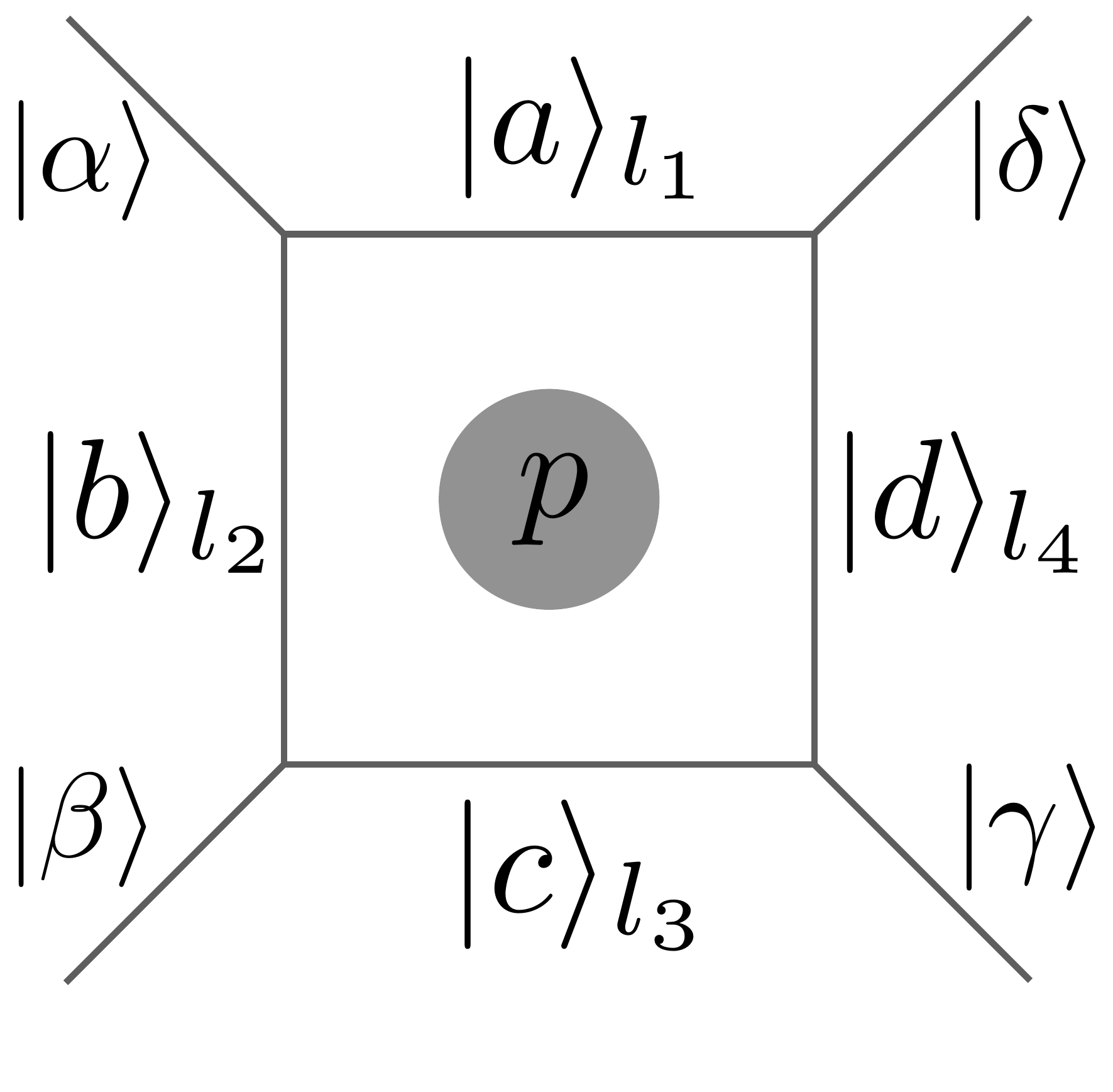}
		\caption{A configuration with a square plaquette $p$ together with four boundary legs $l_1$, $l_2$, $l_3$, and $l_4$, and four external legs. This configuration serves as a minimal example where long-link terms do not commute, see main text for more details.}
		\label{fig:LW-Fibo-LongLink}
	\end{figure}

	As a concrete example for this effect, we focus on the doubled Fibonacci phase. We consider a square plaqutte with four external legs as depicted in Fig.~\ref{fig:LW-Fibo-LongLink}. This choice admits the minimal instance of having noncommuting long-link terms.  We will now explicitly construct the matrix representation of $C_{l_1, l_3}$ and $C_{l_2, l_4}$. To facilitate this procedure, we first note that long-link terms act diagonally on states at external legs, denoted by $\vert \alpha \rangle$, $\vert \beta \rangle$, $\vert \gamma \rangle$, $\vert \delta \rangle$ in Fig.~\ref{fig:LW-Fibo-LongLink}.  In particular, we fix the quantum states of external legs to be $\vert \alpha, \beta, \gamma, \delta \rangle = \vert \tau, \tau, \tau, \tau \rangle$. We label states of the dynamical part of the remaining 7-dimensional Hilbert space, with an orthogonal basis
	\begin{align}
		\{ \vert a, b, c, d \rangle \} = \big\{ & \vert 1, \tau, 1, \tau \rangle, \vert \tau, 1, \tau, 1 \rangle, \vert 1, \tau, \tau, \tau \rangle, \vert \tau, 1, \tau, \tau \rangle, \nonumber \\
		& \vert \tau, \tau, 1, \tau \rangle, \vert \tau, \tau, \tau, 1 \rangle, \vert \tau, \tau, \tau, \tau \rangle \big\}.
	\end{align}
	Here, $a$, $b$, $c$, $d$ label states at $l_1$, $l_2$, $l_3$, and $l_4$, respectively, see Fig.~\ref{fig:LW-Fibo-LongLink}. In this basis, we can write,
	\begin{equation}
	\big[ C_{l_1, l_3} \big] = \frac{1}{1 + \varphi^2} \left(
	\begin{array}{ccccccc}
	1 & \frac{1}{\varphi} & 0 & \frac{1}{\sqrt{\varphi}} & 0 & \frac{1}{\sqrt{\varphi}} & 1 \\
	\frac{1}{\varphi} &  \frac{1}{\varphi^2} & 0 &  \frac{1}{\varphi^{3/2}} & 0 &  \frac{1}{\varphi^{3/2}} &  \frac{1}{\varphi} \\
	0 & 0 & 0 & 0 & 0 & 0 & 0 \\
	\frac{1}{\sqrt{\varphi}} &  \frac{1}{\varphi^{3/2}} & 0 & \frac{1}{\varphi} & 0 &  \frac{1}{\varphi} &  \frac{1}{\sqrt{\varphi}} \\
	0 & 0 & 0 & 0 & 0 & 0 & 0 \\
	\frac{1}{\sqrt{\varphi}} &  \frac{1}{\varphi^{3/2}} & 0 & \frac{1}{\varphi} & 0 &  \frac{1}{\varphi} &  \frac{1}{\sqrt{\varphi}} \\
	1 & \frac{1}{\varphi} & 0 & \frac{1}{\sqrt{\varphi}} & 0 & \frac{1}{\sqrt{\varphi}} & 1
	\end{array} \right) \nonumber
	\end{equation}
	and 
	\begin{equation}
	\big[ C_{l_2, l_4} \big] = \frac{1}{1 + \varphi^2} \left(
	\begin{array}{ccccccc}
	\frac{1}{\varphi^2} & \frac{1}{\varphi} & \frac{1}{\varphi^{3/2}} & 0 & \frac{1}{\varphi^{3/2}} & 0 & \frac{1}{\varphi} \\ 
	\frac{1}{\varphi} & 1 & \frac{1}{\sqrt{\varphi}} & 0 & \frac{1}{\sqrt{\varphi}} & 0 & 1 \\ 
	\frac{1}{\varphi^{3/2}} & \frac{1}{\sqrt{\varphi}} & \frac{1}{\varphi} & 0 & \frac{1}{\varphi} & 0 & \frac{1}{\sqrt{\varphi}} \\ 
	0 & 0 & 0 & 0 & 0 & 0 & 0 \\ 
	\frac{1}{\varphi^{3/2}} & \frac{1}{\sqrt{\varphi}} & \frac{1}{\varphi} & 0 & \frac{1}{\varphi} & 0 & \frac{1}{\sqrt{\varphi}} \\ 
	0 & 0 & 0 & 0 & 0 & 0 & 0 \\ 
	\frac{1}{\varphi} & 1 & \frac{1}{\sqrt{\varphi}} & 0 & \frac{1}{\sqrt{\varphi}} & 0 & 1 \\ 
	\end{array} \right), \nonumber
	\end{equation}
	which indeed gives $[C_{l_1, l_3}, C_{l_2, l_4}] \ne 0$.
	
	To see how the above affects our RSRG scheme, suppose that we wish to decimate the long link term $C_{l_1, l_3}$. Unlike in previous cases, such a decimation not only renormalizes plaquette terms but can potentially generate contributions from link terms. Indeed, we find that already in first-order perturbation theory we obtain nontrivial terms. Explicitly, the following nonvanishing contribution
	\begin{equation}
	C_{l_1, l_3} \big( C_{l_2, l_4} \big) C_{l_1, l_3}
	\end{equation}
	cannot be proportional to $C_{l_1, l_3}$. To see why this is the case, observe that on the subspace $\vert \alpha, \beta, \gamma, \delta \rangle = \vert 1, 1, 1, 1 \rangle$, $C_{l_1, l_3}$ and $C_{l_2, l_4}$ are identical, while on the $\vert \alpha, \beta, \gamma, \delta \rangle = \vert \tau, \tau, \tau, \tau \rangle$ subspace they are different and do not commute.
	
	Although there is great similarity between the  RSRG scheme for the Levin-Wen model that results from the above approach, and the one we presented earlier for discrete gauge theories, they still differ nontrivially. Physically, the noncommutative nature of link terms in the Levin-Wen model means that the operation of generating two distinct anyon pairs does not necessarily commute. We note that in all previous cases, both for untwisted and twisted gauge theories, we did not encounter this issue. Therefore, we can not directly generalize our superuniversality claim to the Levin-Wen case. We leave the interesting question of the IR behavior that results from applying RSRG to these models to a future study.

	\section{Excited-state criticality in Abelian topological phases via RSRG-X}
	\label{sec:RSRG-X}
	Lastly, we consider extending our RSRG procedure beyond ground state physics to quantum phase transitions between trivial and topologically ordered excited eigenstates~\cite{Parameswaran_2017}. To probe the excited state physics, we employ the RSRG for excited states (RSRG-X) scheme, which extends the RSRG method beyond ground state properties~\cite{PhysRevX.4.011052, PhysRevLett.114.217201, PhysRevB.93.104205, PhysRevB.95.024205}. Crucially, we will demonstrate that the set of RSRG-X decimation rules is identical to RSRG rules derived in Sec.~\ref{sec:RSRG} and Sec.~\ref{sec:pert}. In the following, we will focus solely on Abelian lattice gauge theories with a discrete gauge group. The reason for this restriction is twofold. First, non-Abelian gauge symmetry is incompatible with full MBL, owing to the extensive degeneracy of highly excited eigenstates~\cite{PhysRevB.94.224206}. Second, although non-Abelian theories can nevertheless exhibit nontrivial IR behavior in the presence of strong randomness, such as putative ``quantum critical glass'' phases~\cite{PhysRevLett.114.217201, PhysRevB.95.024205},  understanding such structure introduces additional difficulties, as we show below. 
	
	The RSRG-X scheme attempts to construct an approximate eigenstate ``sampled'' from the Hamiltonian with an appropriate ``Boltzmann weight'' corresponding to its energy, just as RSRG attempts to construct an approximate $T=0$ ground state.  To that end, rather than always projecting to the ground state of the subsystem being decimated in an RG step, in RSRG-X one randomly projects onto an excited subspace chosen at random, appropriately weighted by a thermal Boltzmann factor~\cite{PhysRevX.4.011052, PhysRevLett.114.217201,PhysRevB.95.024205} that accounts for the fraction of final eigenstates of the whole system that descend from the specified decimation outcome. 
	
	As it stands, the model of Eq.~\eqref{eq:Ham_topological} has an {extensive} degeneracy to all excited states, except in the simplest case $G=\mathbb{Z}_2$. To split these  degeneracies, we consider the generalized Hamiltonian [see Eq.~\eqref{eq:Ham_pert_topological_mod}] introduced in our stability analysis. 
	Since we are focusing on an Abelian $G$, every irreducible representation is one dimensional, i.e., $d_\Gamma = 1$ for all $\Gamma \in \textrm{irrep}(G)$ and every conjugacy class is a singleton, i.e., $[g] = \{g \}$. Note that every discrete Abelian group $G$ is isomorphic to a direct product of $\mathbb{Z}_n$'s, i.e., $G = \mathbb{Z}_{n_1} \times \mathbb{Z}_{n_2} \times \cdots \times \mathbb{Z}_{n_k}$ for some $n_1, n_2, \cdots, n_k \in \mathbb{Z}^+$. The set of irreducible representations of $G$ also forms a dual Abelian group, generated by a discrete Fourier transform. With the above definitions, the Hamiltonian in Eq.~\eqref{eq:Ham_pert_topological_mod} reduces to
	\begin{equation}
	H = - \sum_{\substack{v \in \mathcal{V},\\ \Gamma \in \textrm{irrep}(G)}} J_v^\Gamma \mathcal{A}_v^\Gamma - \sum_{\substack{l \in \mathcal{L}_\mathcal{V}, \\g \in G}} h_l^g C_l^g,
	\label{eq:H_abelian}
	\end{equation}
	where $\mathcal{C}_l^{[g]}=C_l^g=\vert g \rangle \langle g \vert_l$, and we also allow for long link terms. In addition, we assume that the coupling constants $J_v^\Gamma$ and $h_l^g$ are random. In the following, we derive the RSRG-X rules by constructing unitary transformations which directly map to the RSRG rules previously derived in the context of ground-state properties.
	
	Suppose that the dominant term in the Hamiltonian is a link term $h_l^h C_l^h$. We decimate this term by projecting onto the subspace defined by the projector $C_l^g$ with a group element $g$ that is possibly distinct from $h$.  Let us now introduce the unitary transformation $U_v^g \equiv A_v^{g^{-1}}$, acting locally on a vertex $v$ from which the link $l$ is directed. A link term, $C_l^{g'}$, transforms under $U_v^g$ as
	\begin{equation}
	U_v^g \big( C_l^{g'} \big) (U_v^{g})^\dagger = C_l^{g^{-1} \cdot g'}
	\end{equation}
	and in particular 	$U_v^g \big( C_l^g \big) (U_v^{g})^\dagger = C_l^{e}$, as desired. Moreover, vertex terms, defined on $v$, are invariant under the unitary transformation:
	\begin{align}
		&U_v^g \big( A_v^\Gamma \big) (U_v^{g})^\dagger = \frac{d_\Gamma}{|G|} \sum_{h \in G} \chi_\Gamma^* (h) U_v^g \big( A_v^h \big) (U_v^g)^\dagger \nonumber \\
		&= \frac{d_\Gamma}{|G|} \sum_{h \in G} \chi_\Gamma^* (h) A_v^{g^{-1} \cdot h \cdot g} = \frac{d_\Gamma}{|G|} \sum_{h' \in G} \chi_\Gamma^* (g \cdot h' \cdot g^{-1}) A_v^{h'} \nonumber \\
		&= \frac{d_\Gamma}{|G|} \sum_{h' \in G} \chi_\Gamma^* (h') A_v^{h'} = A_v^\Gamma.
	\end{align}
	where in the second line, we introduced the dummy variable  $h'$ via $h \equiv g \cdot h' \cdot g^{-1}$ and in going from the second line to the third line, we used the fact that the character is a constant function on a conjugacy class. Therefore, applying the unitary transformation $U_v^g$ induces the following mapping:
	\begin{equation}
	C^g_l \big( \mathcal{A}_v^\Gamma \mathcal{A}_{v'}^{\Gamma'} \big) C^g_l \mapsto (U_v^g)^\dagger C^e_l \big( \mathcal{A}_v^\Gamma \mathcal{A}_{v'}^{\Gamma'} \big) C^e_l U_v^g.
	\label{eq:unitary_trans_link}
	\end{equation}
	The above relation maps the excited state projection to an equivalent ground state projection, for which we already derived the corresponding RSRG rules appearing in Eq.~\eqref{eq:vertex_gamma_merge}.
	
	Next, suppose that the dominant term in the Hamiltonian is the vertex term $J_v^{\tilde{\Gamma}} \mathcal{A}_v^{\tilde{\Gamma}}$ for some vertex $v$ and irrep $\tilde{\Gamma}$. We decimate this term by projecting onto the subspace defined by the projector $ \mathcal{A}_v^{\Gamma}$ with the same $v$ but possibly distinct irrep $\Gamma$. As before, we seek to find a local unitary transformation which maps $\mathcal{A}_v^{\Gamma}$ to a vertex term with a trivial representation $\mathcal{A}_v^{\Gamma=1}(=A_v)$. To this end, we take a link $l$ in $\mathcal{L}$ connected to the vertex $v$, which without loss of generality can be written as $l = \langle v, v' \rangle$, i.e., directed from $v$, and consider the unitary transformation, $U_l^{\Gamma} \equiv \sum_{g \in G} \chi_{\Gamma}(g) C_l^g$. This operator represents a unitary transformation since the characters $\chi_\Gamma(g)$ of an Abelian group are always a pure phase and $\sum_{g \in G} C_l^g$ is the identity operator. We first note that the local unitary transformation $U_l^{\Gamma}$ leaves the link terms invariant. Acting with $U_l^{\Gamma}$ on a vertex operators at $v$ gives
	\begin{align}
		U_l^\Gamma \big( A_v^\Lambda \big) (U_l^\Gamma)^\dagger &= \frac{1}{|G|} \sum_{h, g, g' \in G} \chi_\Gamma (g') \chi^*_\Lambda (h) \chi^*_\Gamma (g) C_l^{g'} \big( A_v^h \big) C_l^g \nonumber \\
		&= \frac{1}{|G|} \sum_{h, g \in G} \chi_\Gamma (h \cdot g) \chi^*_\Lambda (h) \chi^*_\Gamma (g) A_v^h C_l^{g} \nonumber \\
		&= \frac{1}{|G|} \sum_{h, g \in G} \chi_\Gamma (h \cdot g) \chi^*_\Lambda (h) \chi_\Gamma (g^{-1}) A_v^h C_l^{g} \nonumber \\
		&= \frac{1}{|G|} \sum_{h, g \in G} \chi_\Gamma (h \cdot g \cdot g^{-1}) \chi^*_\Lambda (h) A_v^h C_l^{g} \nonumber \\
		&= \Big( \frac{1}{|G|} \sum_{h} \chi^*_{\Gamma^{-1} \cdot \Lambda} (h) A_v^h \Big) \Big( \sum_{g \in G} C_l^{g} \Big) \nonumber \\
		&= A_v^{\Gamma^{-1} \cdot \Lambda}.
	\end{align}
	In going from the first line to the second line of the above series of equations, we used $C_l^{g'} A_v^h = A_v^h C_l^{h^{-1} \cdot g'}$ and the fact that $C_l^{g}$ is a projector. To go from the second line to the fifth line we use several properties of the group characters of an Abelian group: (1) $\chi^*_\Gamma (g) = \chi_\Gamma (g^{-1}) = \chi_{\Gamma^{-1}} (g)$, (2) $\chi_\Gamma (g_1 \cdot g_2) = \chi_\Gamma (g_1) \chi_\Gamma (g_2)$, and (3) $\chi_\Gamma (g) \chi_{\Gamma'} (g) = \chi_{\Gamma \cdot \Gamma'} (g)$. Similarly, one can also show that for the complementary vertex $v'$, the following holds, $U_l^\Gamma \big( A_{v'}^\Lambda \big) (U_l^\Gamma)^\dagger = A_{v'}^{\Gamma \cdot \Lambda}$. By applying the transformation  $U_l^\Gamma$ prior to the vertex decimation step [similarly to Eq.~\eqref{eq:unitary_trans_link}], the RSRG-X rules reduce to the RSRG decimation rules derived in Sec.~\ref{sec:pert}. In summary,  we have established an equivalence between RSRG-X and RSRG rules, which implies that the universal properties associated with ground- and excited-states criticality are identical. This result further extends our superuniversality claim.
	
As a concluding remark, we comment on the application of the RSRG-X procedure in the non-Abelian case. When the dimension of the irrep $\Gamma$ is nontrivial, i.e.  $d_{\Gamma} > 1$, during the decimation step a vertex term cannot be simply removed because we have to take into account the inherent degeneracy (quantum dimension) of a non-Abelian anyon. Keeping track of all possible assignments of non-Abelian anyons rapidly renders the RSRG-X procedure challenging to analyze. This observation is in accord with the no-go theorems of Ref.~\onlinecite{PhysRevB.94.224206}. We note that excited states at finite density accommodate a finite density of non-Abelian anyons on plaquettes, reminiscent of interacting anyons in (2+1)D with random couplings considered in Refs.~\onlinecite{PhysRevB.85.161301,PhysRevB.85.224201}.

	\section{Summary and discussion}
	\label{sec:summ}
	In this work, we have studied the fate of confinement-deconfinement phase transitions of two-dimensional discrete lattice gauge theories in the presence of strong quenched disorder. Our key result is that for a large class of models, the confinement transitions are controlled by   an infinite-disorder fixed point. Remarkably, the associated critical properties are predicted to exhibit a `superuniversal' behavior that is independent of the underlying gauge group structure. This is in stark contrast with the clean limit, where the order of the transition and critical properties (in the case of a continuous transition) sensitively depend on the specific form of the gauge symmetry under consideration. 
	
	To derive these results, we have constructed a RSRG scheme, specifically tailored for the renormalization of lattice gauge theories. Through an exact duality mapping between our gauge theory model and the quantum Potts model, we have shown that the respective disorder driven confinement-deconfinement and order-disorder transitions belong to the same universality class. We tested our predictions using a numerically exact large-scale quantum Monte Carlo simulations on the random quantum Potts model for $Q=2$ and $Q=3$. We found numerical evidence for a flow to an infinite-randomness fixed point. At this fixed point we find that, within our numerical precision, the critical properties are independent of $Q$, thereby further substantiating our superuniversality claim.  
	
	We have explored the generality of our RSRG scheme, by examining several extensions of our results to a broader class of models. Specifically, we showed that our RSRG rules naturally carry over to twisted lattice gauge theories within the Dijkgraaf-Witten formalism. On the other hand, for the case of Levin-Wen models, while the resulting set of RSRG rules exhibits some universal features, the specific structure of the theory entails nontrivial modifications that deviate from our original construction. Lastly, for Abelian topological orders, we derived an exact mapping between the ground state (RSRG) and excited state (RSRG-X) decimation rules, allowing us to also treat confinement phase transitions of highly excited topologically ordered states, 
	
	In the remaining, we highlight several future lines of research suggested by this work, some of which were already mentioned in the main text. As previously discussed, the Hamiltonian in Eq.~\eqref{eq:Ham_topological} describes a high-symmetry point in parameter space that is invariant with respect to arbitrary permutations among the different gauge charges. While this choice is natural, terms breaking it cannot be excluded without resorting to fine-tuning. In that regard, it would be interesting to explore the relevance of such perturbations by a numerical simulation of the generalized RSRG rules outlined in Sec.~\ref{sec:pert}. Crucially, if such terms prove to be irrelevant in the RG sense, this would not only indicate that the infinite-disorder fixed point is stable, but also flag the emergence of an enlarged permutation symmetry at criticality. One can also envisage numerical tests of stability using QMC. However, permutation symmetry proved crucial in devising efficient cluster updates based on the Fortuin-Kasteleyn representation; it is unclear if such updates can be devised in general away from the permutation-symmetric point. Nevertheless, we note that for the specific case of $G=S_3$ such a representation is still possible \cite{SZLACHANYI1991273}. 
	
In the SPT/SET setting, the  RSRG framework we have introduced also suggests possible new directions, of which we now flag one in particular. So far we have focused primarily on the superuniversality of scaling properties in the {\it bulk}. Owing to the topological `twist' in their definition, the DW models (resp. their un-gauged magnetic counterparts) represent confinement (resp. symmetry-breaking) phase transitions in systems that host topologically protected gapless surfaces on manifolds with \textit{boundaries}. These surface degrees of freedom have been argued to lead to distinct boundary scaling properties in clean systems in both one and two dimensions \cite{PhysRevLett.118.087201,PhysRevX.7.041048,2019arXiv190506969V}. Very recently, they have also been shown to modify the boundary criticality in one dimensional infinite-randomness critical points of SPT phases~\cite{RV_new_SPT}. In all these instances, the {\it bulk} scaling structure is indistinguishable from the 
nontwisted situation. It may therefore be fruitful to explore the boundary critical behavior of the 2D transitions in this paper, extract scaling exponents, and determine the implications (if any) of superuniversality for this `critical' bulk-boundary correspondence.

Our results  for excited-state criticality between distinct (Abelian) MBL phases also raises the possibility that similar techniques can be used to study transitions between distinct periodically driven (Floquet) phases distinguished either by symmetry-breaking or topology. In this setting, disorder is essential even to stabilize Floquet phase structure, since in the absence of MBL, periodically driven systems would heat to the infinite-temperature Gibbs state where there is no applicable notion of phase structure. Various examples of Floquet-SPTs and Floquet-SETs have been explored~\cite{PhysRevX.6.041001, PhysRevX.6.041070, PhysRevB.95.155126, PhysRevLett.119.123601, PhysRevB.96.245116, PhysRevB.97.245106, PhysRevB.98.174203, Berdanier9491, PhysRevB.97.224302}, including several that have no equilibrium analog. The RSRG techniques used to study 1D eigenstate transitions have been successfully applied to disordered Floquet transitions in 1D systems~\cite{Berdanier9491,PhysRevB.98.174203}, suggesting that a similar extension may be possible in 2D systems. 
	
	In all cases considered in this work, we have explicitly expelled flux excitations by enforcing a zero-flux constraint on each plaquette. Relaxing this restriction allows richer anyon content including nontrivial fluxes and even dyonic bound states. Confinement transitions driven by condensing fluxes  are potentially distinct from those driven by charge condensation. For that reason, we cannot extend our superuniversality claim to an arbitrary anyon condensation pattern. A natural future line of research is therefore to devise a generalized RSRG scheme that can capture these effects, which could potentially result in novel critical behavior. As noted above, for the Abelian case, duality arguments can be used to infer properties of flux driven confinement transitions from the corresponding dual electric-charge description. However, our results do not directly apply in self-dual systems, since, there, electric charge and magnetic flux excitations must condense simultaneously. Understanding the fate of the disorder-driven confinement transition along such self-dual lines is particularly interesting, since even the clean limit is not fully understood \cite{PhysRevB.82.085114}. 

We also comment on links between our results and the efficient representation of quantum states using unitary circuits. Our RG can  be framed as an algorithm for constructing an {\it approximate} ground state by iteratively applying a projection at each RG step (recall that  the projections from different RG steps commute). As the RG progresses, the projections become highly nonlocal in terms of the microscopic (UV) degrees of freedom. A corollary of the RG scheme is that highly entangled (approximate) ground states of disordered gauge theories can be successively disentangled by application of these projections encoded in terms of unitary gates.
	In other words, starting from a trivial product state (where the degrees of freedom encode the states of each decimated cluster), an approximate  gauge theory ground state can be constructed by running the RG in reverse, successively applying unitary gates that implement the projections from each RG step~\cite{PhysRevB.93.104205}. This is reminiscent of the multi-scale entanglement renormalization ansatz (MERA)~\cite{PhysRevLett.99.220405}. At the critical point, the constructed approximate ground state thus becomes {\it asymptotically exact}, although formally this is also the limit where the circuit depth diverges (logarithmically). Although implementing nonlocal gates can be challenging, we flag this as a possible route to constructing approximate ground states in strongly correlated 2D topological phases.
	
Finally, it is interesting to consider possible experimental signatures of the infinite-disorder fixed point in the setting of lattice gauge theories, and in particular when the latter are viewed as low-energy descriptions of strongly correlated quantum materials \cite{PhysRevX.8.031028}. However, directly probing confinement experimentally is challenging, due to the inherently nonlocal description of topological order, which necessitates subtle diagnostics~\cite{Gregor_2011}. Nevertheless, in the presence of matter fields (either dynamical or static) or when topological order is symmetry-enriched, confinement typically intertwines with conventional symmetry-breaking phenomena \cite{PhysRevD.19.3682,RevModPhys.51.659}. In such cases, we expect that critical exponents associated with the scaling behavior of the symmetry-breaking order parameter, which is typically more accessible experimentally, are also controlled by universal properties of the confinement transition. However, the best  probes with which to access the topological phase transition must be tailored to the specific experimental setting; we defer such a study to future work.
		
	\begin{acknowledgements}

We acknowledge  D.~Huse for a correspondence on excited-state criticality. B.K. is supported by KIAS Individual Grant PG069402 at the Korea Institute for Advanced Study and the National Research Foundation of Korea (NRF) grant funded by the Korea government (MSIT) (No. NRF-2020R1F1A1075569). S.A.P. acknowledges support from the  European Research Council (ERC) under the European Union Horizon 2020 Research and Innovation Programme (Grant Agreement No.~804213-TMCS), from EPSRC Grant EP/S020527/1, and from the U.S. National Science Foundation Grant DMR-1455366 during the early stages of this project.  R.V. is supported by the U.S. Department of Energy, Office of Science, Basic Energy Sciences, under Early Career Award No. DE-SC0019168, and the Sloan Foundation through a Sloan Research Fellowship. A.P. is supported by NSF DMR-1653007. S.G. acknowledges support from the Israel Science Foundation, Grant No. 1686/18. Computational resources were provided by the KISTI National Supercomputing Center (KSC-2019-CRE-0187) and the Intel Labs Academic Compute Environment.
	\end{acknowledgements}

\appendix
	
\section{Group Cohomology and Topological Dijkgraaf-Witten Models}
	\label{app:top-via-group-coho}

In this appendix, we review several topological aspects of the group cohomology $[\omega] \in H^3\big(G, U(1) \big)$ that provide a more topological intuition and will be relevant to our discussion below. Explicitly, we describe how to assign the $U(1)$ phase factor to a given 3-manifold and describe its invariance under the so-called $1$-$4$ and $2$-$3$ Pachner moves~\cite{PACHNER1991129, PhysRevB.87.155115}. 
 
Although the detailed mathematical steps in order to prove the various equivalence relations and determine the different RG rules are complicated, they are straightforward to implement using standard techniques of topological quantum field theory {\it once} a graphical argument is provided. Accordingly, throughout this work we simply give the relevant graphical arguments, and provide the background necessary to understand why they work. We emphasize that the graphical arguments themselves are nontrivial, and are intimately linked to the strong-disorder RG decimation procedure. A key aspect of Dijkgraaf-Witten models is that they are restricted to theories defined on planar triangulations. As emphasized in the main text, the phase factor of each vertex term, is associated to a ``tent'' 3-manifold constructed from that vertex (see Fig.~\ref{fig:DW}). We now discuss how to assign and manipulate these phase factors, as they will be central to our analysis.

	 In the appendices only, we denote a link $l$ (tetrahedron $T$) by an ordered set of vertices, $l = [v_1, v_2]$  ($T = [v_1, v_2, v_3, v_4]$), which is {\it not} necessarily ordered according to predefined vertex indices. We will typically omit the orientation of links whenever the result holds independent of the vertex ordering. We denote the group element on a link $l=[v_1,v_2]$ as $g_{v_1 v_2}$, where $g_{v_1 v_2} = ( g_{v_2 v_1})^{-1}$ holds, and $T[\{g \}]$ as the tetrahedron $T$ equipped with the set of group elements $\{g \}$ on links satisfying the zero-flux constraints on each plaquette. 
	
\begin{figure}[!t]
		\includegraphics[width=0.8\columnwidth]{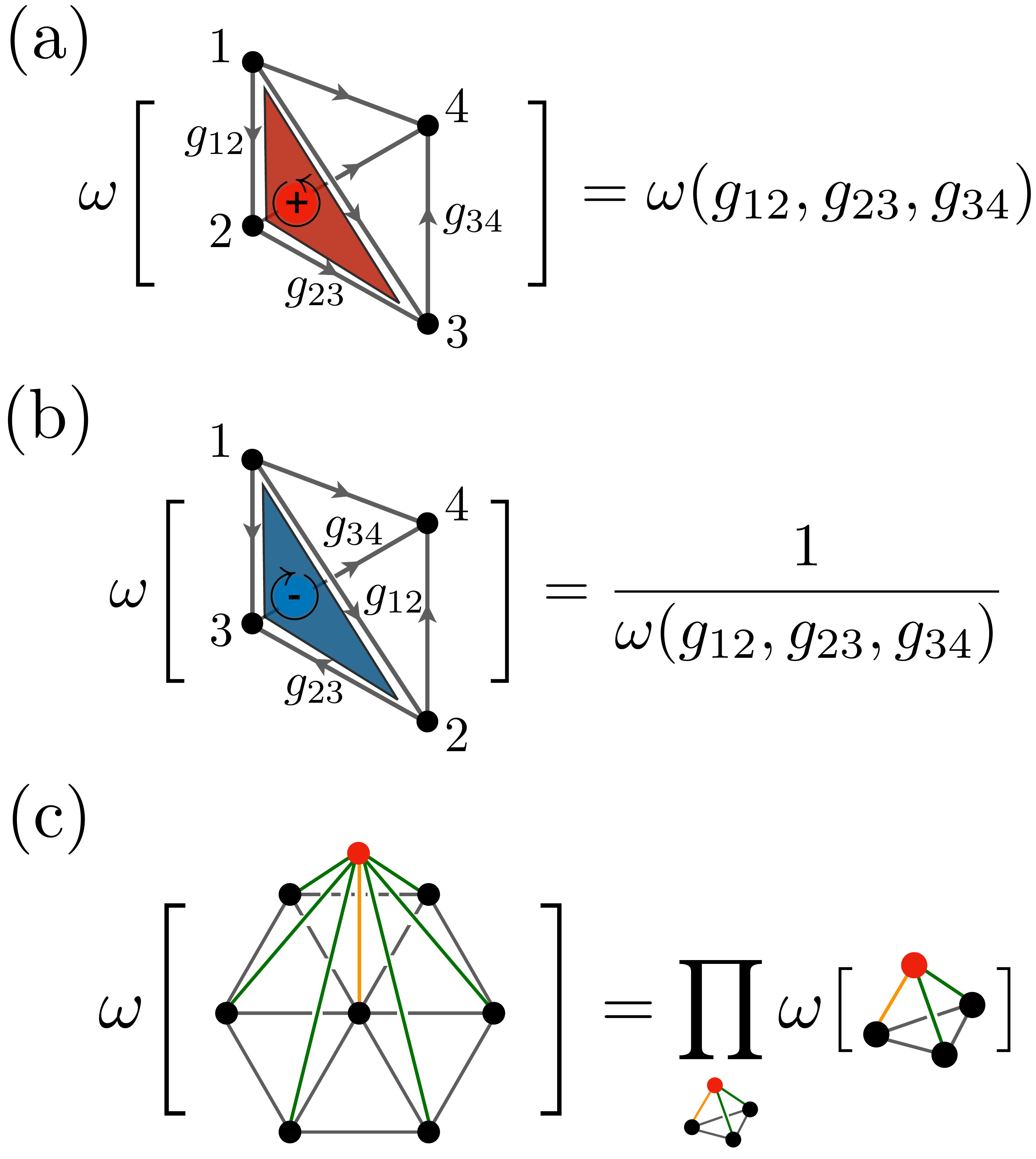}
		\caption{We assign $U(1)$ phase factors for tetrahedrons with (a) positive orientation and (b) negative orientation according to the group elements on links. We denote each vertex as its vertex index, which induces the link orientations and the tetrahedron orientation. (c) The $U(1)$ phase factor associated with a triangulated 3-manifold is given by the product of the $U(1)$ phase factors of individual tetrahedrons in the triangulation.} 
		\label{fig:top-via-group-coho}
	\end{figure}

	For each tetrahedron, we assign a $U(1)$ phase factor according to its orientation and the group elements on its links, as shown in Figs.~\ref{fig:top-via-group-coho} (a) and \ref{fig:top-via-group-coho} (b). Note that two tetrahedrons in Fig.~\ref{fig:top-via-group-coho} are mirror images of each other, and accordingly, their respective $U(1)$ phase factors are a complex conjugate pair. The $U(1)$ phase factor of a 3-manifold composed of tetrahedrons is defined by the product of the $U(1)$ phase factors over all tetrahedrons, as shown in Fig.~\ref{fig:top-via-group-coho} (c) . 
	
	\begin{figure}[!t]
		\includegraphics[width=0.8\columnwidth]{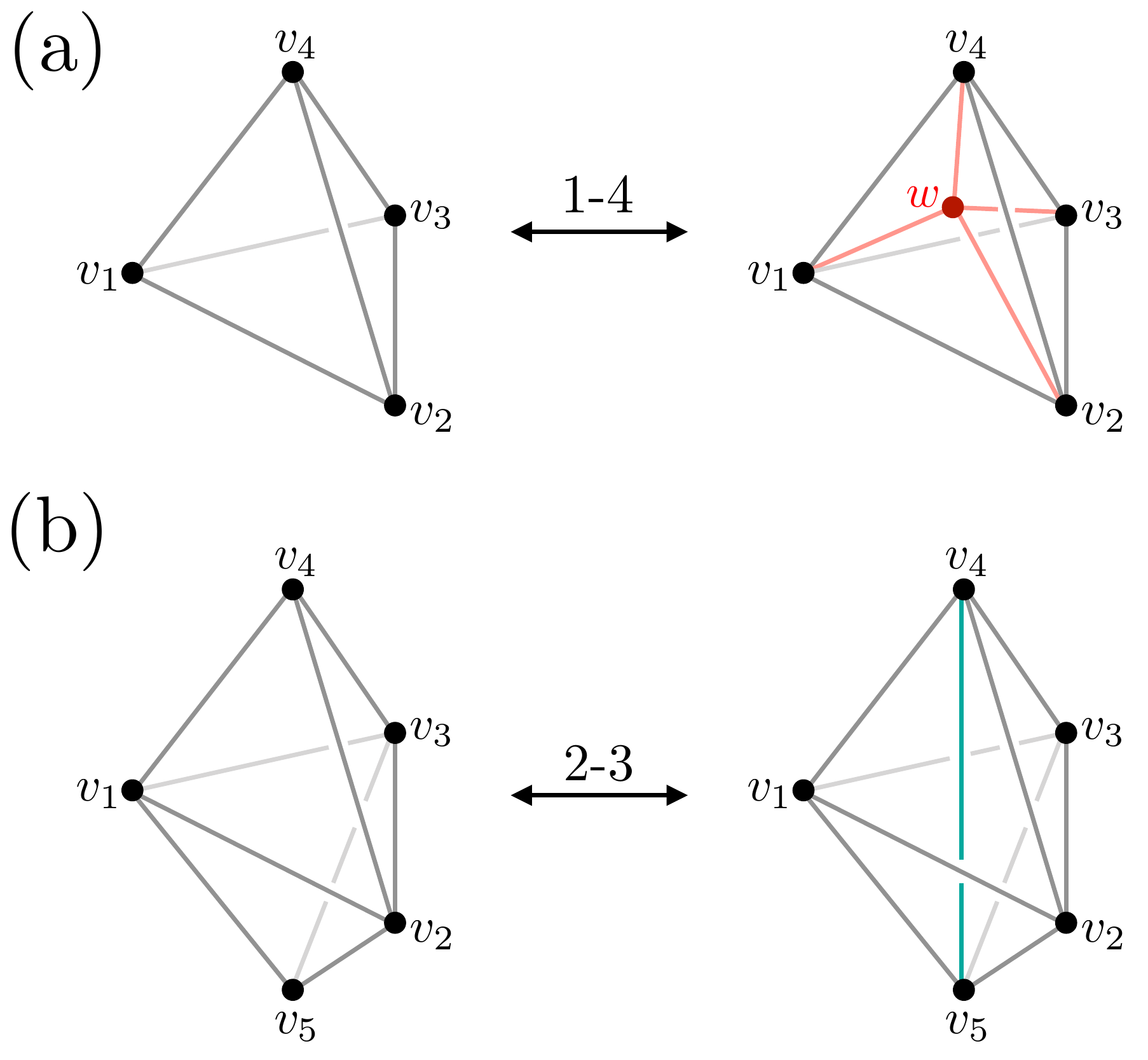}
		\caption{(a) $1$-$4$ Pachner move. The 3-manifold on the left is a single tetrahedron $[v_1, v_2, v_3, v_4]$ and the 3-manifold on the right is obtained by gluing four tetrahedrons $[w, v_1, v_2, v_3]$, $[w, v_1, v_2, v_4]$, $[w, v_1, v_3, v_4]$, and $[w, v_2, v_3, v_4]$. (b) $2$-$3$ Pachner move. The 3-manifold on the left is obtained by gluing two tetrahedrons $[v_1, v_2, v_3, v_4]$ and $[v_1, v_2, v_3, v_5]$ and the 3-manifold on the right is obtained by gluing three tetrahedrons $[v_1, v_2, v_4, v_5]$, $[v_1, v_3, v_4, v_5]$, and $[v_2, v_3, v_4, v_5]$. The $U(1)$ phase factors associated with the 3-manifolds on the left and right are identical in both $1$-$4$ and $2$-$3$ Pachner moves.} 
		\label{fig:Pachner}
	\end{figure}

	The $U(1)$ phase factor is invariant under  the so-called $1$-$4$ and $2$-$3$ ``Pachner moves'', depicted  in Fig.~\ref{fig:Pachner}. The proof directly follows from the defining properties of group cohomology:
	\begin{equation}
	\frac{\omega(g_0, g_1, g_2) \, \omega (g_0, g_1 \cdot g_2, g_3) \, \omega(g_1, g_2, g_3)}{\omega (g_0 \cdot g_1, g_2, g_3) \, \omega(g_0, g_1, g_2 \cdot g_3)} = 1,
	\end{equation}
	where $g_0, g_1, g_2, g_3 \in G$, and the zero flux constraint imposed on every plaquette. We remark that in the $1$-$4$ Pachner move, no averaging over a group element, say, $g_{v_1 \omega}$ is required.  Using a series of $2$-$3$ Pachner moves, one can show that the vertex operators appearing in the DW model satisfy $A_v^{g, \omega} A_v^{g', \omega} = A_v^{g \cdot g', \omega}$ and $[A_v^{g,\omega}, A_{v'}^{g', \omega}] = 0$ for all distinct vertices $v$, $v'$ and group elements $g, g' \in G$.
	
	\section{Lattice Isomorphisms}
	\label{app:lattice-iso}
	In this section, we introduce a set of lattice isomorphisms, denoted by $\hat{T}$. Each lattice isomorphism defines a mapping between the planar graphs $\mathcal{G} \to\mathcal{G}' $ that is accompanied by unitary transformations between Hilbert spaces,  $\hat{T}:\mathcal{H}_{\mathcal{G}}\to\mathcal{H}_{\mathcal{G}'}$ and Hamiltonian terms, $H_{\mathcal{G}'}=\hat{T}H_{\mathcal{G}}\hat{T}^{-1}$. In our discussion below, we first define the basic local building blocks used to construct various lattice isomorphisms and then demonstrate how they are implemented during the various link and vertex term decimation steps. Due to the inherent difference between untwisted and twisted gauge theories, we discuss each case separately, starting with the former. Note that the lattice isomorphisms (since they are fully invertible) may be implemented in either direction; as will be evident, this freedom is crucial in defining a sensible RSRG decimation scheme.
	
	\begin{figure}[!h]
		\includegraphics[width=0.8\columnwidth]{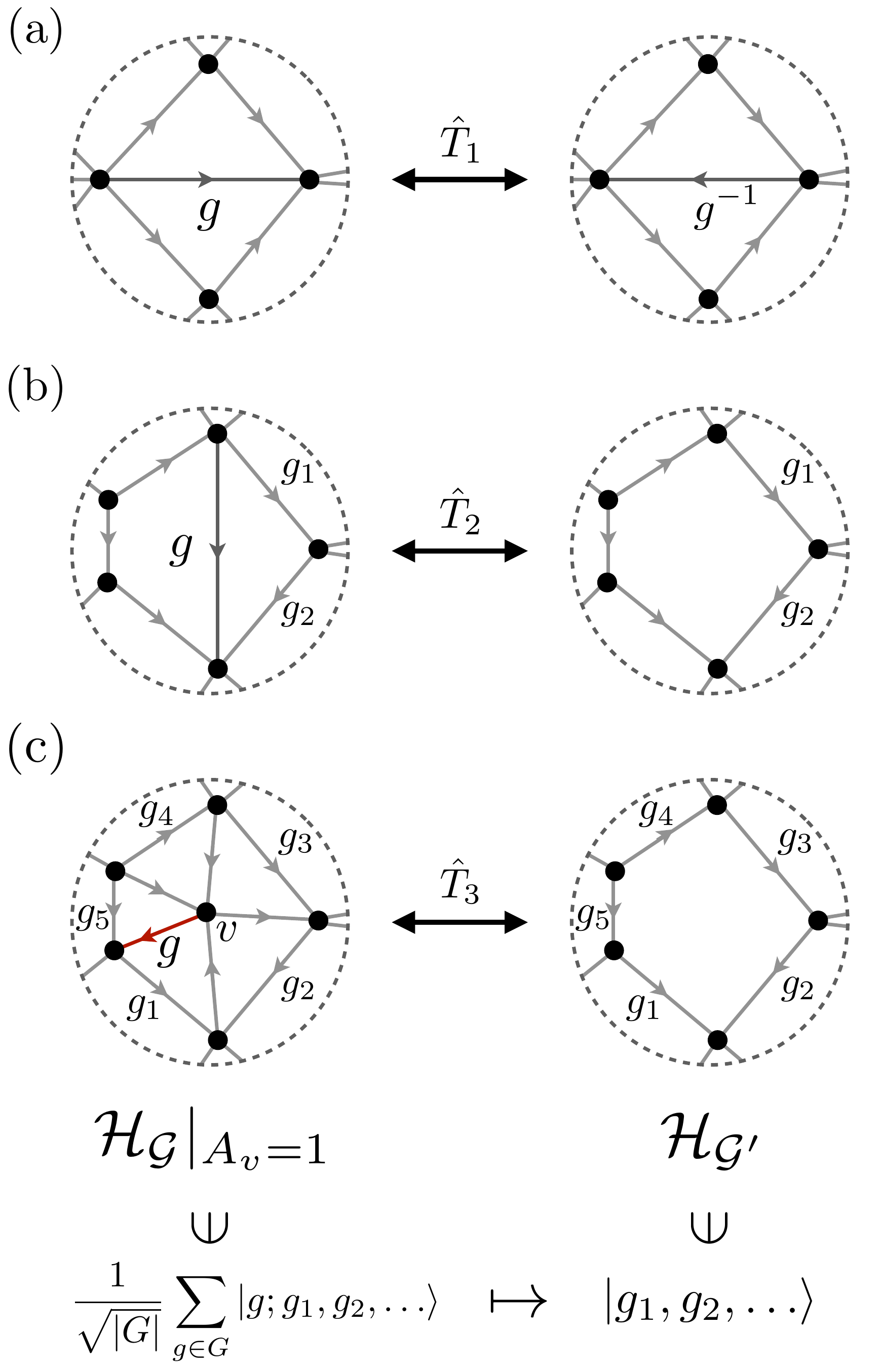}
		\caption{(a) Lattice isomorphism $\hat{T}_1$ amounts to inverting the direction of a link. Accordingly, the basis element of the corresponding link gets inverted. (b) Lattice isomorphism $\hat{T}_2$ amounts to removing/adding a link. Due to gauge constraints, $\hat{T}_2$ does not change the dimension of the Hilbert space. For example, the basis element $g$ is completely fixed by $g_1$ and $g_2$, the states on neighboring links, via $g = g_1 \cdot g_2$. (c) Lattice isomorphism $\hat{T}_3$ relates the lattice with a vertex $v$ (left column) and the lattice without the vertex $v$ and the links connected to $v$ (right column). $\hat{T}_3$ induces a mapping between $A_v = 1$ subspace of the Hilbert space on the left lattice and the Hilbert space on the right lattice. To find such mapping, we pick a representative link (colored link) and define the preferred basis in the left lattice. The preferred basis is mapped to the computational basis on the right lattice via $\hat{T}_3$.} 
		\label{fig:LI}
	\end{figure}
	
	\subsection{Untwisted Gauge Theory}
	In the context of  untwisted gauge theories, we consider three lattice isomorphisms, defined pictorially in Fig.~\ref{fig:LI}. The first lattice isomorphism, $\hat{T}_1$, amounts to reversing the direction of a link $l$ and simultaneously changing the link variable on $l$, from $g$, to its inverse, $g^{-1}$. This operation trivially maps link terms and vertex terms defined in $\mathcal{G}$ to their corresponding terms in $\mathcal{G}'$. 
	
	\begin{figure}[!t]
		\includegraphics[width=1.0\columnwidth]{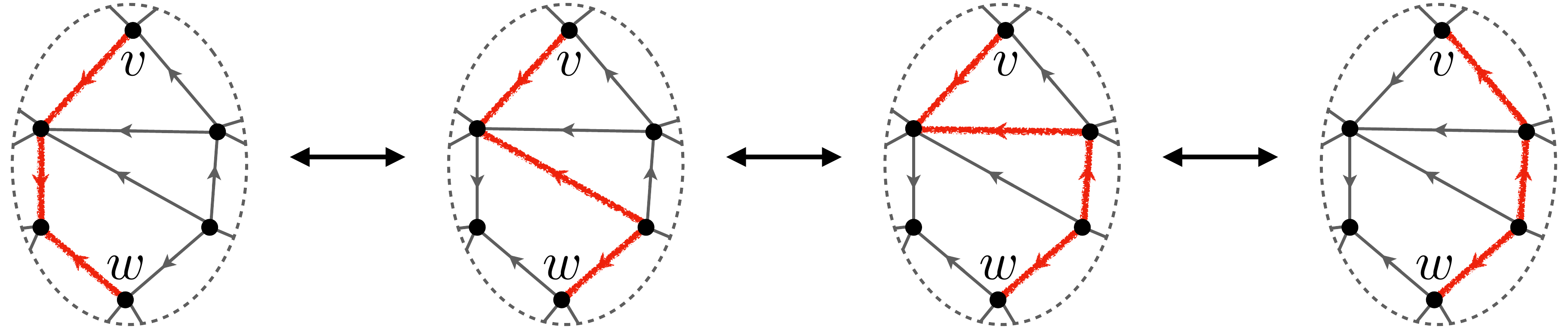}
		\caption{A series of support deformations of a long link term, where the red links denote the support of (or equivalently, the directed path associated with) the link term. The deformations can be implemented by invoking the zero-flux constraint on each plaquette. Note that the end points are fixed during the deformation.} 
		\label{fig:link-deform}
	\end{figure}
	
	The second lattice isomorphism $\hat{T}_2$ [see Fig.~\ref{fig:LI} (b)] amounts to removing (adding) a link, $l$, from (to) $\mathcal{G}$. At first sight, this operation may appear to reduce (enlarge) the Hilbert space dimension. However, by virtue of the zero-flux constraint, the state at $l$ is completely fixed by the states on neighboring links. It is therefore convenient to label $\mathcal{H}_\mathcal{G}$ using the degrees of freedom on links other than $l$, e.g., $g_1$ and $g_2$ in Fig.~\ref{fig:LI} (b). With this choice,  $\hat{T}_2$ induces a direct mapping between the computational bases of $\mathcal{G}$ and  $\mathcal{G}'$.  
	 
	Next, we must specify how Hamiltonian terms transform under $\hat{T}_2$. Vertex terms in $\mathcal{G}$ simply map to the corresponding vertex terms in $\mathcal{G}'$. This result can be verified directly by comparing between the operators $A_v^g \hat{T}_2$ (first mapping to $\mathcal{G}'$ and only then applying $A_v^g$ in $\mathcal{G'}$) and $\hat{T}_2 A_v^g $ (first applying $A_v^g$ in $\mathcal{G}$ and only then mapping to $\mathcal{G}'$). Indeed, a straightforward calculation reveals that the above two operations yield precisely the same quantum states. In the case of the link terms, we have to account for instances where the associated directed path $\pi$ passes through the removed link $l$. The defining path $\pi$ can be freely deformed locally such that no link terms passes through $l$; see Fig.~\ref{fig:link-deform}. Following these deformations, we can safely remove $l$ from the planar graph and directly map between link terms in $\mathcal{G}$ and $\mathcal{G}'$.
	
	The third lattice isomorphism, $\hat{T}_3$, involves removing a vertex $v$ and its adjacent links from $\mathcal{G}$, as described in Fig.~\ref{fig:LI} (c). Note that here, unlike for the $\hat{T}_2$ isomorphism, since $\mathcal{G}$ has an additional vertex relative to $\mathcal{G}'$, $\mathcal{H}_{\mathcal{G}}$ has additional degrees of freedom relative to  $\mathcal{H}_{\mathcal{G}'}$, even after taking into account the zero-flux constraints.
	For this reason, we must supplement this move with an additional constraint, $A_v = 1$, imposed on the Hilbert space $\mathcal{H}_{\mathcal{G}}$. Projecting to the subspace with $A_v = 1$  arises naturally during a vertex decimation step. The subspace defined by imposing $A_v=1$ on $\mathcal{H}_{\mathcal{G}}$ is isomorphic to the Hilbert space $\mathcal{H}_{\mathcal{G}'}$ on the target graph $\mathcal{G'}$, where the vertex $v$ and its adjacent links were removed.
	
	Operationally, we define $\hat{T}_3 : \mathcal{H}_\mathcal{G} \vert_{A_v = 1} \to \mathcal{H}_{\mathcal{G}'}$ by choosing a preferred basis for states belonging to the $A_v=1$ subspace, and mapping it to the computational basis in $\mathcal{G}'$. To define the preferred basis, we first pick a representative link $l$ among the links adjacent to $v$. (By possibly employing $\hat{T}_1$, we assume that $l$ is directed from $v$.) The zero-flux constraints fixes the rest of the link variables. Thus, we can label states in $\mathcal{H}_\mathcal{G}$ as $\vert g; g_1, g_2, \ldots \rangle$, where $g$ is the state on $l$ and $\{ g_1, g_2, \ldots \}$ are the collection of states on links that are not connected to $v$. We define the preferred basis as 
	\begin{equation}
	\frac{1}{\sqrt{G}} \sum_{g \in G} \vert g; g_1, g_2, \cdots \rangle,
	\label{eq:Av=1-untwisted}
	\end{equation}
	which indeed spans the $A_v = 1$ subspace. We note that Eq.~\eqref{eq:Av=1-untwisted} is also an eigenstate of $A_v^h$, under which $\vert g; g_1, g_2, \cdots \rangle$ maps to $\vert h \cdot g; g_1, g_2, \cdots \rangle$,  for all $h \in G$ with the corresponding eigenvalue $1$. Under $\hat{T}_3$, the preferred basis in Eq.~\eqref{eq:Av=1-untwisted} is simply mapped to $\vert g_1, g_2, \ldots \rangle$, the computational basis states in $\mathcal{H}_{\mathcal{G}'}$.
	
	We now argue that $\hat{T}_3$ correctly maps both link and vertex terms. We must assume that the Hamiltonian does not contain link terms that either end or begin on a link connected to $v$. In the context of our RSRG scheme, this assumption is always satisfied during a vertex term decimation, since such link terms are never a part of the renormalized Hamiltonian. On the other hand, long-link terms whose  defining path $\pi$ passes thorough links adjacent to $v$ are allowed. For such long-link terms, we employ the procedure described in Fig.~\ref{fig:link-deform}, which deforms the path $\pi$, so that it does not act directly on links adjacent to $v$. Following the above step, we conclude that link terms on $\mathcal{G}$ are directly mapped to their corresponding ones $\mathcal{G}'$ via $\hat{T}_3$. 
	
	Moving to vertex terms, we need to examine only terms, $A_w^h$, defined on a vertex $w$ adjacent to $v$ and an arbitrary group element $h \in G$. As before, we compare between the operations $\hat{T}_3 A_w^h$ and $A_w^h \hat{T}_3$, acting on the preferred basis states, previously defined in Eq.~\eqref{eq:Av=1-untwisted}. A straight forward calculation, using the rearrangement theorem, shows that the above two operations yield the identical result. This means that every vertex term in $\mathcal{G}$ is simply mapped to the corresponding vertex term in $\mathcal{G}'$.
	
	\begin{figure}[!t]
		\includegraphics[width=1.0\columnwidth]{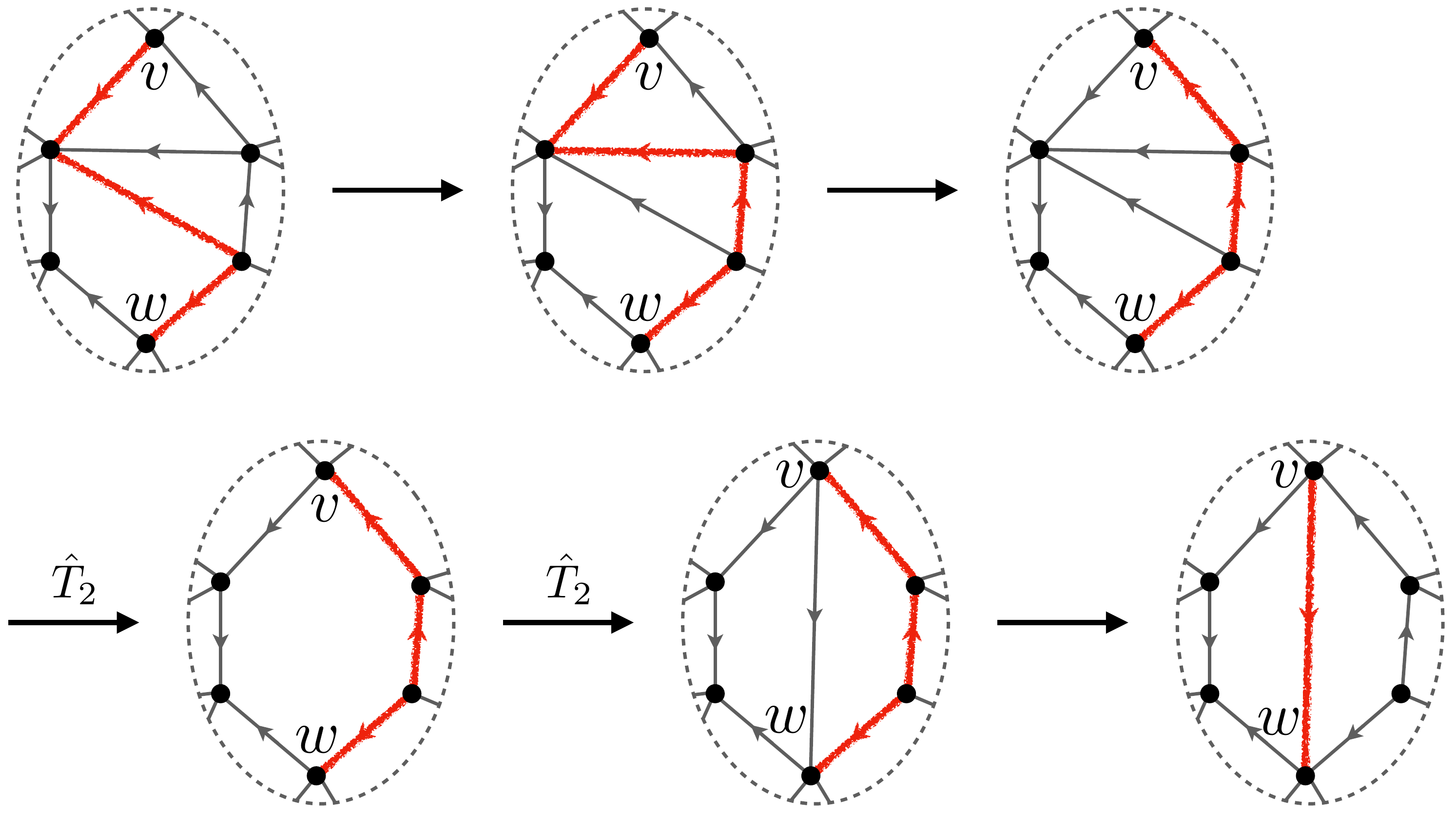}
		\caption{A procedure which maps a long-link term to a short-link term. We employ a series of support deformation of a link term and a series of lattice isomorphism $\hat{T}_2$ during the procedure.} 
		\label{fig:long-link-to-link}
	\end{figure}
	
	In Fig.~\ref{fig:long-link-to-link}, we demonstrate an important application of lattice isomorphism, used in the main text during a of long-link decimation step. Explicitly, we show that a long-link term $C_l^g$ for some long-link $l = \langle v, w \rangle$ and $g \in G$ can be mapped to a short-link term by first deforming the defining path of $C_l^g$ and following it with a series of $\hat{T}_2$ lattice isomorphisms.
	
	\subsection{Twisted Gauge Theory}
	\label{app:lattice-iso-DW}
	We now turn our attention to constructing lattice isomorphisms in the context of DW models~\cite{PhysRevB.87.125114, PhysRevB.87.155115}. Crucially, in this case, we must take into account the transformation of the $U(1)$ phase factors, originating from a group cohomology element $[\omega] \in H^3 \big( G, U(1) \big)$, under the various lattice isomorphisms.  
	
	In the DW model, reversing a link orientation during a $\hat{T}_1$ lattice isomorphism, translates into exchanging a pair of vertex indices. However, such a reordering can be consistently implemented in only a few special cases~\cite{dijkgraaf1990topological} and hence to maintain full generality, we will {\it not} employ $\hat{T}_1$ in the following discussion.
	
	\begin{figure}[!t]
		\includegraphics[width=1.0\columnwidth]{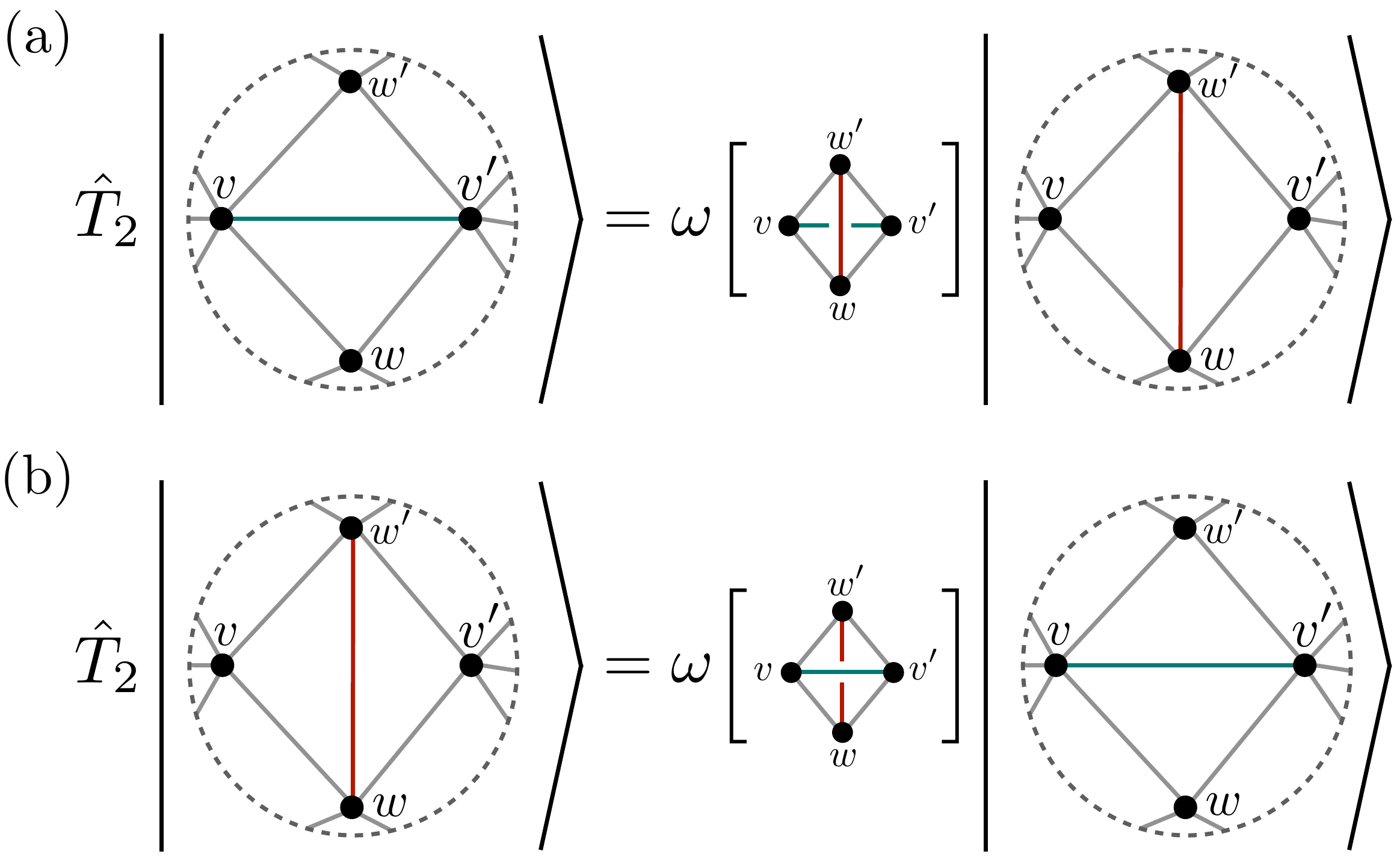}
		\caption{$\hat{T}_2$ lattice isomorphism acting on a computational basis state. Here, group elements (which are not shown explicitly) are placed on each link and satisfy the zero-flux constraint on each plaquette. Note that group elements on colored links are completely fixed by the group elements on neighboring links. Compared to the untwisted case, $\hat{T}_2$ carries additional $U(1)$ phase factor in the twisted case.} 
		\label{fig:link-LI-DW}
	\end{figure}
	
	The second lattice isomorphism $\hat{T}_2$ involves deforming the link configuration. Unlike the untwisted case, we cannot freely add or remove a link from the planar graph $\mathcal{G}$, as this operation can potentially generate nontriangular plaquettes, which are disallowed in DW models. Instead, we redefine $\hat{T}_2$, so that the resulting link configuration is compatible with the triangulation structure,  as shown in Figs.~\ref{fig:link-LI-DW} (a) and \ref{fig:link-LI-DW} (b). As before, due to the zero-flux constraints, states on modified links are completely fixed by states on neighboring links. Unlike before, $\hat{T}_2$ must be accompanied by a $U(1)$ phase factor, which is pictorially defined in Figs.~\ref{fig:link-LI-DW} (a) and \ref{fig:link-LI-DW} (b). Note that the two operations appearing in Fig.~\ref{fig:link-LI-DW}  are inverses of each other since the two associated tetrahedrons are mirror images of each other. 
	
	\begin{figure}[!t]
		\includegraphics[width=0.9\columnwidth]{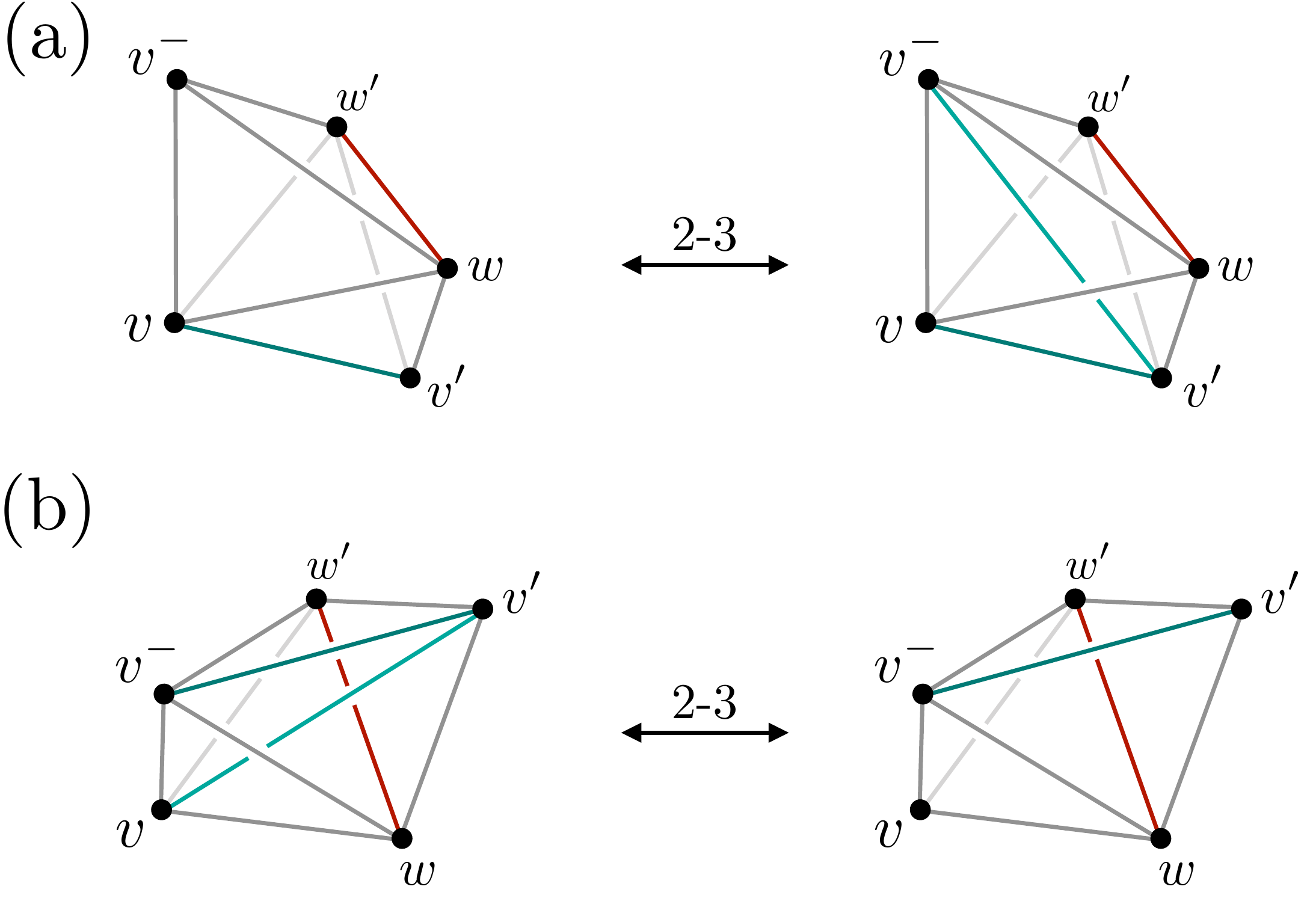}
		\caption{The 3-manifold in the left (right) column appears in the phase factor of $A_v^{h, \omega} \hat{T}_2$ ($\hat{T}_2 A_v^{h, \omega}$) in Figs.~\ref{fig:link-LI-DW} (a) and \ref{fig:link-LI-DW} (b). More explicitly for case (a), the left panel consists of  two tetrahedrons $[v,v',w,w']$ and $[v^-,v,w,w']$. The former accounts from the phase factor accompanying the $\hat{T}_2$ transformation and the latter arises from the tent construction. The right panel compromises three tetrahedrons, $[v^-,v',w,w']$, $[v^-,v,v',w]$, and $[v^-,v,v',w']$. The first tetrahedron originates from the  $\hat{T}_2$ transformation and the last two tetrahedrons originate from the tent construction. Similar considerations lead to the 3-manifold construction corresponding to (b). In each case, using a $2$-$3$ Pachner move, the phase factors of $A_v^{h, \omega} \hat{T}_2$ and $\hat{T}_2 A_v^{h, \omega}$ acting on a computational basis state can be shown to be identical.} 
		\label{fig:T2-LI-DW}
	\end{figure}
	
	We now argue that $\hat{T}_2$ properly maps the Hamiltonian terms. By following the path deformation procedures used in the untwisted case, link terms in $\mathcal{G}$ are mapped to the corresponding link terms in $\mathcal{G}'$ under $\hat{T}_2$. In the case of the vertex terms, it is crucial to check whether the phase factors associated with vertex terms map faithfully under $\hat{T}_2$. Since $\hat{T}_2$ modifies the lattice locally, we need to consider only the vertex operators acting on $v$ in Figs.~\ref{fig:link-LI-DW} (a) and \ref{fig:link-LI-DW} (b). Specifically, we compare the $U(1)$ phase factors associated with the operations $A_v^{h, \omega} \hat{T}_2$ and $\hat{T}_2 A_v^{h, \omega}$ for an arbitrary $h \in G$ when acting on a computational basis element in $\mathcal{H}_\mathcal{G}$. Since all other $U(1)$ phase factors (coming from the remaining plaquettes) trivially cancel with each other, we need to compare only the $U(1)$ phase factors assigned to the 3-manifolds shown in Fig.~\ref{fig:T2-LI-DW}. Using a $2$-$3$ Pachner move presented in Appendix~\ref{app:top-via-group-coho}, it follows immediately that the two associated $U(1)$ phase factors are identical. Note that the additional phase factor introduced in Fig.~\ref{fig:link-LI-DW} precisely compensates for the otherwise nonvanishing phase difference.
	
	\begin{figure}[!t]
		\includegraphics[width=1.0\columnwidth]{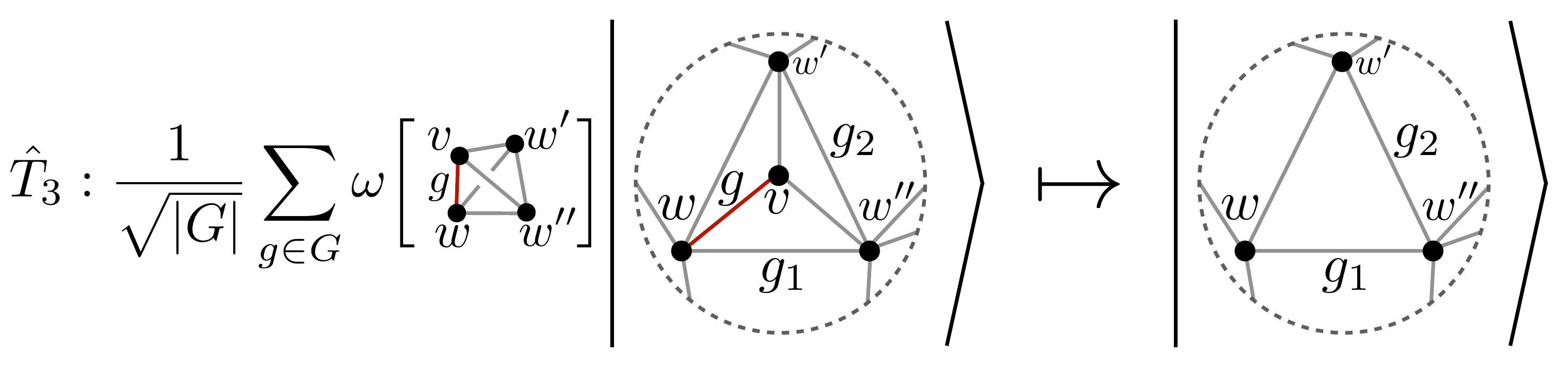}
		\caption{Lattice isomorphism $\hat{T}_3$ is defined for a vertex $v$. We map the preferred basis vector, shown in the left column, on $A_v^{\omega} = 1$ subspace on the graph with the vertex $v$ to the computational basis $\vert g_1, g_2, \cdots \rangle$ on the graph where the vertex $v$ and adjacent links are removed, shown in the right column.} 
		\label{fig:T3-DW}
	\end{figure}
	
	\begin{figure}[!t]
		\includegraphics[width=1.0\columnwidth]{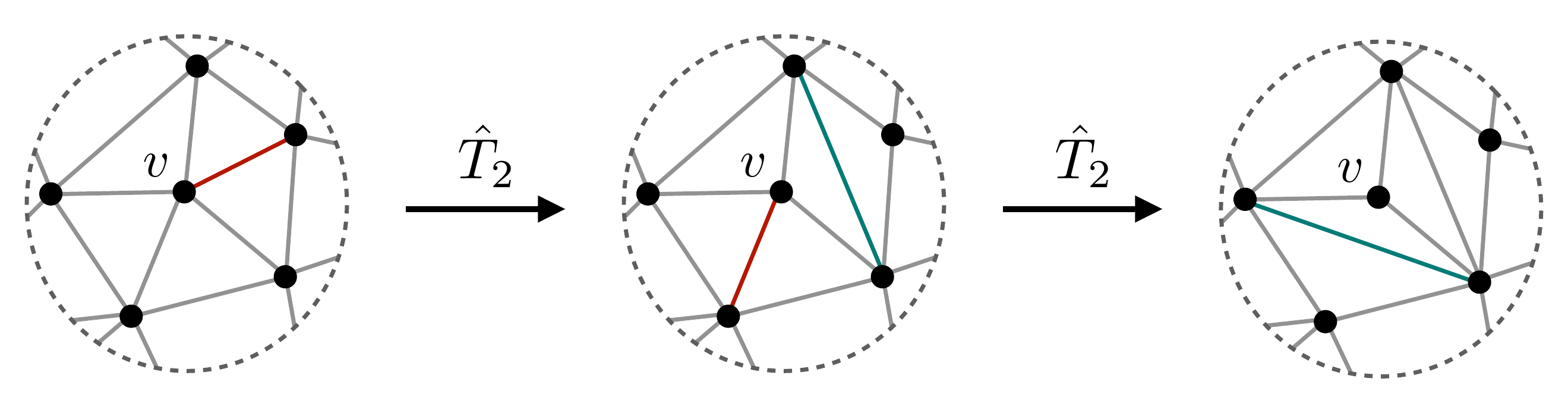}
		\caption{Using a series of $\hat{T}_2$ lattice isomorphisms, one can isolate a vertex $v$ in such a way that it is surrounded by a triangle. We colored the links which are affected by $\hat{T}_2$.} 
		\label{fig:vertex-isolation-DW}
	\end{figure}
	
	Finally, we define the third lattice isomorphism $\hat{T}_3$, which amounts to removing a vertex $v$ and the adjacent links from $\mathcal{G}$. Since only triangular plaquettes are allowed in the DW models, we assume that the vertex is surrounded by a triangle, such as the one in the left column of Fig.~\ref{fig:T3-DW}. In general, we can isolate  a vertex in this manner by employing a series of $\hat{T}_2$ lattice isomorphisms as depicted in Fig.~\ref{fig:vertex-isolation-DW}. 
	
\begin{figure}[!t]
		\includegraphics[width=1.0\columnwidth]{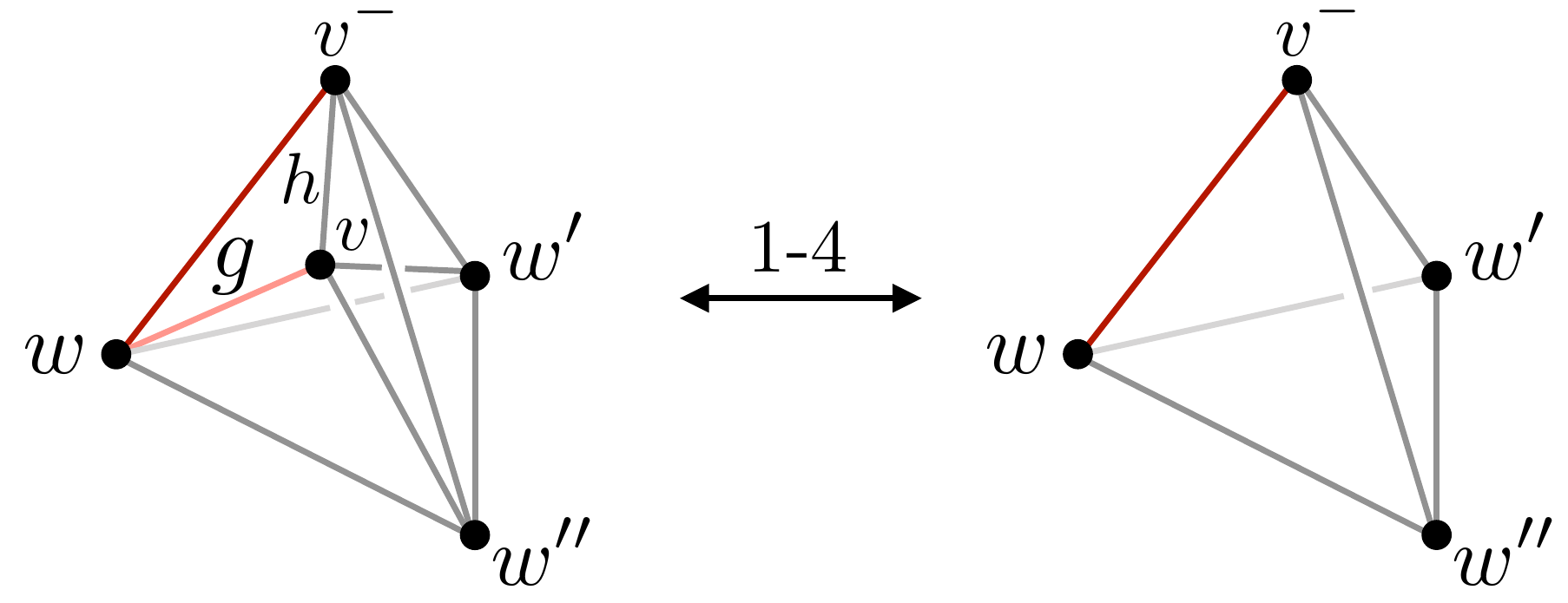}
		\caption{Graphical proof of Eq.~\eqref{eq:Av_1_proof}. The 3-manifold on the left is obtained by gluing the tetrahedron associated with the $U(1)$ phase factors appearing in the amplitude of $\vert g; g_1, g_2 \rangle$ in Eq.~\eqref{eq:Avw-basis} and the 3-manifold from the tent construction of $A_v^{h, \omega}$. The 3-manifold on the right is associated with the phase factor of the amplitude of $\vert h \cdot g; g_1, g_2 \rangle$ in Eq.~\eqref{eq:Avw-basis}. The $U(1)$ phase factors associated with the above two 3-manifolds are identified by a $1$-$4$ Pachner move.}
		\label{fig:Av-basis-proof-DW}
	\end{figure}

	Similarly to the untwisted case, $\hat{T}_3$ induces an isomorphism between $\mathcal{H}_{\mathcal{G}} \vert_{A_v^\omega = 1}$ and $\mathcal{H}_{\mathcal{G}'}$. To define $\hat{T}_3$, we first choose a preferred basis for the $A_v^\omega=1$ subspace of $\mathcal{H}_\mathcal{G}$. $\hat{T}_3$ maps the preferred basis to the computational basis in $\mathcal{H}_{\mathcal{G}'}$. To this end, we first pick a representative link, that we denote $[v, w]$, among the links adjacent to $v$.\footnote{The representative link is fixed once and for all, and in particular is independent of the group elements on links.}  As before, we denote a computational basis state in $\mathcal{H}_\mathcal{G}$ as $\vert g; g_1, g_2, \ldots \rangle$, where $g$ is the state on the representative link $[v,w]$ and $\{g_1, g_2, \ldots\}$ are the states on links that are not adjacent to $v$ (states defined on adjacent links other than $[v, w]$ are fixed by the zero-flux constraints). The preferred basis is constructed by taking superpositions of $\vert g; g_1, g_2, \ldots \rangle$ with amplitudes fixed by the group cohomology element:
	\begin{equation}
	\frac{1}{\sqrt{G}} \sum_{g \in G} \omega \big[ T[\{ g, g_1, g_2 \}] \big] \vert g; g_1, g_2, \ldots \rangle ,
	\label{eq:Avw-basis}
	\end{equation}
	where $T = [v, w, w', w'']$ is the tetrahedron shown in Fig.~\ref{fig:T3-DW}. We first demonstrate that the preferred basis vectors are eigenstates of $A_v^{h, \omega}$ with an eigenvalue $1$ for all $h \in G$, and hence  belong to $A_v^{\omega} = 1$ subspace. To this end, we first note that, as with the untwisted case, acting with $A_v^{h, \omega}$ shuffles the computational basis only in Eq.~\eqref{eq:Avw-basis}, $\vert g; g_1, g_2, \ldots \rangle \mapsto \vert h \cdot g; g_1, g_2, \ldots \rangle$. In the twisted case, we also have to make sure that the additional $U(1)$ phases transform correctly; under the operation of $A_v^{h, \omega}$, we accumulate the phase factor $e^{i\phi_h(\omega)}$, which must satisfy the condition 
	\begin{equation}
	\omega\big[ T[\{ h \cdot g, g_1, g_2 \}]=e^{i\phi_h(\omega)} \omega \big[ T[\{ g, g_1, g_2 \}] \big] ,
	\label{eq:Av_1_proof}
	\end{equation} 
in order for Eq.~\eqref{eq:Avw-basis} belongs to $A_v^{h, \omega}=1$ subspace. The above identity is proven pictorially in Fig.~\ref{fig:Av-basis-proof-DW} using a $1$-$4$ Pachner move. 
	
	\begin{figure}[!t]
		\includegraphics[width=1.0\columnwidth]{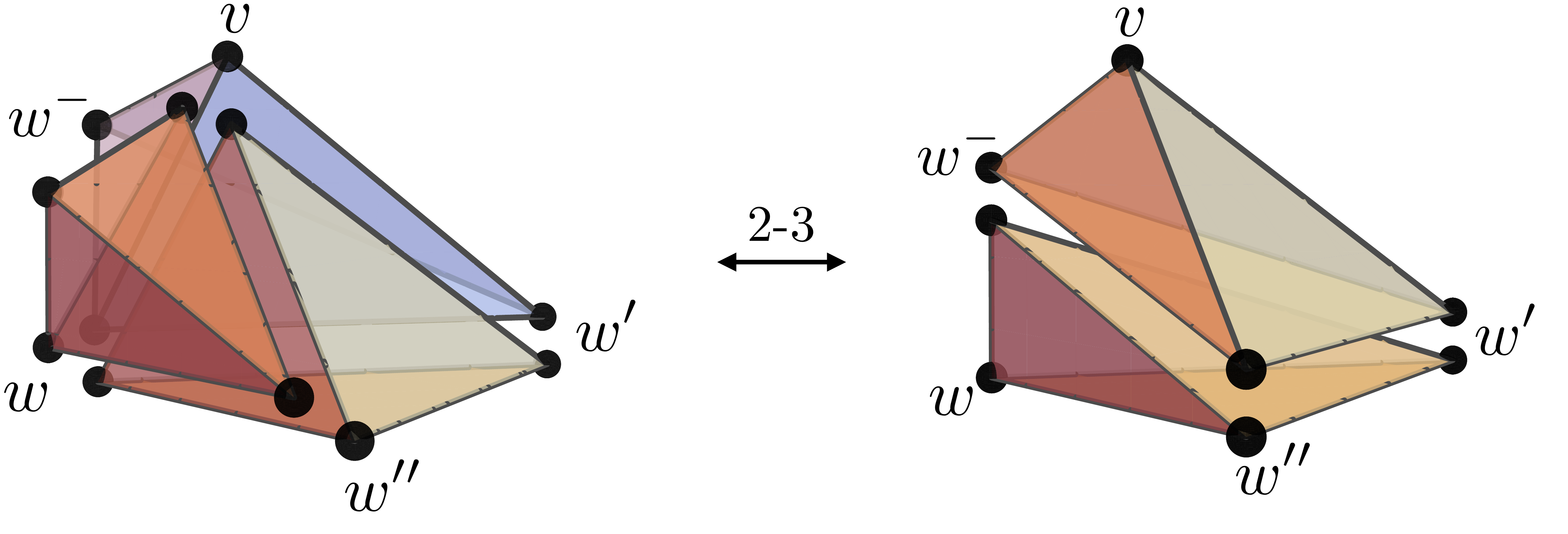}
		\caption{Graphical proof of Eq.~\eqref{eq:phase_single_vertex}. The 3-manifold on the left is given by the gluing of tetrahedron $[v,w,w',w'']$, originating from the phase factor of $\vert g; g_1, g_2 \rangle$  in Eq.~\eqref{eq:Avw-basis}, with two relevant tetrahedrons, $[w^{-},w,v,w']$ and $[w^{-},w,v,w'']$, originating from the tent construction of $A_w^{h, \omega}$ (shown as three distinct tetrahedrons, for clarity). The associated $U(1)$ phase factor for the latter two tetrahedrons corresponds to $e^{i \phi_h^\mathcal{G} (\omega)}$ in Eq.~\eqref{eq:Avhw_T_3}. The 3-manifold on the right is obtained by gluing the tetrahedron $[v, w^-, w', w'']$, originating from the phase factor of $\vert h \cdot g; h \cdot g_1, g_2 \rangle$ in Eq.~\eqref{eq:Avw-basis}, and the tetrahedron $[w^-, w, w', w'']$, with which the phase factor $e^{i \phi_h^{\mathcal{G}'} (\omega)}$ in Eq.~\eqref{eq:T_3_Avhw} is associated. Two 3-manifolds are related by a $2$-$3$ Pachner move and, importantly, the group elements on links of $[w^-, w, w', w'']$ are independent of $g$.} 
		\label{fig:Aw-identify-DW}
	\end{figure}
	
	Next, we show that $\hat{T}_3$ also correctly maps the Hamiltonian terms. We first consider the mapping of link terms. As before, if necessary, we employ a preliminary path deformation step such that all link terms never pass through removed links. Following this operation, link terms appearing in $\mathcal{G}$ are directly mapped to the corresponding ones in $\mathcal{G}'$, similarly to the untwisted case. As for vertex terms, we need to consider only vertex terms $A_w^\omega$, where $w$ is a vertex adjacent to $v$ (see Fig.~\ref{fig:T3-DW}), since other vertex terms are unaffected by $\hat{T}_3$. In the calculation below we ignore phase factors arising from plaquettes untouched by the mapping $\hat{T}_3$, since they trivially cancel out. We begin by considering the case where we first apply the isomorphism $\hat{T}_3$ and only then act with $A_w^{h, \omega}$ for some $h \in G$. The resulting state is,
	\begin{equation}
	e^{i\phi^\mathcal{G'}_h(\omega)} \vert h \cdot g_1, g_2, \ldots \rangle ,
	\label{eq:T_3_Avhw}
	\end{equation}
	where the phase factor is evaluated in $\mathcal{G'}$ after the removal of the vertex $v$ and hence include a single relevant tetrahedron $[w^{-},w,w',w'']$. Next, acting first with $A_w^{h, \omega}$  on a preferred basis state appearing in Eq.~\eqref{eq:Avw-basis} gives
\begin{align}
\frac{1}{\sqrt{G}} \sum_{g \in G} &\frac{e^{i \phi^{\mathcal{G}}_h (\omega)} \omega \big[ T[\{ g, g_1, g_2 \}] \big]}{\omega \big[ T[\{h \cdot g, h \cdot g_1, g_2 \}] \big]} \nonumber \\
& \times \omega \big[ T[\{h \cdot g, h \cdot g_1, g_2 \}] \big] \vert h \cdot g; h \cdot g_1, g_2, \ldots \rangle ,
\label{eq:Avhw_T_3}
\end{align}
	where $e^{i \phi_h^{\mathcal{G}} (\omega)}$ is evaluated by a tent construction on $\mathcal{G}$ comprising two relevant tetrahedrons $[w^-,w,v,w']$ and $[w^-,w,v,w'']$, $\omega \big[ T[\{ g, g_1, g_2 \}] \big]$ originates from the $U(1)$ weights in Eq.~\eqref{eq:Avw-basis}, and for convenience we multiply and divide by $\omega \big[ T[\{h \cdot g, h \cdot g_1, g_2 \}] \big]$ to comply with the $U(1)$ weights associated with the states $\vert h \cdot g; h \cdot g_1, g_2, \ldots \rangle $. Finally, we apply the $\hat{T}_3$ transformation on Eq.~\eqref{eq:Avhw_T_3}. In order for the two operations to agree and Eq.~\eqref{eq:Avhw_T_3} belongs to $A_v^\omega=1$ subspace, we must have that,
	\begin{equation}
	e^{i\phi^\mathcal{G'}_h(\omega)} = \frac{e^{i \phi^{\mathcal{G}}_h (\omega)}\omega \big[ T[\{ g, g_1, g_2 \}] \big] } { \omega \big[ T[\{h \cdot g, h \cdot g_1, g_2 \}] \big]} ,
	\label{eq:phase_single_vertex}
	\end{equation}
which, crucially, is  $g$-independent. We give a graphical proof of Eq.~\eqref{eq:phase_single_vertex} in Fig.~\ref{fig:Aw-identify-DW}, using a 2-3 Pachner move.
	
	\begin{figure}[!t]
		\includegraphics[width=1.0\columnwidth]{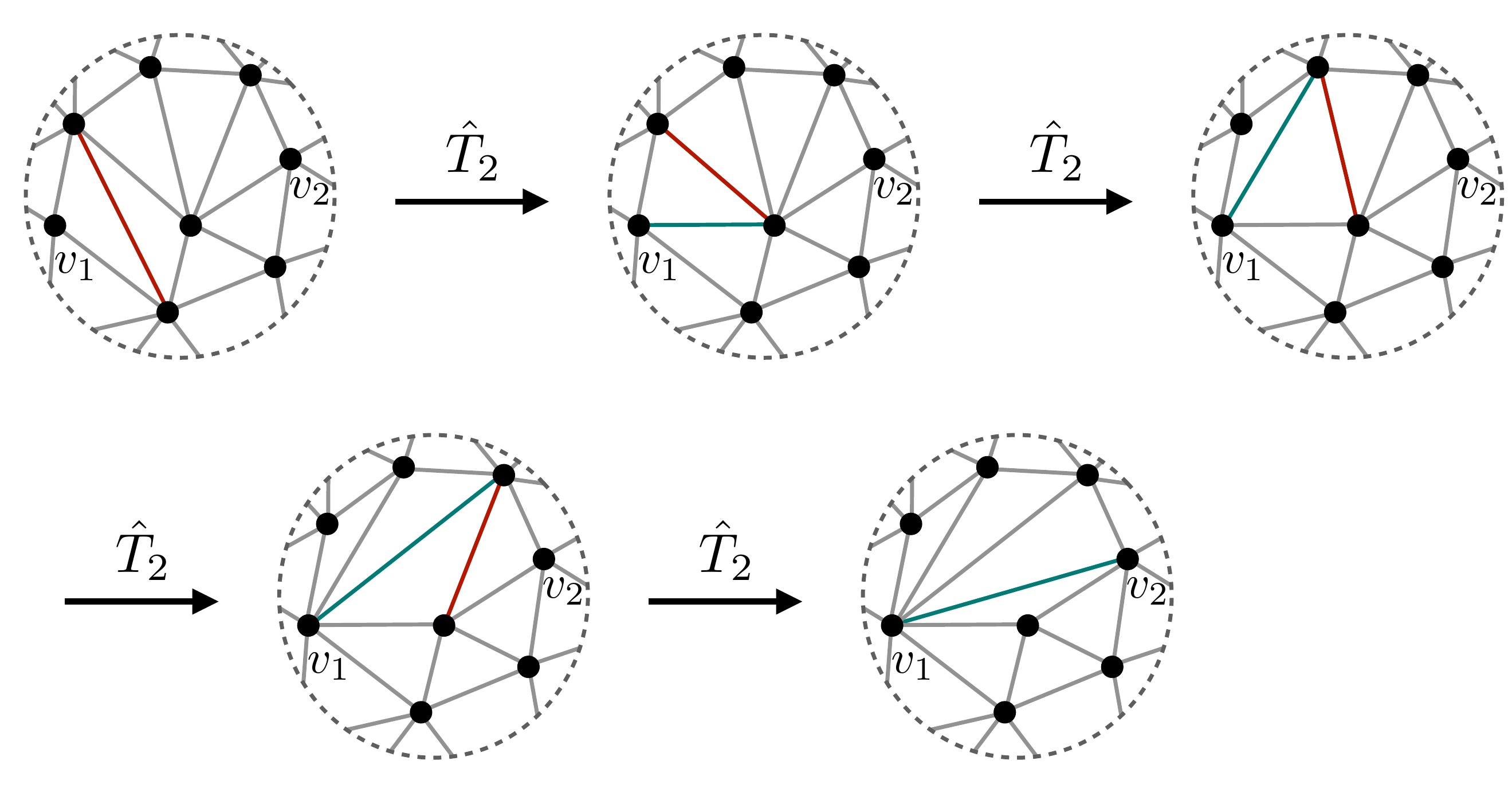}
		\caption{A series of $\hat{T}_2$ lattice isomorphisms that moves two distant vertices $v_1$ and $v_2$ until they are nearest neighbors. We colored the links which are affected by $\hat{T}_2$.} 
		\label{fig:localize-vs-DW}
	\end{figure}
	
	\begin{figure}[!t]
		\includegraphics[width=0.5\columnwidth]{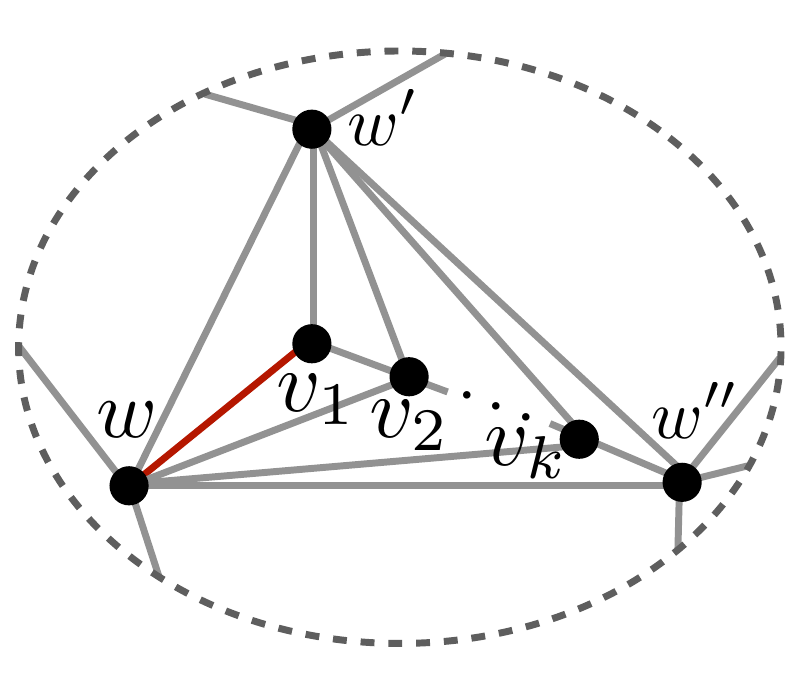}
		\caption{A configuration where the vertices $v_1$, $v_2$, $\cdots$, and $v_k$ associated with the vertex operator $A_{v_1 v_2 \ldots v_k}^{\omega}$ are located locally in space and surrounded by the triangle formed by vertices $w$, $w'$, and $w''$. Since the vertex operator $A_{v_1 v_2 \ldots v_k}^{\omega}$ is generated from link term projections, all the group elements on links $[v_1, v_2]$, $[v_2, v_3]$, $\cdots$, and $[v_{k-1}, v_k]$ are trivial.} 
		\label{fig:k-vertices-DW}
	\end{figure}
	
	Finally, we extend $\hat{T}_3$ to remove not only a single vertex but also a set of vertices associated with a generalized vertex term $A_{v_1 v_2 \ldots v_k}^\omega = \frac{1}{|G|} \sum_{g \in G} A_{v_1}^{g, \omega} A_{v_2}^{g, \omega} \cdots A_{v_k}^{g, \omega}$. In order for the DW models to be well-defined after the vertices removal, we first bring a set of vertices $\{v_1, v_2, \cdots, v_k\}$ locally in space by employing a series of $\hat{T}_2$ lattice isomorphisms as described in Fig.~\ref{fig:localize-vs-DW}. Next, we use $\hat{T}_2$ lattice isomorphisms to make a configuration shown in Fig.~\ref{fig:k-vertices-DW}, where the vertices in $\{v_1, v_2, \cdots, v_k\}$ are surrounded by a triangle. This configuration can be constructed by following the procedure described in Fig.~\ref{fig:vertex-isolation-DW}. 
	
	\begin{figure}[!t]
		\includegraphics[width=1.0\columnwidth]{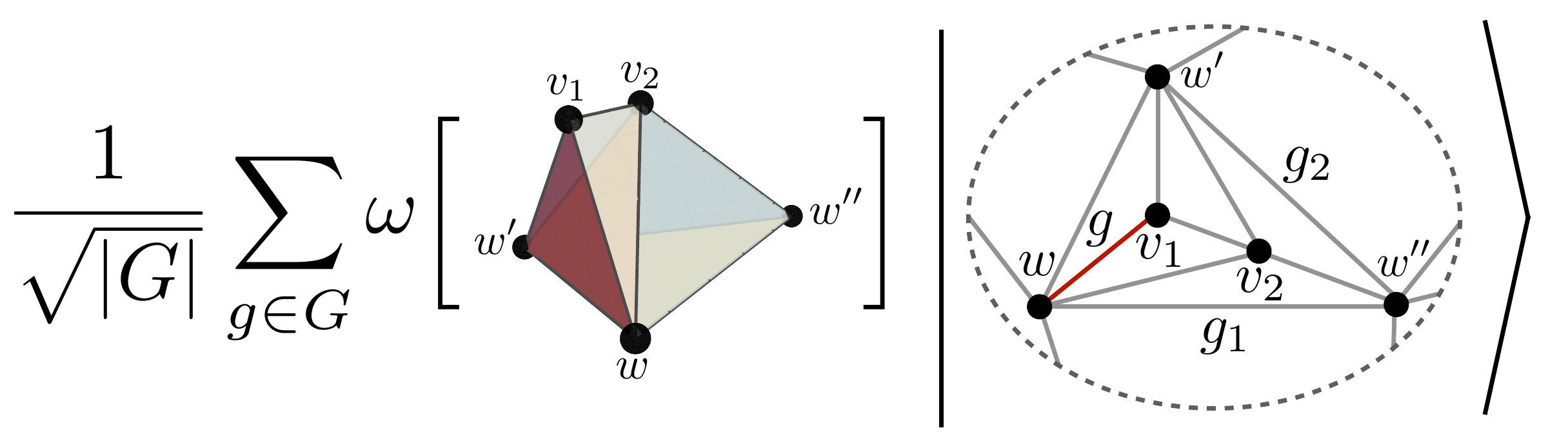}
		\caption{The preferred basis vector for $A_{v_1 v_2}^\omega=1$ subspace. As in the single vertex case, we fix the representative link, denoted as $[v_1, w]$ in this case, at the outset of the calculation. The phase factor is given by the $U(1)$ phase factor of a 3-manifold obtained by gluing two tetrahedrons. For $k$ vertices, the corresponding 3-manifold is obtained by gluing $k$ tetrahedrons.} 
		\label{fig:Av1v2-subspace-DW}
	\end{figure}
	
	\begin{figure*}[!t]
		\includegraphics[width=1.8\columnwidth]{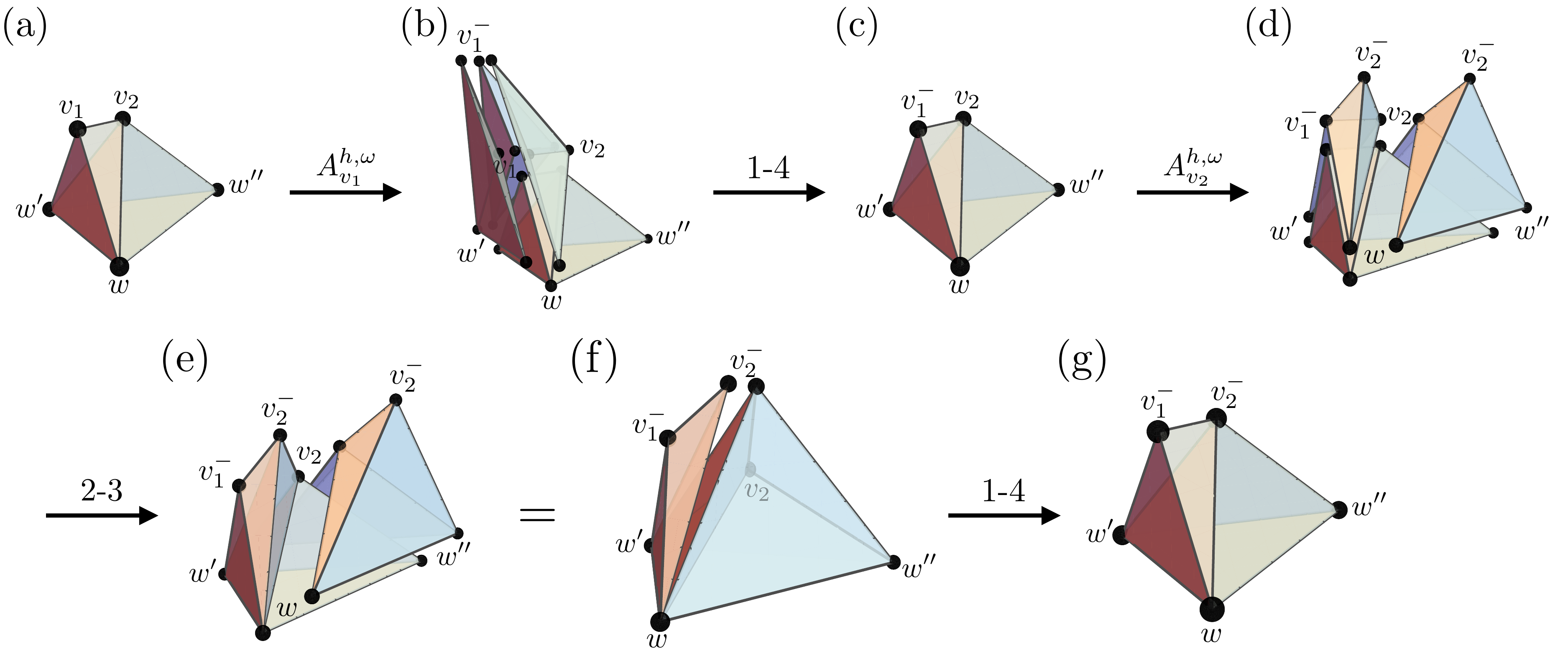}
		\caption{ A graphical proof that the preferred basis state appearing in Eq.~\eqref{eq:Av1v2w-basis} belongs to the subspace $A_{v_1,v_2}^{h, \omega}=1$. (a) $U(1)$ phase factor associated with a specific component of the preferred basis,  $\omega \big[ M[\{g, g_1, g_2\}] \big]$, where $g$ is the state on $[v_1,w]$. (b) Applying $A_{v_1}^{h, \omega}$ adds three tetrahedrons associated with the tent construction. (c) Applying a 1-4 Pachner move simplifies four tetrahedrons into a single one. (d) Applying $A_{v_2}^{h, \omega}$ adds four tetrahedrons associated with the tent construction. (e)  Applying a 2-3 Pachner  move to the left most tetrahedrons. (f) Detaching the left most tetrahedron and gluing together the remaining four tetrahedrons. (g) Applying a 1-4 Pachner move on the right most tetrahedrons, arriving at the final step, which equals $\omega \big[ M[\{h \cdot g, g_1, g_2\}] \big]$, as required. Note that while the states on links $[v_1, v_2]$ and $[v_1^-, v_2^-]$ are trivial, the state on $[v_1^-, v_2]$ is {\it not} trivial in general.} 
		\label{fig:Av1v2-proof-DW}
	\end{figure*}
	
	For concreteness, we also examine the $\hat{T}_3$ lattice isomorphism for the two-vertex case. It is straightforward to generalize this construction to an arbitrary number of vertices. As before, we first identify the preferred basis spanning the $A_{v_1 v_2}^\omega = 1$ subspace, defined in Fig.~\ref{fig:Av1v2-subspace-DW}. Note that since the vertex operator $A_{v_1 v_2}^\omega$ is generated from the link term $C_{[v_1, v_2]}$ decimation, the state on the link $[v_1, v_2]$ is trivial. We denote the preferred basis as
	\begin{equation}
	\frac{1}{\sqrt{|G|}} \sum_{g \in G} \omega \big[ M[\{g, g_1, g_2\}] \big] \vert g; g_1, g_2, \ldots \rangle ,
	\label{eq:Av1v2w-basis}
	\end{equation}
	where $g$ is the basis element on the representative link $[v_1, w]$, $\{g_1, g_2, \ldots\}$ is the set of states on links that are not adjacent to either $v_1$ or $v_2$, and $M$ is the three manifold, obtained by gluing two tetrahedrons, shown in Fig.~\ref{fig:Av1v2-subspace-DW}. We now show that the preferred basis \eqref{eq:Av1v2w-basis} indeed belongs to $A_{v_1 v_2}^\omega =1$ subspace by proving that it is an eigenstate of $A_{v_1}^{h, \omega} A_{v_2}^{h, \omega}$ with the eigenvalue $1$. As in the proof for Eq.~\eqref{eq:Avw-basis}, we show that the amplitude of $\vert g; g_1, g_2, \ldots \rangle$ in Eq.~\eqref{eq:Av1v2w-basis} maps to the amplitude of $\vert h \cdot g; g_1, g_2, \ldots \rangle$ in Eq.~\eqref{eq:Av1v2w-basis} under $A_{v_1}^{h, \omega} A_{v_2}^{h, \omega}$. The proof is given in Fig.~\ref{fig:Av1v2-proof-DW}. 
	
	\begin{figure*}[!t]
		\includegraphics[width=1.8\columnwidth]{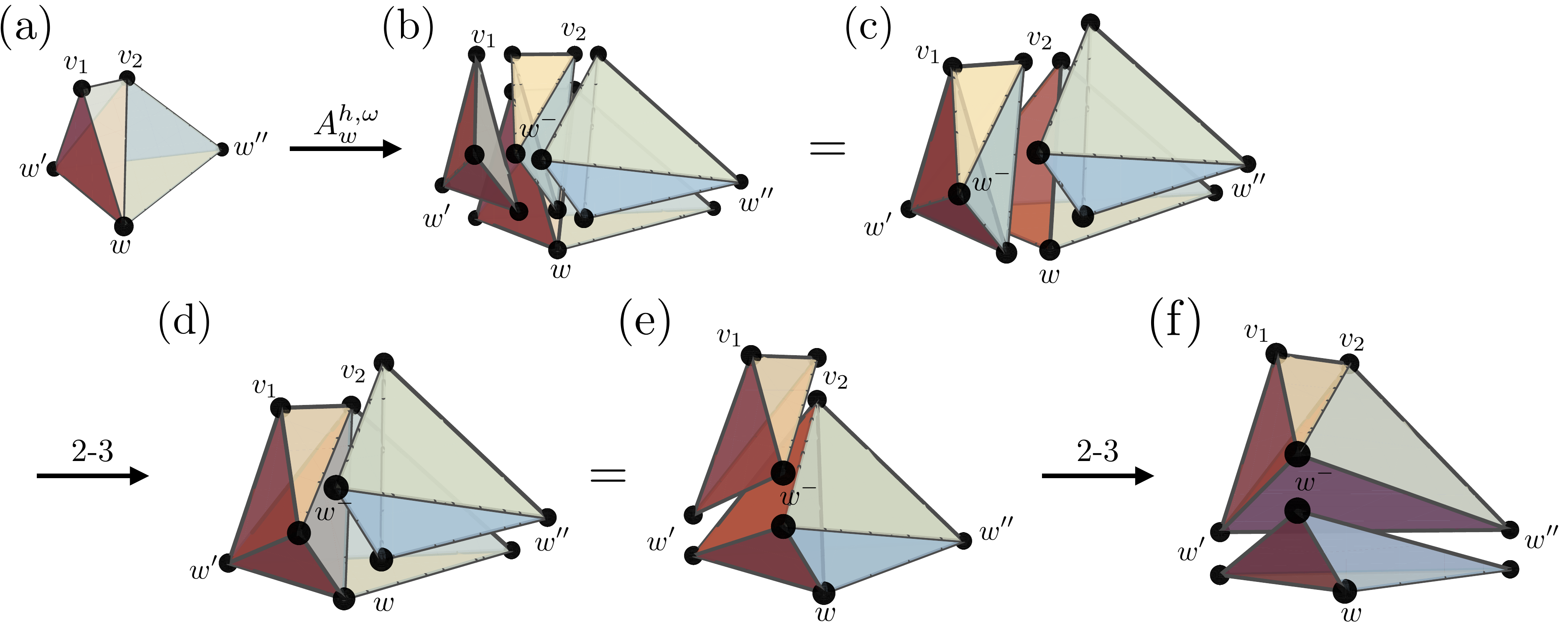}
		\caption{
			A graphical proof that the two operations $A_{v_1,v_2}^{h,\omega}\hat{T}_3$ and $\hat{T}_3 A_{v_1,v_2}^{h,\omega}$ result in the same quantum state on $\mathcal{G}'$. (a) $U(1)$ phase factor associated with a specific component of the preferred basis,  $\omega \big[ M[\{g, g_1, g_2\}] \big]$, where $g$ is the state on $[v_1,w]$. (b) Applying $A_{w}^{h,\omega}$ results in three additional (relevant)  tetrahedrons. (c) Gluing together the left most three tetrahedrons. (d) Applying a 2-3 Pachner move to the three tetrahedrons mentioned in (c). (e) Gluing together the right most three tetrahedrons. (f) Applying a 2-3 Pachner move on the three tetrahedrons mentioned in (e). This concludes our proof that, as in the case of Eq.~\eqref{eq:phase_single_vertex}, the relation  $e^{i\phi^\mathcal{G'}_h(\omega)} = \frac{e^{i \phi^{\mathcal{G}}_h (\omega)}\omega \big[ M[\{ g, g_1, g_2 \}] \big] } { \omega \big[ M[\{h \cdot g, h \cdot g_1, g_2 \}] \big]}$ holds.} 
		\label{fig:Aw-Av1v2-proof-DW}
	\end{figure*}
	
	Finally, we show that the $\hat{T}_3$ lattice isomorphism associated with two-vertex removal correctly maps link and vertex terms in $\mathcal{G}$ to the corresponding ones in $\mathcal{G}'$. Note that the link terms are mapped to each other as in the single vertex removal. As for vertex terms, we need to consider only terms acting on vertices adjacent to $v_1$ and $v_2$. To this end, we show that the matrix representations of vertex operators with respect to the preferred basis in Fig.~\ref{fig:Av1v2-subspace-DW} and the computational basis of $\mathcal{H}_{\mathcal{G}'}$ are identical. For the vertex operator $A_{w''}^{h, \omega}$ acting on $w''$ in Fig.~\ref{fig:Av1v2-subspace-DW}, such identification follows from the procedure presented in Fig.~\ref{fig:Aw-identify-DW}, since we only need to take two tetrahedrons from the tent construction of $A_{w''}^{h, \omega}$ into account. For the vertex term $A_w^{\omega}$ acting on the vertex $w$ in Fig.~\ref{fig:Av1v2-subspace-DW}, the proof of the identification is shown in Fig.~\ref{fig:Aw-Av1v2-proof-DW}. The same construction works for the vertex operator $A_{w'}^\omega$.

	\section{Miscellany on Levin-Wen Models}
	\label{appendix:LW}
	
	\begin{figure}[!t]
		\includegraphics[width=1.0\columnwidth]{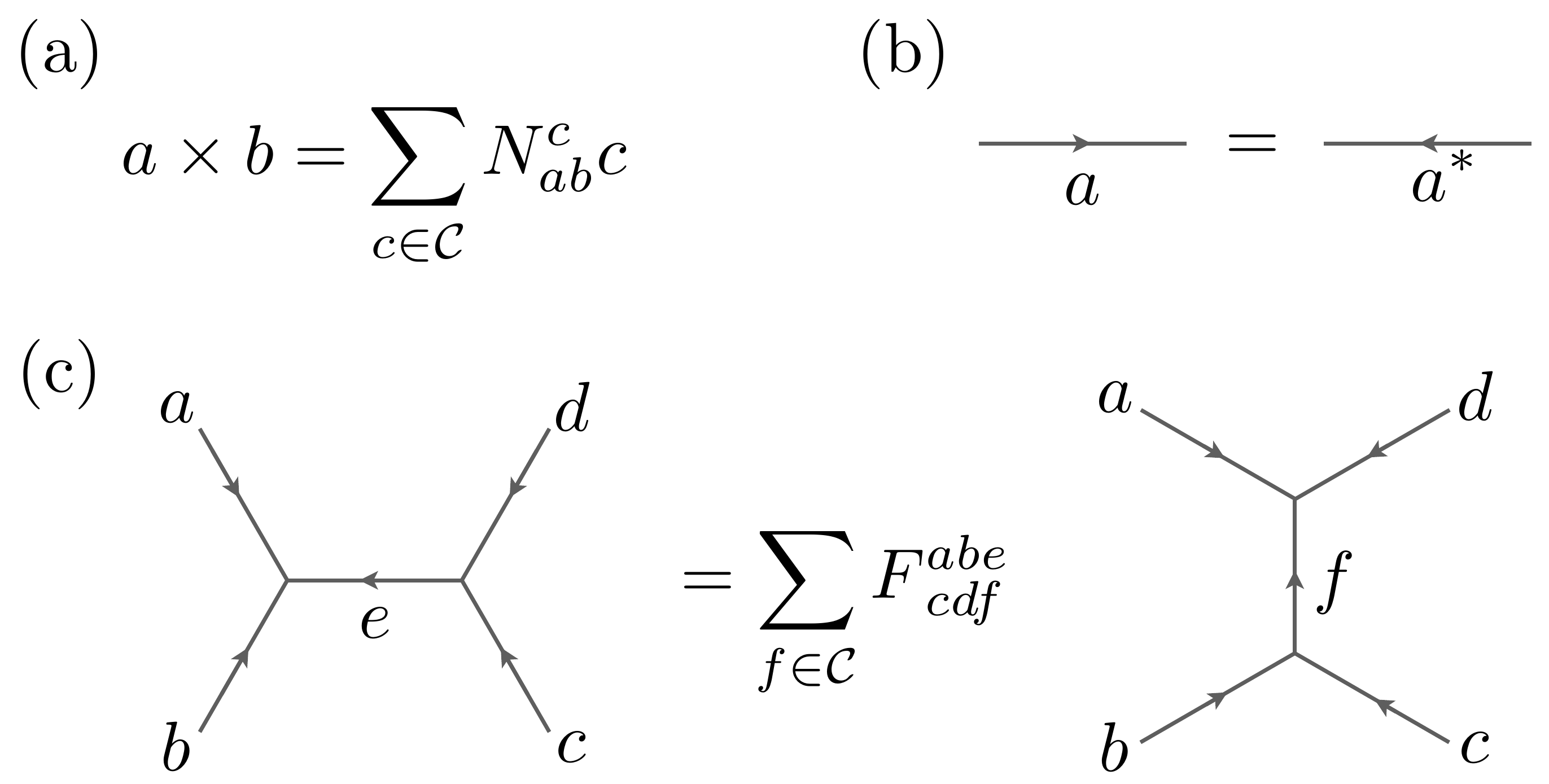}
		\caption{Input data for the Levin-Wen model. (a) The fusion rule among anyons in $\mathcal{C}$. (b) If we invert the direction of a link, we take anti-particle of the corresponding anyon label. (c) $F$-matrix associated with the change in the fusion basis.}
		\label{fig:UFC}
	\end{figure}

	\begin{figure}[!t]
		\includegraphics[width=1.0\columnwidth]{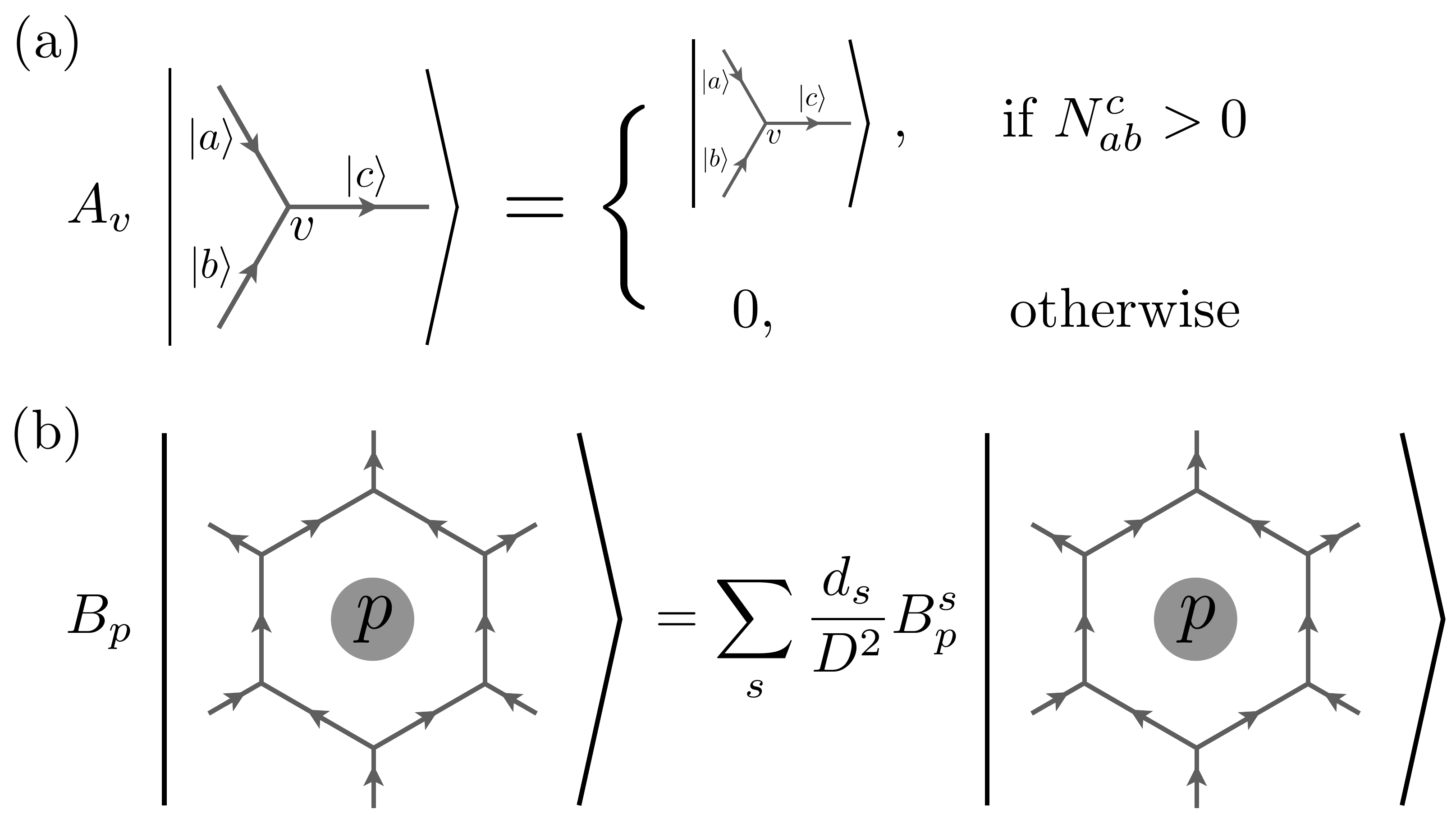}
		\caption{(a) Definition of the vertex term $A_v$ in the Levin-Wen model. $A_v$ projects onto the configuration where three anyons on the adjacent links are compatible with the fusion rule. (b) Definition of the plaquette term $B_p$ in the Levin-Wen model. $B_p$ is given by the summation of $\frac{d_s}{D^2} B_p^s$, where $B_p^s$ operator is defined in Fig.~\ref{fig:LW-B_p^s}.}
		\label{fig:LW-basics}
	\end{figure}
	
	In this section, we summarize the conventions we used for the Levin-Wen model in Sec.~\ref{sec:LW} for completeness of our presentation. While we mainly follow the convention used in the original reference, Ref.~\onlinecite{PhysRevB.71.045110}, there also exist minor deviations, such as using $1$ to denote the trivial anyon. 
	
	The input data for the Levin-Wen model is a unitary fusion category~\cite{wang2010topological}, equivalently a triple $(\mathcal{C}, N, F)$, where $\mathcal{C}$ is the finite set of anyons, $N$ is the fusion coefficients, and $F$ is the collection of unitary $F$-matrices. One can fuse anyons in $\mathcal{C}$ via the fusion rule: $a \times b = \sum_{c \in \mathcal{C}} N_{ab}^c c$. Here, we assume that the fusion rule is commutative, i.e., $N_{ab}^c = N_{ba}^c$, and moreover the fusion coefficient $N_{ab}^c \in \mathbb{Z}^+$ is multiplicity-free, i.e., $N_{ab}^c = 0, 1$ holds. The multiplicity-free condition can easily be dropped; however, we impose the condition for the sake of simplicity. In $\mathcal{C}$, there exists a unique element called the trivial anyon, denoted as $1$, satisfying $a \times 1 = a$ for all $a \in \mathcal{C}$. For each $a \in \mathcal{C}$, there exists unique anyon $a^* \in \mathcal{C}$, the anti-particle of $a$, such that $N_{a a^*}^1 = 1$ holds. Finally, the collection of unitary $F$-matrices is responsible for the basis change appeared in Fig.~\ref{fig:UFC} (c). The collection of $F$-matrices has to satisfy the consistency conditions called the pentagon equations~\cite{wang2010topological}. Furthermore, we assume the so-called tetrahedron symmetries [the second line in Eq.~(9) in Ref.~\onlinecite{PhysRevB.71.045110}] of the $F$-matrices, which in turn simplifies the expressions in the Levin-Wen Hamiltonian.
	
	Similar to lattice gauge theories, the local degrees of freedom lives on every link in the Levin-Wen model. The local Hilbert space is a $|\mathcal{C}|$-dimensional Hilbert space where the computational basis is labeled by anyons in $\mathcal{C}$. Similar to the lattice isomorphism $\hat{T}_1$ of the quantum double model introduced in Appendix~\ref{app:lattice-iso}, we can freely invert the direction of a link $l$ at the expense of mapping the corresponding computational basis $\vert a \rangle_l$ to $\vert a^* \rangle_l$ as described in Fig.~\ref{fig:UFC} (b). The many-body Hilbert space of the Levin-Wen model is given by the tensor product of local Hilbert spaces defined on links. Finally, the Levin-Wen Hamiltonian is given by a sum of commuting projectors
	\begin{equation}
	H_\textrm{LW} = - \sum_v J_v A_v - \sum_p J_p B_p ,
	\end{equation}
	where $A_v$ and $B_p$ are defined in Figs.~\ref{fig:LW-basics} (a) and \ref{fig:LW-basics} (b). In the main text, instead of working on the full Hilbert space, we impose $A_v=1$ on every vertex $v$ as the ``gauge constraint'', and consider the Hamiltonian on the restricted Hilbert space where the gauge constraints are imposed.

	\section{Quantum Monte Carlo Simulations}
	\label{append:QMC}
	
	\begin{figure}[!t]
		\includegraphics[width=0.8\columnwidth]{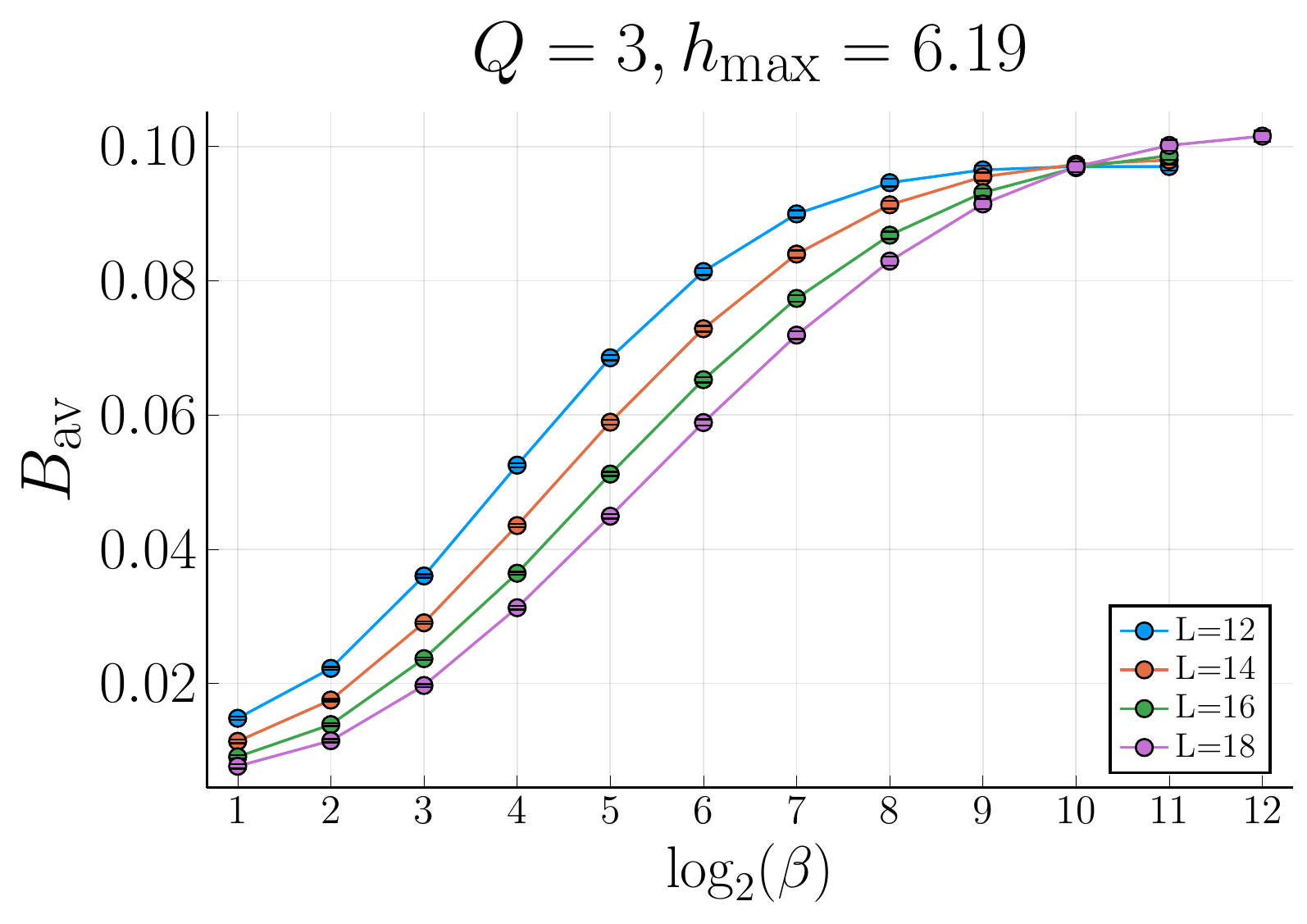}
		\caption{The disorder-average Binder ratio as a function of $\beta$ at $h_{\rm max} = 6.19$ and $Q=3$ for various system sizes.} 
		\label{suppl_Fig_1}
	\end{figure}
	
	\begin{figure}[!t]
		\includegraphics[width=0.85\columnwidth]{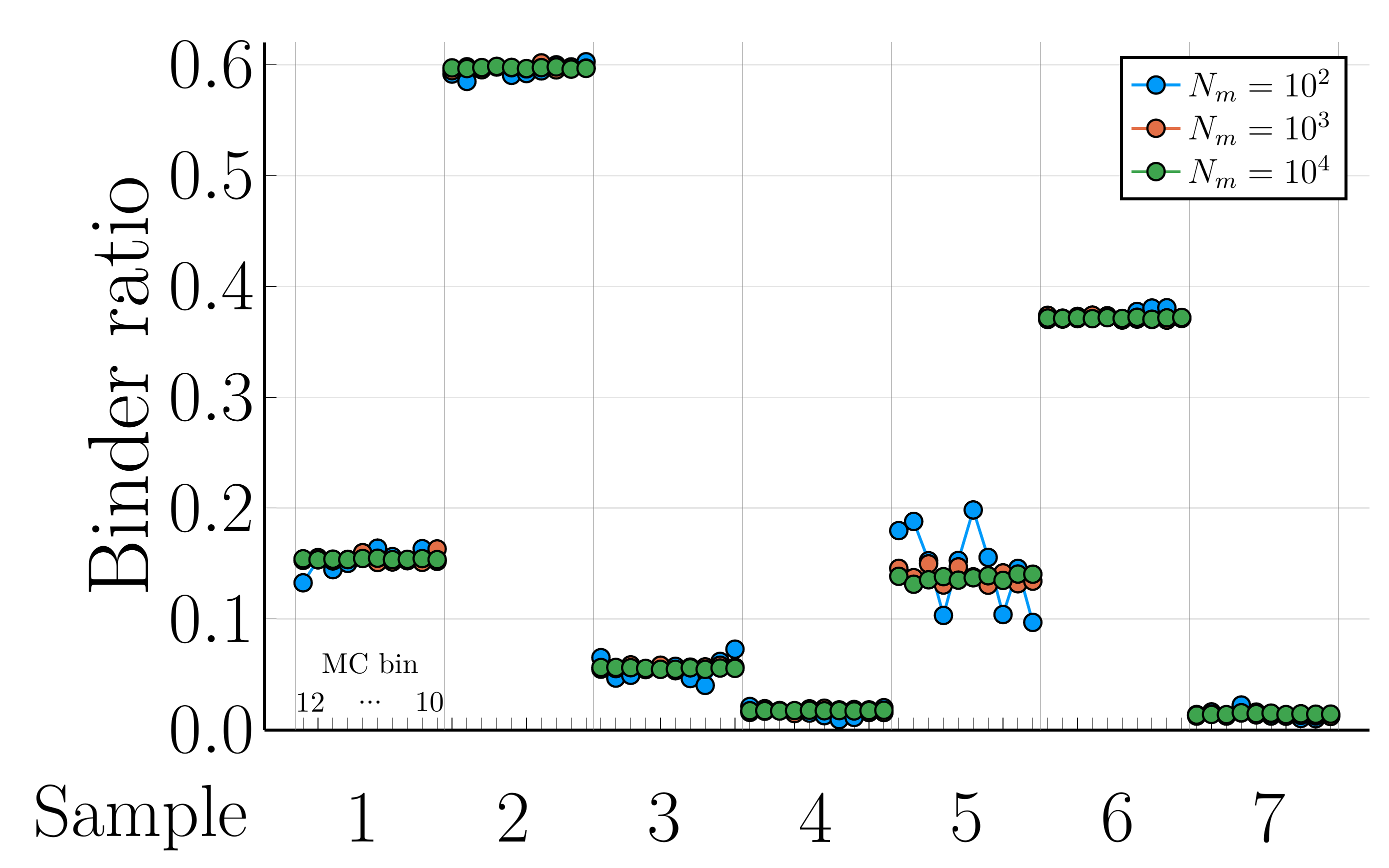}
		\caption{The Binder ratio calculated for system size $L=12$ and $Q=3$ at $h_\textrm{max} = 6.24$ and $\beta_\textrm{max} = 2^{12}$. Different columns correspond for several samples (disorder realizations). For each sample, we show simulation results with different total number of measurements $N_m = 10^2$, $10^3$, and $10^4$ divided $10$ bins.} 
		\label{suppl_Fig_2}
	\end{figure}
	
	\begin{figure}[!t]
		\includegraphics[width=1.0\columnwidth]{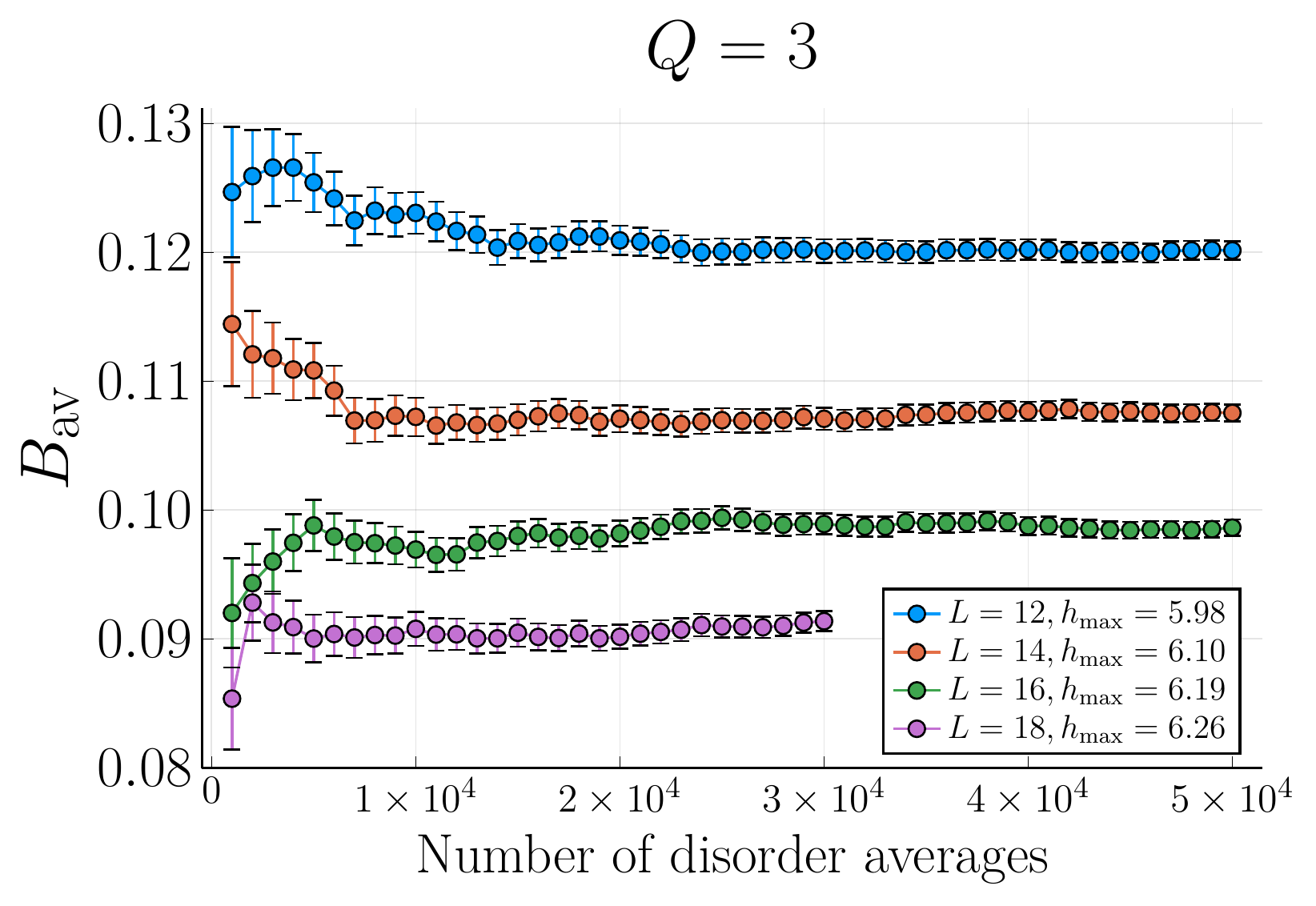}
		\caption{Convergence of the disorder-averaged binder ratio $B_\textrm{av}$ as a function of the number of disorder averages for $Q=3$ for various system sizes $L$ and the field strength $h_\textrm{max}$. } 
		\label{suppl_Fig_3}
	\end{figure}
	In the section, we discuss several technical aspects of our quantum Monte Carlo (QMC) simulation that are relevant to studying the infinite-randomness fixed point ~\cite{PhysRevLett.114.155301}. The expected ``$z=\infty$'' scaling requires a careful monitoring of the convergence to zero-temperature result. To that end, we employ the so-called $\beta$-doubling scheme~\cite{Sandvik_2002, PhysRevLett.114.155301, PhysRevLett.114.255701}, in which one progressively lowers the simulation temperature, akin to simulated annealing, in order to reliably access ground-state properties. To showcase this procedure, in Fig.~\ref{suppl_Fig_1}, we present the disorder-averaged Binder ratio $B _\textrm{av}$ for $Q=3$ as a function of $\beta$ for various system sizes at $h_\textrm{max}/J_\textrm{max}=6.19$. In all cases, we observe a convergence to the zero-temperature result at the largest $\beta$ considered.
	
	An accurate error estimation entails considering the combined effect of disorder averaging (sample to sample variability) and the inherent statistical uncertainty of Monte Carlo sampling. We found that the former dominates our statistical errors. As an example, in Fig.~\ref{suppl_Fig_2}, we demonstrate this behavior, where indeed observe that fluctuations corresponding the average Binder ratio computed for different quenched disorder realization are far larger than statistical fluctuations. This behavior is compatible with the importance of rare-events in the vicinity of the infinite randomness fixed point. For this reason, we also carefully monitor the convergence of disorder averaging as a function of the number of disorder realizations, as shown in Fig.~\ref{suppl_Fig_3}. In all cases, we have taken at least $20,000$ and up to $50,000$ independent disorder realizations.
	
\bibliography{2D_Random_Potts}
	
\end{document}